\documentclass[%
 aip,
 amsmath,amssymb,
 reprint,
]{revtex4-2}

\usepackage[dvipsnames,monochrome]{xcolor}
\usepackage{tabularx}
\usepackage{graphicx}
\usepackage{dcolumn}
\usepackage{bm}
\usepackage[utf8]{inputenc}
\usepackage[T1]{fontenc}
\usepackage{mathptmx}
\usepackage{etoolbox}
\usepackage{array}

\usepackage{placeins}
\usepackage[caption=false]{subfig} 
\usepackage{epsf}
\usepackage{epsfig}
\usepackage{psfrag}
\usepackage{wrapfig}

\usepackage{mathrsfs}
\usepackage{amsthm}
\usepackage{stmaryrd}
\usepackage{bbding}
\usepackage{amsfonts}
\usepackage{esint}

\usepackage{multirow}
\usepackage{diagbox}
\usepackage{verbatim}
\usepackage{makecell}
\usepackage{float}
\usepackage{makeidx}
\usepackage{setspace}

\usepackage[ruled,linesnumbered]{algorithm2e}

\usepackage{hyperref}

\newtheorem{remark}{Remark}[section]

\makeatletter
\def\@email#1#2{%
 \endgroup
 \patchcmd{\titleblock@produce}
  {\frontmatter@RRAPformat}
  {\frontmatter@RRAPformat{\produce@RRAP{*#1\href{mailto:#2}{#2}}}\frontmatter@RRAPformat}
  {}{}
}%
\makeatother

\begin{document}
	
\preprint{AIP/123-QED}

\title{A Two-Temperature Gas-Kinetic Scheme for Hypersonic Non-Equilibrium Flow Computations}
	
		
	
	


\author{Xingjian Gao}
\affiliation{
	Shaanxi Key Laboratory of Environment and Control for Flight Vehicle, 
	State Key Laboratory for Strength and Vibration of Mechanical Structures, 
	School of Aerospace Engineering, 
	Xi'an Jiaotong University, Xi'an 710049, China
}

\author{Xing Ji}
\email{jixing@xjtu.edu.cn}
\affiliation{
	Shaanxi Key Laboratory of Environment and Control for Flight Vehicle, 
	State Key Laboratory for Strength and Vibration of Mechanical Structures, 
	School of Aerospace Engineering, 
	Xi'an Jiaotong University, Xi'an 710049, China
}

\author{Hualin Liu}
\affiliation{
	College of Sciences, China Jiliang University, Hangzhou, China
}

\author{Gang Chen}
\affiliation{
	Shaanxi Key Laboratory of Environment and Control for Flight Vehicle, 
	State Key Laboratory for Strength and Vibration of Mechanical Structures, 
	School of Aerospace Engineering, 
	Xi'an Jiaotong University, Xi'an 710049, China
}
\date{\today}

\begin{abstract}
Accurate aerodynamic and aerothermodynamic predictions are crucial for numerous hypersonic applications. 
This paper proposes a gas-kinetic scheme (GKS) coupled with
a two-temperature kinetic model, which distinguishes between the translational-rotational and 
vibrational modes of temperature. Compared with one-temperature model and the translational-rotational 
multi-temperature model, the proposed model provides a more physically accurate simulation of 
real gas effects when vibrational energy modes of air are excited. 
On the other hand, it is computationally simpler than multi-temperature model with independent 
translational, rotational and vibrational modes. The scheme is implemented on both structured and 
unstructured grids. To further improve the robustness for strong shock and rarefaction waves, 
the discontinuity feedback factor is employed instead of traditional limiters. 
Numerical verifications are conducted on one-dimensional shock structure, two-dimensional (2D) hypersonic flow 
over a cylinder, 2D hypersonic flow over a wedge and 2D Edney Type IV shock/shock interaction. 
Compared with experimental data, the reference results from direct simulation Monte Carlo (DSMC) method 
and Navier--Stokes (NS) solvers, the present method demonstrates accurate prediction of the thermally 
non-equilibrium shock wave structures and hypersonic flow fields.

\end{abstract}

\maketitle


\maketitle

\section{Introduction}
The development of space exploration requires addressing critical 
challenges in hypersonic aerodynamics~\cite{Anderson1989}. During atmospheric reentry of hypersonic 
vehicles, velocities reach several kilometers per second~\cite{Bose2013} 
or even exceed 10 km/s~\cite{Olynick1999}. Under these conditions,
the bow shock formed at the vehicle nose elevates post-shock gas temperatures to thousands of Kelvin,
occasionally surpassing 10,000 K. The surrounding flowfield exhibits complex high-temperature phenomena 
involving internal energy excitation (translational, rotational, vibrational, and electronic modes), 
dissociation, ionization, electronic energy-level transitions, and radiation.
Besides gas-phase processes, surface-related phenomena—including surface catalysis, oxidation, and 
ablation—also play significant roles. These processes critically influence aerodynamic 
characteristics (e.g., stability, thrust, and lift-to-drag ratio) and aerodynamic heating~\cite{Li2017}, 
collectively referred to as high-temperature gas effects.

Due to the simultaneous occurrence of thermodynamic internal energy excitation and chemical reactions 
in high-speed flows, strong coupling interactions arise among these processes. The relaxation of various 
energy modes, chemical reactions, and flow evolution typically occur on comparable time scales~\cite{Kane2022}. 
According to Damköhler number estimates, such conditions often fall into the regime of non-equilibrium. 
Therefore, the coupled behavior of flow and high-temperature gas effects is referred to as 
thermochemical non-equilibrium~\cite{Dong1996,Wang2017,Hao2018}.
Thermodynamic equilibrium assumes all internal energy modes follow a Boltzmann distribution 
based on translational temperature. However, at high altitudes with low air density and 
reduced molecular collision frequencies, finite-rate energy excitation and chemical reactions occur 
with timescales comparable to flow timescales, resulting in thermal non-equilibrium states~\cite{Candler2019}. 
To model thermochemical non-equilibrium in hypersonic flows, multi-temperature models are widely 
adopted~\cite{Appleton1964}. These models assign separate temperatures to distinct energy modes: translational, 
rotational, vibrational, and electronic. Park later simplified the 
multi-temperature model into a two-temperature model, where translational and rotational energies are governed 
by a single temperature $T_{tr}$, while vibrational and electronic energies are described by 
$T_{ve}$~\cite{Park1989-1,Park1989-2}. This model has gained widespread adoption in engineering applications. 

Conventional CFD methods for thermochemical non-equilibrium rely on macroscopic governing equations, 
such as the Navier--Stokes (NS) equations, which are typically solved using finite difference~\cite{Li2023}, 
finite volume~\cite{Li2022}, or finite element methods~\cite{He2023}.
However, these approaches often assume local thermodynamic equilibrium and face difficulties in accurately 
capturing strong non-equilibrium effects in hypersonic flows.
As an alternative, the gas-kinetic scheme (GKS), originally proposed by Xu based on mesoscopic kinetic 
theory~\cite{Xu1998}, provides a physically consistent and numerically robust framework. Unlike traditional 
continuum-based methods, GKS directly evolves the gas distribution function derived from the 
Bhatnagar--Gross--Krook (BGK) model, naturally incorporating non-equilibrium transport and multiscale effects.

Over the years, GKS has been successfully extended to simulate a wide range of flow regimes—from low to 
high Mach numbers—and has demonstrated strong capabilities in handling shock structures, rarefied effects, 
and complex geometries~\cite{Xiao2019,Liu2019,Pan2020,Li2020,Pan2017,Xu2014,Xu2004,Xu2017}. 
Extensive numerical validations have been conducted to demonstrate the accuracy and reliability of 
the method~\cite{Hou2018,Li2011,Xu2002,Xu1995}. In the context of 
hypersonic flows, GKS has shown strong capability in modeling both continuum and near-continuum 
regimes~\cite{Li2009}, and its inherent coupling mechanism allows for accurate prediction of heat 
fluxes~\cite{Tang2005,Liao2007,Ong2006,Li2005}. For turbulence simulations, the high-order GKS~\cite{Li2010} 
serve as promising tools for direct numerical simulation (DNS) of 
turbulent flows~\cite{Cao2019,Cao2019-3T,Cao2021,Kumar2013}. 
Moreover, GKS can be effectively coupled with conventional engineering turbulence models~\cite{Righi2014}, 
and has demonstrated excellent performance in practical turbulent flow 
simulations~\cite{Li2010-2,Li2014,Tan2018,Tan2011}.

For thermal non-equilibrium flows, GKS can be coupled with multi-temperature thermodynamic models. 
Prior developments include a multiple translational temperature GKS~\cite{Xu2005-Trans,Xu2007-Trans}, 
a translational-rotational multi-temperature GKS~\cite{Xu2004-2T,Xu2006-2T,Xu2008-2T,Cao2018-2T} and 
a translational-rotational-vibrational multi-temperature GKS~\cite{Cai2008-3T,Liu2021-3T,Cao2022-3T}. 
The multiple translational temperature GKS models the translational energies along different directions 
using distinct temperature components. The translational-rotational multi-temperature GKS distinguishes 
between translational and rotational energies by representing them with separate translational and 
rotational temperatures. Furthermore, the translational-rotational-vibrational multi-temperature GKS 
incorporates vibrational energy, which is characterized using an additional vibrational temperature. 
These models successfully simulate shock structures, low-density nozzle flows, flat-plate boundary 
layers, and shock/shock interactions. 

Different from the above schemes, this paper proposes a two-temperature GKS that 
distinguishes only between translational-rotational and vibrational energy modes. The translational and 
rotational temperatures are assumed to be in equilibrium, while the vibrational temperature is treated 
as a separate thermodynamic mode. Only the relaxation process from the vibrational temperature toward 
the translational-rotational temperature is considered, and the inter-mode energy exchange is modeled 
using a source term based on the Landau--Teller--Jeans-type relaxation model~\cite{Benettin1997}.
The reason for adopting this model lies in the nature of thermal non-equilibrium in rarefied hypersonic 
flows, which is primarily manifested as a discrepancy between vibrational and translational temperatures. 
In contrast, the rotational temperature tends to rapidly equilibrate with the translational one. This 
physical insight justifies the engineering prevalence of two-temperature models. In fact, the well-known 
Park model also assumes equilibrium between translational and rotational temperatures and remains one of 
the most widely used high-temperature models in practical applications.

It should be noted that this study considers only thermal non-equilibrium effects and neglects chemical 
non-equilibrium by adopting a chemically frozen-flow assumption, i.e., chemical reactions are not included. 
Therefore, for cases with sufficiently high temperatures to trigger chemical reactions (approximately above 
2000 K for air), the applicability of the present method may become limited, which is further examined in this 
study.

The paper is organized as follows. Section~2 introduces the extended gas-kinetic model and the corresponding 
macroscopic governing equations for diatomic gases in two dimensions. Section~3 describes the numerical 
methodology, including the finite volume framework, gas-kinetic solver, spatial reconstruction, 
time integration, and boundary conditions. Section~4 presents hypersonic validation cases, including the
1D shock structure, 2D hypersonic flow over a cylinder, 2D hypersonic flow over a wedge and 
2D Edney Type IV shock/shock interaction. Conclusions are drawn in the final section.

\section{Gas-kinetic models and macroscopic governing equations}

In this section, the extended kinetic model and its derived macroscopic equations in 
two dimensions for diatomic gases are presented. 

\subsection{Non-equilibrium translational-rotational and vibrational temperature model}
The Boltzmann equation describes the behavior of a many-particle kinetic system
through the evolution of a single-particle gas distribution function.
The right-hand side represents binary molecular collisions, which are valid over a wide range of pressures.
The Bhatnagar--Gross--Krook (BGK) model is usually applied for 
the simplification of the collision term in Boltzmann equations~\cite{Bhatnagar1954}. 
In equilibrium flows, all energy modes (translational, rotational, and vibrational) are assumed to share a 
common temperature. However, this assumption becomes inaccurate for non-equilibrium flows 
because of the different temperatures for the translational, rotational and vibrational energy modes. 
In this subsection, we propose a BGK model in which translational and rotational energies are assumed to 
be equilibrated, while the vibrational energy remains in non-equilibrium. Although BGK models for 
non-equilibrium vibrational energy have been introduced in earlier studies, this is the first formulation 
that distinguishes between a translational-rotational equilibrium and vibrational non-equilibrium within a 
BGK framework, and couples it with the Park two-temperature model in the context of GKS.
For the non-equilibrium two-temperature diatomic gas flow, 
the above-mentioned BGK model can be extended in the following form:
\begin{equation}\label{bgk_nonequ}
\frac{\partial f}{\partial t} + u\frac{\partial f}{\partial x} + v\frac{\partial f}{\partial y} 
= \frac{{f}^{eq} - f}{\tau } + \frac{g - {f}^{eq}}{Z_v\tau } 
= \frac{{f}^{eq} - f}{\tau } + {Q}_{s},
\end{equation}
where $f$ is the distribution function, defined as the number density of molecules at the position $(x, y)$ 
and particle velocity $(u, v)$ at time $t$, and $g$ denotes the local equilibrium state, represented as a 
Maxwellian distribution constructed from local macroscopic quantities.
To model thermal non-equilibrium, an intermediate equilibrium distribution $f^{eq}$ is introduced, 
characterized by two distinct temperatures: a translational-rotational temperature and a 
vibrational temperature.
$\tau=\mu/p$ is the characteristic relaxation time ($\mu$ can be computed by Sutherland's
law or by a power law), $Q_s$ is inelastic collision operator, accounts for the energy exchange between 
translational-rotational and vibrational modes, serving as a source in the macroscopic two-dimensional 
flow evolution equations. The coefficient \(Z_v\) is termed the vibrational collision number. 
It is a dimensionless multiplier relating the vibrational relaxation time \(\tau_v\) 
to the characteristic relaxation time \(\tau\) of the total energy, 
i.e., \(\tau_v = Z_v\,\tau\); thus \(Z_v\) scales the overall relaxation timescale 
to the vibrational mode.
The left-hand side of the equation represents the free transport of molecules in 
physical space, while the right-hand side models the relaxation process due to particle collisions.
The intermediate equilibrium state $f^{eq}$ is expressed as follows:
\begin{equation}\label{f^eq}
	\begin{aligned}
{f}^{eq} = &\rho {\left( \frac{{\lambda }_{tr}}{\pi }\right) }^{(K_r+3)/2} 
{e}^{-{\lambda }_{tr}\left\lbrack  {{\left( u - U\right) }^{2} + {\left( v - V\right) }^{2}
+{\xi }_{t}^{2}+{\xi }_{r}^{2} }\right\rbrack}  \\
&{\left( \frac{{\lambda }_{v}}{\pi }\right) }^{K_v/2}{e}^{-{\lambda }_{v}{\xi }_{v}^{2}}.
	\end{aligned}
\end{equation}
Here, $\rho$ is the density, and $(U, V)$ are the macroscopic fluid velocities in the $x$- and $y$- directions. 
where $\lambda_{tr} = m/2kT_{tr}$ is related to the translational-rotational temperature $T_{tr}$
The parameters $\lambda_{v} = m/2kT_{v}$ accounts for the vibrational temperature $T_v$. For two-dimensional 
non-equilibrium diatomic gas, the internal variable \( \xi \) accounts for the translational, 
rotational, and vibrational modes, and has the expression \( \xi^2 = \xi_t^2 + \xi_r^2 + \xi_v^2 \). Here, \( \xi_t \), 
\( \xi_r \), and \( \xi_v \) correspond to translational (in the \( z \)-direction), rotational, and vibrational 
energies, with \( K_t \), \( K_r \), and \( K_v \) degrees of freedom, respectively.
For two-dimensional non-equilibrium diatomic gas, $K_t=1$, $K_r=2$, and $K_v$ are 
determined by the vibrational-energy equation~\cite{Bird1994}. 
\begin{equation}\label{K_v}
	K_{v}=\frac{2\theta_{v}/T_{v}}{e^{\theta_{v}/T_{v}}-1},
\end{equation}
where $\theta_v$ is the vibrational characteristic temperature. For nitrogen, $\theta_v=3393~\mathrm{K}$ 
is used in this paper. Notice that the specific heat ratio is not constant, 
and the BGK solver must compute it locally in each time step for each cell: 
\begin{equation}\label{gamma}
\gamma  = \frac{3 + {K}_{r} + {K}_{v} + 2}{3 + {K}_{r} + {K}_{v}}.
\end{equation}
The RHS collision term in Eq.~\eqref{bgk_nonequ} consists of two terms corresponding to elastic and inelastic 
collisions, respectively, where the relaxation process becomes $f \to f^{eq} \to g$. 
In the elastic collision stage, internal energy exchange between translational-rotational and vibrational 
modes is prohibited. Over a time scale $\tau$, the gas relaxes from its initial non-equilibrium state to 
an intermediate equilibrium state $f^{eq}$, where translational and rotational energies follow a Maxwellian 
distribution characterized by a translational-rotational temperature $T_{tr}$, and the vibrational energy 
follows a Maxwellian distribution at a vibrational temperature $T_v$.

Subsequently, during the inelastic collision stage, energy exchange between translational-rotational and 
vibrational modes occurs over a time scale $Z_v \tau$, and the distribution function further relaxes to 
the final equilibrium state $g$, where all energy modes share a common equilibrium temperature and follow 
a full Maxwellian distribution. The coefficient $Z_v$ is termed the vibrational collision number. 

The relation between mass $\rho$, momentum $(\rho U, \rho V)$, total energy 
$\rho E$, and vibrational energy $\rho E_v$ with the distribution function $f$ is given by
\begin{equation}\label{macro&micro}
\mathbf{W} = \left( \begin{matrix} \rho \\  {\rho U} \\  {\rho V} \\  {\rho E} \\ 
	 \rho {E}_{v} \end{matrix}\right)  = \int {\psi }_{\alpha }{f\mathrm{d}u\mathrm{d}
	 v\mathrm{d}{\xi}_{t}\mathrm{d}{\xi}_{r}\mathrm{d}{\xi}_{v}},\alpha  = 1,2,3,4,5.
\end{equation}
\textcolor{blue}{
$\mathbf{W}$ represents the matrix composed of all the aforementioned conserved quantities together with the 
vibrational energy.}
The vibrational energy $\rho E_v$ can be calculated using the relation $\rho E_v=\frac{K_v}{2}\rho RT_v$. 
The integration is performed over the entire velocity space and internal degrees of freedom space, 
with limits from $-\infty$ to $+\infty$.
The detailed formulas for the moment calculation are in \hyperref[appendixA]{Appendix A}.
$\psi_\alpha$ is the component of the vector for moments as follows:
\begin{equation}
	\begin{aligned}
\psi_\alpha  &= {\left( {\psi }_{1},{\psi }_{2},{\psi }_{3},{\psi }_{4},{\psi }_{5}\right) }^{T} \\
&= {\left( 1,u,v,\frac{1}{2}\left( {u}^{2} + {v}^{2} + {\xi }_{t}^{2} + {\xi }_{r}^{2} 
+ {\xi }_{v}^{2}\right) 
,\frac{1}{2}{\xi }_{v}^{2}\right) }^{T}.
	\end{aligned}
\end{equation}
Using the formulas in \hyperref[appendixA]{Appendix A}, 
the flux expressions can be further obtained as follows for subsequent use.
\begin{equation}\label{Flux}
	\mathbf{F} = \int u{\psi }_{\alpha }{f\mathrm{d}u\mathrm{d}
		 v\mathrm{d}{\xi}_{t}\mathrm{d}{\xi}_{r}\mathrm{d}{\xi}_{v}},\alpha  = 1,2,3,4,5.
\end{equation}
As a separate vibrational temperature $T_v$ is introduced, the constraint of
vibrational energy relaxation has to be imposed on the above
extended kinetic model to self-consistently determine all
unknowns. However, since only mass, momentum, and total energy are conserved during particle 
collisions, while vibrational energy undergoes exchange with translational-rotational modes, 
the original compatibility condition for the collision term is no longer strictly satisfied. 
Instead, a modified compatibility condition is imposed, where the vibrational energy relaxation 
appears as a non-conservative source term in the macroscopic equations.
\begin{equation}\label{cc}
\int(\frac{f^{eq}-f}{\tau}+Q_{s})\psi_{\alpha}\mathrm{d}u\mathrm{d}
v\mathrm{d}{\xi}_{t}\mathrm{d}{\xi}_{r}\mathrm{d}{\xi}_{v}=\mathbf{S}=(0,0,0,0,s)^{T}, 
\end{equation}
where $\alpha=1,2,3,4,5$.
The source term for the vibrational energy $s$ is from the energy
exchange between translational-rotational and vibrational energies during
inelastic collision. Which is modeled through the Landau--Teller--Jeans-type relaxation model as follows,
\begin{equation}\label{source}
s=\frac{(\rho E_v)^{eq}-\rho E_v}{Z_v\tau}.
\end{equation}
The equilibrium energy $(\rho E_{v})^{eq}$ is determined by the
assumption $T_{tr} = T_v = T^{eq}$ such that
\begin{equation}\label{source_cal}
	\begin{aligned}
	\rho E_{v}^{eq}&=\frac{K_{v}}{2}\rho RT^{eq},\\
	T^{eq}&=\frac{(3+K_{r}) T_{tr}+K_{v} T_{v}}{3+K_{r}+K_{v}}.
    \end{aligned}
\end{equation}
The product of the particle collision time $\tau$ and the vibrational collision number $Z_v$ represents the 
relaxation time for the vibrational energy to equilibrate with the translational-rotational energy. 
By default, the value of $Z_v$ is calculated using the following empirical expression~\cite{Wang2017-UGKS}:
\begin{equation}
Z_v=\frac{3+K_{r}}{3+K_{r}+K_{v}}  \hspace{3pt} \frac{c_1}{T_{tr}^\omega} 
\exp \left({\frac{c_2}{T_{tr}^{1/3}}}\right),
\end{equation}
where $c_1 = 100$, $c_2 = 100$, and $\omega = 0.75$. A sensitivity analysis of these parameters 
will be conducted in the subsequent numerical examples.

In order to simulate the flow with any realistic Prandtl number, a modification of the heat flux in the 
energy transport is used in GKS, which is also implemented in the present study.

In the two-temperature GKS, the equilibrium state $g$ 
has been superseded by the intermediate equilibrium state $f^{eq}$ in 
computational implementation. To maintain notation consistency with one-temperature GKS literature, 
the symbol $g$ is hereby formally redefined as $g \equiv f^{eq}$ for all subsequent derivations.

\subsection{Macroscopic equation corresponding to the current method}
Based on the intermediate state given by Eq.~\eqref{f^eq}, with vibrational energy exchange frozen,  
the first-order Chapman--Enskog expansion of the non-equilibrium distribution function $f$ yields the 
following expression~\cite{Chapman1970}:
\begin{equation}
f=f^{eq}+\epsilon f_{1}=f^{eq}-\tau(\frac{\partial f^{eq}}{\partial t}+u\frac{\partial f^{eq}}
{\partial x}+v\frac{\partial f^{eq}}{\partial y}).
\end{equation}
where $\epsilon$ is a small dimensionless quantity.
The corresponding macroscopic non-equilibrium translational-rotational and vibrational 
two-temperature macroscopic equations in two dimensions can be
derived as follows, and the detailed derivation is provided in \hyperref[appendixB]{Appendix B}. 
\begin{equation}\label{NS}
    \frac{\partial W}{\partial t}+\frac{\partial F}{\partial x}+\frac{\partial G}{\partial y}
    =\frac{\partial F_{v}}{\partial x}+\frac{\partial G_{v}}{\partial y}+S,
\end{equation}
with
\begin{equation}
	\begin{gathered}
    W=\begin{pmatrix}\rho\\ \rho U\\ \rho V\\ \rho E\\\rho E_v \end{pmatrix}
    F=\begin{pmatrix}\rho U\\ \rho U^2+p\\ \rho U V\\ \left (  \rho E+p \right )U\\
		\rho E_v U\end{pmatrix}
    G=\begin{pmatrix}\rho V\\ \rho U V\\ \rho V^2+p\\ (\rho E+p)V\\\rho E_v V\end{pmatrix} \\
    F_v=\begin{pmatrix}0\\ \tau_{xx}\\ \tau_{xy}\\ U\tau_{xx}+V\tau_{xy}+q_x\\U\tau_{tr-v}+q_{vx}
	\end{pmatrix}
    G_v=\begin{pmatrix}0\\ \tau_{yx}\\ \tau_{yy}\\ U\tau_{yx}+V\tau_{yy}+q_y\\V\tau_{tr-v}+q_{vy}
	\end{pmatrix},
    \end{gathered}
\end{equation}
where $\rho E = \frac{1}{2} \rho(U^2 + (3+K_r)RT_{tr}+K_v RT_v)$ is the total energy and
$\rho E_{v} = \frac{K_v}{2}\rho RT_v$ is the vibrational energy. The pressure $p$ is related to 
the translational-rotational temperature as $p = \rho RT_{tr}$. In particular, 
the viscous normal stress terms are
\begin{equation}
    \begin{aligned}
    \tau_{xx} =&\tau p[2\frac{\partial U}{\partial x}
    -\frac{2}{3+K_r}(\frac{\partial U}{\partial x}+\frac{\partial V}{\partial y})] \\
    &-\frac{\rho K_v}{2(K_r+K_v+3)Z_v}(\frac{1}{\lambda_{tr}}-\frac{1}{\lambda_{v}}), \\
    \tau_{yy} =&\tau p[2\frac{\partial V}{\partial y}
    -\frac{2}{3+K_r}(\frac{\partial U}{\partial x}+\frac{\partial V}{\partial y})] \\
    &-\frac{\rho K_v}{2(K_r+K_v+3)Z_v}(\frac{1}{\lambda_{tr}}-\frac{1}{\lambda_{v}}), 
    \end{aligned}
\end{equation}
with the viscous shear stress component given by
\begin{equation}
	\tau_{xy}=\tau_{yx}=\tau p(\frac{\partial U}{\partial y}+\frac{\partial V}{\partial x}), 
\end{equation}
and the heat conduction components are expressed as
\begin{equation}
    \begin{gathered}
    q_{x}=\tau p [\frac{K_v}{4}\frac{\partial}{\partial x}(\frac{1}{\lambda_v})
    +\frac{5+K_r}{4}\frac{\partial}{\partial x}(\frac{1}{\lambda_{tr}})], \\
    q_{y}=\tau p[\frac{K_v}{4}\frac{\partial}{\partial y}(\frac{1}{\lambda_{v}})
    +\frac{5+K_r}{4}\frac{\partial}{\partial y}(\frac{1}{\lambda_{tr}})].
    \end{gathered}
\end{equation}
The following terms contribute to the governing equation of vibrational energy $\rho E_v$:
\begin{equation}
	\begin{gathered}
    \tau_{tr-v}=\frac{(K_r+3)\rho K_v}{4(K_r+K_v+3)Z_v}(\frac{1}{\lambda_{tr}}-\frac{1}{\lambda_v}), \\
    q_{vx}=\tau p \frac{K_v}{4}\frac{\partial}{\partial x}(\frac{1}{\lambda_v}), \\
    q_{vy}=\tau p \frac{K_v}{4}\frac{\partial}{\partial y}(\frac{1}{\lambda_v}).
    \end{gathered}
\end{equation}
The source term in Eq.~\eqref{NS} is defined in Eq.~\eqref{source}.
\begin{remark}
To further illustrate the advantages of the proposed TR-V 2T GKS compared with Park's macroscopic two-temperature 
model, we next focus on the formulation of the vibrational energy equation. From the above derivation, the 
vibrational energy equation in the present two-temperature GKS can be expressed as:
\begin{equation}
\frac{\partial}{\partial t}\left(\rho E_{v}\right)+\nabla\cdot\left(\rho E_{v}\mathbf{u}\right)=
\nabla\cdot\left(\mathbf{u}\tau_{t-v}+\kappa_{v}\nabla T_{v}\right)+\mathbf{S},
\end{equation}
where
\begin{equation}\label{kappa_v_GKS}
\kappa_{v}=\mu R_{g}\left(\frac{\theta_{v}}{T_{v}}\right)\frac{1}{\exp\left(\frac{\theta_{v}}{T_{v}}\right)-1}.
\end{equation}
In contrast, the vibrational energy equation employed in conventional Navier-Stokes solvers coupled with Park's model 
takes the form:
\begin{equation}
\frac{\partial}{\partial t}\left(\rho E_{v}\right)+\nabla\cdot\left(\rho E_{v}\mathbf{u}\right)=
\nabla\cdot\left(\kappa_{v}\nabla T_{v}\right)+\mathbf{S},
\end{equation}
where
\begin{equation}
\kappa_{v}=1.37\mu R_{g}\left(\frac{\theta_{v}}{T_{v}}\right)^{2}
\frac{\exp\left(\frac{\theta_{v}}{T_{v}}\right)}{\left({\exp\left(\frac{\theta_{v}}{T_{v}}\right)-1}\right)^2}.
\end{equation}

A direct comparison highlights the key differences. In Park's model, vibrational relaxation enters solely as an 
empirical source term in the energy equation, while the heat conduction is modeled through prescribed vibrational 
conductivity correlations. By contrast, in the TR-V 2T GKS, the separation of translational-rotational and 
vibrational temperatures is embedded directly into the kinetic collision operator and distribution function. As a 
result, the derived vibrational energy equation naturally contains, in addition to the conventional source term, an 
extra relaxation contribution within the divergence term, 
as well as modified transport coefficients (see Eqs.~\ref{kappa_v_GKS}).

These terms are not introduced in an ad-hoc manner but arise consistently from the underlying kinetic formulation. 
This feature ensures a closer physical connection between microscopic relaxation processes and macroscopic transport, 
representing a significant advancement of the present TR-V 2T GKS framework beyond Park's classical two-temperature 
model.
\end{remark}

\section{Numerical Method}
\subsection{Gas-kinetic scheme on the framework of finite volume method}
First, take the moments of the BGK model Eq.~\eqref{bgk_nonequ} in the velocity and 
internal state spaces. Then integrate it over a control volume $\Omega_i$, we obtain
\begin{equation}\label{integrate_bgk}
\begin{split}
{\int }_{{\Omega }_{i}}\int \left( {{f}_{t} + \mathbf{u} \cdot  \nabla f}\right) 
\psi \mathrm{d}\Xi \mathrm{d}V = \\ {\int }_{{\Omega }_{i}}\int 
(\frac{{f}^{eq} - f}{\tau } + \frac{g - {f}^{eq}}{Z_v\tau })  \psi \mathrm{d}\Xi \mathrm{d}V ,
\end{split}
\end{equation}
where $\mathrm{d}\Xi$ denotes 
$\mathrm{d}u \mathrm{d}v \mathrm{d}\xi_t \mathrm{d}\xi_r \mathrm{d}\xi_v$ 
and $\mathrm{d}V$ is the integration of control volume. 
It should be noted that $\nabla f$ represents the divergence of $f$ in physical space, 
which is independent of $(\boldsymbol{u}, \boldsymbol{\xi})$. Therefore, we obtain:
\begin{equation}
\int \left( {\mathbf{u} \cdot  \nabla f}\right) \psi \mathrm{d}\Xi  = \int \nabla  
\cdot  \left( {\mathbf{u}f}\right) \psi \mathrm{d}\Xi  
= \nabla  \cdot  \int \mathbf{u}{f\psi }\mathrm{d}\Xi.
\end{equation}
Based on Eq.~\eqref{macro&micro}, Eq.~\eqref{Flux}, and the modified compatibility condition in 
Eq.~\eqref{cc}, the integral form is obtained from Eq.~\eqref{integrate_bgk}:
\begin{equation}\label{integral_fvm}
{\int }_{{\Omega }_{i}}{\mathbf{W}}_{t}\mathrm{\;d}V + {\int }_{{\Omega }_{i}}\nabla  
\cdot  \mathbf{F}\mathrm{d}V = {\int }_{{\Omega }_{i}} \mathbf{S} \mathrm{\;d}V.
\end{equation}
The integral form in Eq.~\eqref{integral_fvm} is discretized using the finite volume method (FVM),
\begin{equation}
\frac{\mathrm{d}{\overline{\mathbf{W}}}_{i}}{\mathrm{\;d}t} =  - \frac{1}{\left| {\Omega }_{i}
\right| }{\int }_{{\Omega }_{i}}\nabla  \cdot  \mathbf{F}\mathrm{d}V + \mathbf{S},
\end{equation}
where $|{\Omega }_{i}|$ is the volume of the control volume 
and $\overline{\mathbf{W}_i}$ represents the cell-averaged conserved variables in cell $i$.
From Gauss's theorem, the semi-discrete form of FVM is written as
\begin{equation}\label{semi-discrete fvm}
\begin{aligned}
\frac{\mathrm{d}{\overline{\mathbf{W}}}_{i}}{\mathrm{\;d}t} &= \mathcal{L}\left( 
	{\mathbf{W}}_{i}\right)  =  - \frac{1}{\left| {\Omega }_{i}\right| }
	{\oint}_{\partial {\Omega }_{i}}\mathbf{F} \cdot  \mathbf{n}\mathrm{d}s + \mathbf{S}  \\
	&= - \frac{1}{\left| {\Omega }_{i}\right| }\mathop{\sum }
	\limits_{{p = 1}}^{{N}_{f}}{\int }_{{\Gamma }_{ip}}\mathbf{F} \cdot  
	{\mathbf{n}}_{p}\mathrm{\;d}s + \mathbf{S},	
\end{aligned}
\end{equation}
where $\mathcal{L}\left( {\mathbf{W}}_{i}\right)$ represents the cell residual, 
$\partial {\Omega }_{i}$ denotes the boundary of the control volume, 
$\mathrm{d}s$ is the corresponding boundary element (surface area in 3D or line length in 2D), 
and $\mathbf{n}_p$ is the unit outward normal vector of the interface.
$\partial {\Omega }_{i}$ is expressed as the union of all its boundary faces, as given below.
\begin{equation}
	\partial {\Omega }_{i} = \mathop{\bigcup }\limits_{{p = 1}}^{{N}_{f}}{\Gamma }_{ip},
\end{equation}
where ${\Gamma }_{ip}$ is the neighboring interface of the cell ${\Omega }_{i}$, 
${N}_{f}$ is the number of cell interfaces.
Numerical method is used to evaluate the surface integral of fluxes,
\begin{equation}
	{\int }_{{\Gamma }_{ip}}\mathbf{F} \cdot  {\mathbf{n}}_{p}\mathrm{\;d}s = \left| 
		{\Gamma }_{ip}\right| \mathbf{F}\left( {{\mathbf{x}}_{p},t}\right)  \cdot  {\mathbf{n}}_{p},
\end{equation}
where $\left|{\Gamma }_{ip}\right|$ is the area of the mesh face.

\subsection{Gas-kinetic solver}
A finite volume method is used to solve the BGK-type model. The general integral solution of $f$ 
in Eq.~\eqref{bgk_nonequ} at a cell interface $(x_{i+1/2},y_j)$ at time $t$ is expressed as
\begin{equation}
	\begin{aligned}
f&\left( {{x}_{i + 1/2},{y}_{j},t,u,v,{\xi }_{t},{\xi }_{r},{\xi }_{v}}\right)  \\
&= \frac{1}{\tau } \int_{0}^{t} g\left( {{x}^{\prime },{y}^{\prime },
{t}^{\prime },u,v,{\xi }_{t},{\xi }_{r},{\xi }_{v}}\right) {e}^{-\left( {t - {t}^{\prime }}\right) 
/\tau }d{t}^{\prime } \\
&+ {e}^{-t/\tau }{f}_{0}\left( {{x}_{i + 1/2} - {ut},{y}_{j} - {vt},
u,v,{\xi }_{t},{\xi }_{r},{\xi }_{v}}\right),
    \end{aligned}
\end{equation}
where $(x' =x_{i+1/2}-u(t-t'),y' =y_j-v(t-t'))$ is the trajectory of particle
motion, and $f_0$ is the initial gas distribution function at the beginning of each time step.

For viscous flow, the physical collision time $\tau$ is defined as 
\begin{equation}
\tau= \frac{\mu}{p},
\end{equation}
where $\mu$ is the dynamic viscosity. To properly capture discontinuities with additional
numerical dissipation, the numerical collision time is modified as
\begin{equation}\label{tau}
\tau =\frac{\mu }{p} + C\frac{\left| {p}_{L} - {p}_{R}\right| }{\left| {p}_{L} + {p}_{R}\right|}\Delta t,
\end{equation}
where $C$ is set to 5.0 in the computation. $p_L$ and $p_R$ denote the 
pressures on the left- and right-hand sides at the cell interface, 
which reduces to $\tau = \mu/p$ in smooth flow regions. $\Delta t$ is the 
time step determined according to the Courant--Friedrichs--Lewy (CFL) condition.
$\mu$ is the dynamic viscosity coefficient given by Sutherland's law~\cite{Sutherland1893}
\begin{equation}
\mu  = {\mu }_{ref}{\left( \frac{T}{{T}_{ref}}\right) }^{1.5} 
\left( \frac{{T}_{ref} + S}{T + S}\right),
\end{equation}
where $\mu_{ref} = 1.656\times 10^{-5}~\mathrm{Pa}\cdot\mathrm{s}$ for nitrogen, with 
$T_{ref} = 273.11~\mathrm{K}$ and $S = 104.7~\mathrm{K}$.

The initial gas distribution function $f_0$ can be constructed as
\begin{equation}\label{f}
f_0=\begin{cases}g^l(1-t(a^lu+b^lv)-\tau(a^lu+b^lv+A^l)),x\leq0\\
	g^r(1-t(a^ru+b^rv)-\tau(a^ru+b^rv+A^r)),x>0\end{cases},
\end{equation}
where $g^l$ and $g^r$ are related to the macroscopic values reconstructed at the two sides of a cell
interface. The microscopic slopes $a^{l,r},b^{l,r},A^{l,r}$ can be calculated using the 
macroscopic slopes. The specific calculations of microscopic slopes are shown in 
\hyperref[appendixC]{Appendix C}. The above equation can be simplified as
\begin{equation}
f_0=\begin{cases}f_0^l,\quad x\leq0\\
	f_0^r,\quad x>0\end{cases},
\end{equation}
the above equation can be further simplified as follows:
\begin{equation}
f_0=f_0^l\mathbb{H}(x_1)+f_0^r(1-\mathbb{H}(x_1)),
\end{equation}
where $\mathbb{H}(x_1)$ is the Heaviside function. After determining the kinetic part $f_0$, 
the intermediate equilibrium state $g$ part can be expressed as
\begin{equation}\label{g}
\begin{aligned}
\frac{1}{\tau } &\int_{0}^{t} g\left( {{x}^{\prime },{y}^{\prime },
{t}^{\prime },u,v,{\xi }_{t},{\xi }_{r},{\xi }_{v}}\right) {e}^{-\left( {t - {t}^{\prime }}\right) 
/\tau }d{t}^{\prime } \\
& =C_1g^c+C_2a^c (u +v) g^c+C_3 A^c g^c,
\end{aligned}
\end{equation}
where the coefficients $a^c, A^c$ are defined from the expansion of the intermediate equilibrium 
state $g^c$. The coefficients $C_m(\mathrm{where}\ m = 1, 2, 3)$ in Eq.~\eqref{g} are given by
\begin{equation}
C_{1}=1-e^{-t/\tau},C_{2}=(t+\tau)e^{-t/\tau}-\tau,C_{3}=t-\tau+\tau e^{-t/\tau}.
\end{equation}
The details of the calculation of each microscopic term's coefficients $a^{l,r,c}$ and $A^{l,r,c}$ 
in Eq.~\eqref{f} and Eq.~\eqref{g} from macroscopic quantities are given in \hyperref[appendixC]{Appendix C}.
Then the second-order time dependent gas distribution function at a cell interface is
\begin{equation}\label{f_all}
	\begin{aligned}
f&\left( {{x}_{i + 1/2},{y}_{j},t,u,v,{\xi }_{t},{\xi }_{r},{\xi }_{v}}\right) 
= \left( {1 - {e}^{-t/\tau }}\right) {g}^{c} \\
&+ \left( {\left( {t + \tau }\right) {e}^{-t/\tau } 
- \tau }\right) ({a}^{c}u+{b}^{c}v){g}^{c} + \left( {t - \tau  + \tau {e}^{-t/\tau }}\right) 
{A}^{c}{g}^{c} \\
&+{e}^{-t/\tau }{g}^{l}\left\lbrack  {1 - \left( {\tau  + t}\right) 
({a}^{l}u+{b}^{l}v) - \tau {A}^{l}}\right\rbrack  \mathbb{H}\left( {u}_{1}\right) \\ 
&+{e}^{-t/\tau }{g}^{r}\left\lbrack  {1 - \left( {\tau  + t}\right) ({a}^{r}u+{b}^{r}v)
- \tau {A}^{r}} \right\rbrack  \left( {1 - \mathbb{H}\left( {u}_{1}\right) }\right).
\end{aligned}
\end{equation}
The gas distribution function \(f\) is substituted into Eq.~\eqref{Flux} to obtain the flux \(\mathbf{F}\)
in the semi-discrete finite volume formulation of Eq.~\eqref{semi-discrete fvm}.
\begin{remark}
The Prandtl number can take different values for different gases,
e.g., $Pr \approx 0.73$ for air. However, a fixed Prandtl number $Pr = 1$ is provided by
the BGK model. In this thesis, the Prandtl number is fixed by modifying the heat flux 
$q$ in the GKS flux Eq.~\eqref{Flux} according to the method in~\cite{Xu2001},
\begin{equation}
	{\mathbf{F}}_{\rho E}^{new} = {\mathbf{F}}_{\rho E} + \left( {\frac{1}{Pr} - 1}\right) q,
\end{equation}
where $\mathbf{F_{\rho E}}$ refers to the total-energy component of the flux in
Eq.~\eqref{Flux}, and $\mathbf{F_{\rho E}^{new}}$ represents its corrected form,
and the time-dependent heat flux is given by
\begin{equation}
	q=\int\frac{1}{2}(u_1-U_1)[(u_i-U_i)^2+\xi^2]fd\Xi.
\end{equation}
\end{remark}
As mentioned in the previous section, the microscopic slopes $a^{l,r},b^{l,r},A^{l,r}$ can be 
calculated using the macroscopic slopes. The corresponding macroscopic flow variables must be 
reconstructed within each cell. In this study, a second-order reconstruction is employed.
To achieve second-order spatial accuracy, the scheme employs a linear reconstruction of flow 
variables within each computational cell. This reconstruction is applied to both structured and 
unstructured grids. For a given cell \( i \), the value of a conservative variable \( \phi \) at the 
face center \( \mathbf{x}_f \) is reconstructed as
\begin{equation}
	\phi_f = \phi_i + \nabla \phi_i \cdot (\mathbf{x}_f - \mathbf{x}_i),
\end{equation}
where \( \phi_i \) is the cell-averaged value, \( \nabla \phi_i \) is the gradient within the cell, 
\( \mathbf{x}_i \) denotes the cell centroid, and \( \mathbf{x}_f \) is the position vector of the 
face center.

The procedures for gradient computation and slope limiting depend on the grid topology. On structured 
grids, gradients are approximated using central differencing, and the reconstructed slopes are limited 
using both the Van Leer limiter and the discontinuity feedback factor (DFF) to suppress non-physical 
oscillations and enable adaptive order reduction near discontinuities. On unstructured grids, 
the Green--Gauss method is used to compute gradients, and the DFF is applied to ensure stability 
and maintain monotonicity.

In the following sections, the detailed reconstruction strategies for structured and unstructured grids 
are presented separately. All derivations below are presented for 2D cases.

\subsection{Spatial reconstruction on structured grid}
On structured grids, the reconstruction of macroscopic variables is performed using limited directional 
slopes derived from neighboring cell-averaged values. The Van Leer limiter yields a first-order accurate 
slope, which, when used in a linear reconstruction framework, enables second-order accurate interface 
values in smooth flow regions. Meanwhile, the limiter prevents spurious oscillations 
by reducing the slope near strong gradients, thereby contributing to both accuracy and robustness. 
For a conservative variable \( \phi \), the one-sided slopes in the \(x\)-direction are computed as
\begin{equation}
s = \frac{\phi_{i+1} - \phi_i}{x_{i+1} - x_i}, \quad
r = \frac{\phi_i - \phi_{i-1}}{x_i - x_{i-1}},
\end{equation}
where \( \phi_i \) denotes the cell-averaged value in cell \(i\), and \( x_i \) is the coordinate of the 
cell center, and $s$, $r$ represent the one-sided slopes on the right and left sides of 
the cell, respectively. The Van Leer limiter is then defined as
\begin{equation}
L(s, r) = \left( \operatorname{sign}(s) + \operatorname{sign}(r) \right) \cdot \frac{sr}{|s| + |r|},
\end{equation}
where \( \operatorname{sign}(\cdot) \) denotes the sign function, which returns 1 for positive inputs, 
$-1$ for negative inputs, and 0 otherwise. This formulation ensures that the limited slope vanishes 
when \( s \) and \( r \) have opposite signs, thus avoiding the creation of new extrema in regions 
with strong gradients or discontinuities. 
The limited linear reconstruction within cell \(i\) is then given by:
\begin{equation}
\phi(x) = \phi_i + L(s, r) \cdot (x - x_i),
\end{equation}
which provides a directional approximation of the variable distribution along the \( x \)-axis.

In two-dimensional cases, however, the evaluation of numerical fluxes at face centers additionally 
requires the directional derivative in the \( y \)-direction. To obtain this, a similar 
reconstruction is applied, but instead of using neighboring cell centers, the reconstruction 
stencil is formed by the face center under consideration and its neighboring face centers along 
the \( y \)-direction. These face-centered values are themselves obtained from the prior reconstruction 
in the \( x \)-direction. This two-stage reconstruction process ensures consistent and accurate  
directional slope evaluation at each face center for flux calculation.

To further enhance robustness in the presence of strong discontinuities, the DFF is applied as a 
multiplicative correction to the limited slope. The DFF
\( \alpha \in [0,1] \) is computed based on jumps in pressure and Mach number across cell interfaces, 
serving as a smoothness indicator. When strong discontinuities are detected, \( \alpha \) approaches 
zero and the reconstruction is degraded to first-order; in smooth regions, \( \alpha \to 1 \), 
and no additional limiting is applied. The final limited slope is thus given by
\begin{equation}
\tilde{L}(s, r) = \alpha_i^{\star} \cdot L(s, r).
\end{equation}
This combined limiting strategy allows the scheme to maintain second-order accuracy in smooth regions, 
while automatically degrading to first-order accuracy near discontinuities, thereby ensuring both 
accuracy and robustness.

\subsection{Spatial reconstruction on unstructured grid}
The present method is also applicable to unstructured quadrilateral grids, enabling flexible mesh 
generation for complex geometries.
On unstructured grids, the spatial gradients of macroscopic variables are computed using the 
Green--Gauss method. 
The method provides first-order accurate gradients, it enables second-order spatial 
accuracy at cell interfaces when combined with linear reconstruction. 
For a conservative variable $\phi$ that is piecewise continuously differentiable within the control 
volume and across its boundaries, the Green--Gauss theorem gives:
\begin{equation}
\int_{\Omega_i} \nabla \phi \,\mathrm{d}A = \oint_{\partial \Omega_i} \phi\, \mathbf{n}_p \mathrm{d}l,
\end{equation}
where $\mathbf{n}_p$ is the outward unit normal vector at each face of the control volume. 
The differential $\mathrm{d}A$ represents a cell area element, and 
$\mathrm{d}l$ represents a boundary length element.
The cell-center value is a second-order accurate approximation to the cell average,
which allows the cell-averaged gradient to be estimated as
\begin{equation}
\nabla \phi_i = \frac{1}{\left| \Omega_i \right|} \oint_{\partial \Omega_i} \phi\, \mathbf{n}_p\, 
\mathrm{d}l + \mathcal{O}(\Delta^2),
\end{equation}
where $\left| \Omega_i \right|$ is the area of the control volume $\Omega_i$ in two dimensions. 
In practical implementation, the line integral is evaluated using midpoint approximations:
\begin{equation}
\oint_{\partial \Omega_i} \phi\, \mathbf{n}_p\, \mathrm{d}l \approx \sum_{p=1}^{N_f} \tilde{\phi}
_{\text{mid},p}\, \mathbf{n}_p\, \Delta l_p + \mathcal{O}(\Delta^3),
\end{equation}
where $N_f$ is the number of faces of cell $i$, $\Delta l_p$ is the length of face $p$, and 
$\tilde{\phi}_{\text{mid},p}$ is the value of $\phi$ at the midpoint of face $p$. 
This gives the gradient estimate:
\begin{equation}
\nabla \phi_i = \frac{1}{\left| \Omega_i \right|} \sum_{p=1}^{N_f} \tilde{\phi}_{\text{mid},p}\, 
\mathbf{n}_p\, \Delta l_p + \mathcal{O}(\Delta).
\end{equation}
The key step is the evaluation of the face-centered value $\tilde{\phi}_{\text{mid},p}$.  
Assume that cells $\Omega_1$ and $\Omega_2$ lie on both sides of face $f$, with cell-centered values 
$\phi_1$ and $\phi_2$, respectively. 
In this study, the face-centered value is approximated by a simple arithmetic average:
\begin{equation}
\tilde{\phi}_{\text{mid},p} = \frac{1}{2}(\phi_1 + \phi_2),
\end{equation}
which corresponds to the assumption that the face center lies at the midpoint between the two adjacent 
cell centers. This condition is typically met in mildly distorted unstructured meshes. 
When combined with a properly reconstructed slope in each cell, this approximation provides 
a straightforward and robust second-order accuracy. Although it may not strictly maintain 
second-order precision on highly skewed or irregular grids, it is still widely adopted in 
engineering computations due to its simplicity and acceptable performance in most practical applications.

To improve robustness in the presence of strong discontinuities, the computed gradient is further 
modified by applying the DFF as follows:
\begin{equation}
	\widetilde{\nabla \phi_i} = \alpha_i \nabla \phi_i.
\end{equation}

\subsection{Discontinuity feedback factor}
To deal with possible discontinuities in the flow field, Ji \textit{et al.}~\cite{Ji2021} proposed 
an indicator to measure the strength of interface discontinuities, based on reconstructed
values of the interface, which is called the discontinuity feedback factor (DFF). When discontinuities 
are detected in the reconstruction stencil, the DFF causes the high-order polynomial to automatically 
degrade to first-order accuracy.
thus improving the robustness of the algorithm. 
For a two-dimensional quadrilateral grid, the DFF \(\alpha_{i}\) is first computed for each 
targeted cell \(\Omega_{i}\) as
\begin{equation}
    \alpha_{i} = \prod_{n=1}^{4} \alpha_n,
\end{equation}
where \(\alpha_n\) is the discontinuity feedback factor at the center of the \(n\)th interface of 
cell \(\Omega_{i}\).

To further improve the discontinuity detection, an additional step is applied: the final DFF 
used in the reconstruction is calculated as the product of \(\alpha_{i}\) over all cells within 
the reconstruction stencil centered at \(\Omega_{i}\). This means the overall DFF at cell 
\(\Omega_{i}\) is
\begin{equation}
    \alpha^{\star}_{i} = \prod_{j \in \mathcal{S}(i)} \alpha_j,
\end{equation}
where \( \mathcal{S}(i) \) denotes the set of indices of all cells in the reconstruction stencil 
centered at cell \( \Omega_i \).
Then, the updated slope is then modified as
\begin{equation}
	{\widetilde{\nabla\mathbf{W}}}_{i}^{n + 1} = {\alpha }^\star_{i}{\nabla\mathbf{W}}_{i}^{n + 1}.
\end{equation}
The discontinuity feedback factor at the center of interface is defined as
\begin{equation}
	{\alpha_n } = \frac{1}{1+D^2_n},
\end{equation}
with
\begin{equation}
	D = \frac{\left| {p}^{l} - {p}^{r}\right| }{{p}^{l}} 
	+ \frac{\left| {p}^{l} - {p}^{r}\right| }{{p}^{r}} + {\left( {\mathrm{{Ma}}}_{n}^{l} 
	- {\mathrm{{Ma}}}_{n}^{r}\right) }^{2} + {\left( {\mathrm{{Ma}}}_{t}^{l} 
	- {\mathrm{{Ma}}}_{t}^{r}\right) }^{2},
\end{equation}
where $p^l,p^r$ denote the left and right pressure of the center of interface $x_n$, 
$\mathrm{Ma}^l_n$ and $\mathrm{Ma}^l_t$ represent the left-side Mach numbers defined based on the 
normal and tangential velocities, respectively. They can take negative values to indicate flow 
direction relative to the interface orientation. 
$\mathrm{Ma}^r_n$ and $\mathrm{Ma}^r_t$ are the corresponding right-side values.
For smooth flows, $a \to 1$, meaning no additional limiting is applied to the reconstruction, when strong 
discontinuities are present, $a \to 0$, and the reconstruction is reduced to first-order accuracy.

\subsection{Time integration and local time stepping}

The spatial discretization is given by the semi-discrete finite volume formulation shown in 
Eq.~\eqref{semi-discrete fvm}.
For time advancement, an explicit single-step scheme is employed:
\begin{equation}
	\overline{\mathbf{W}}_i^{n+1} = \overline{\mathbf{W}}_i^{n} + \int_{t^n}^{t^{n+1}} 
	\mathcal{L}(\mathbf{W}_i(t)) \, dt,
\end{equation}
where \(\overline{\mathbf{W}}_i^{n}\) denotes the cell-averaged conservative variables at time 
step \(n\), and \(\mathcal{L}(\mathbf{W}_i(t))\) represents the spatial residual operator, 
as formulated in the integral discretization of Eq.~\eqref{semi-discrete fvm}.
This method is straightforward and efficient, and when combined with a time-accurate 
gas-kinetic flux function, it achieves second-order temporal accuracy.

Traditionally, a global time stepping method is adopted, where a uniform time step \(\Delta t\) is 
determined by the most restrictive stability condition across the entire computational domain:
\begin{equation}
\Delta t = \min_{i} \left( \frac{\mathrm{CFL} \cdot \Delta x_i}{|\mathbf{u}_i| + c_i} \right),
\end{equation}
where \(\mathrm{CFL}\) is the Courant--Friedrichs--Lewy number controlling stability, \(\Delta x_i\) 
denotes the characteristic length of cell \(\Omega_i\), \(\mathbf{u}_i\) is the local flow velocity, 
and \(c_i\) is the local speed of sound. 
For structured grids, \(\Delta x_i\) is taken as the minimum grid spacing in the \(x\)- and \(y\)-
directions, i.e.,
\begin{equation}
\Delta x_i = \min(\Delta x, \Delta y),
\end{equation}
while for unstructured grids, it is defined as the ratio of the cell area to the length of the longest edge:
\begin{equation}
\Delta x_i = \frac{\left| \Omega_i \right|}{L^{\max}_i}.
\end{equation}
Although the global time stepping method is stable, it can be overly restrictive, 
as the smallest cells control the time step, thereby slowing down convergence.

To enhance convergence efficiency for steady-state simulations, a local time stepping (LTS) strategy is 
applied. In LTS, each cell independently advances with its own local time step:
\begin{equation}
\Delta t_i = \frac{\mathrm{CFL} \cdot \Delta x_i}{|\mathbf{u}_i| + c_i}.
\end{equation}
This approach allows larger time steps in cells where stability constraints are less restrictive, thereby 
accelerating convergence while maintaining numerical stability.

Within the GKS framework, the flux function inherently couples space and time through integration of 
the time-evolving gas distribution function over a fixed time interval \([t^n, t^{n+1}]\). 
This interval is determined by a global time step \(\Delta t\), which is used uniformly in the flux 
evaluation to maintain consistency across cell interfaces. 
However, since each cell advances with its own local time step \(\Delta t_i\) under LTS, the 
update of the cell-averaged conservative variables must reflect this local progression. 
As a result, the net contribution of the spatial residual \(\mathcal{L}(\mathbf{W}_i)\) is rescaled 
according to the ratio \(\Delta t_i / \Delta t\), leading to the following update formula:
\begin{equation}
\overline{\mathbf{W}}_i^{n+1} = \overline{\mathbf{W}}_i^{n} + 
\frac{\Delta t_i}{\Delta t}\int_{t^n}^{t^{n+1}} \mathcal{L}(\mathbf{W}_i(t)) \, dt.
\end{equation}
This treatment enables each cell to evolve efficiently with its own stability-constrained time step, 
while preserving the globally coupled flux structure derived from the time-accurate gas-kinetic formulation.

\subsection{Wall boundary condition}
In the near-continuum regime, as the flow becomes increasingly rarefied, intermolecular collisions near the 
wall become insufficient to equilibrate gas molecules with the wall conditions. This invalidates the classical 
no-slip boundary condition and gives rise to slip conditions, in which the gas velocity and temperature near 
the wall differ from those of the wall. These slip boundary conditions are adopted in the present study.

Maxwell was the first to derive the slip boundary condition, as discussed in Ref.~\cite{Lockerby2004}. 
For an isothermal wall, the temperature gradient can be neglected, and the simplified form of the Maxwell 
slip condition is given by:
\begin{equation}
U_s=U_0-U_w=A\left(\frac{2-\sigma}{\sigma}\right)
\lambda\left.\frac{\partial u_t}{\partial n}\right|_{n=0},
\end{equation}
where $U_s$ is the velocity slip, $U_0$ represents the tangential velocity of the fluid at the wall, 
and $U_w$ is the wall velocity. For a stationary wall, $U_w=0$. 
$A =\sqrt{2/\pi}$ is a constant of proportionality, 
$\sigma$ is the tangential momentum accommodation coefficient, 
$u_t$ is the velocity in the surface tangential direction, 
$\lambda$ is the mean free-path, which is calculated from typical gas flow properties as~\cite{Maccormack1989} 
\begin{equation}
\lambda=\frac{2\mu}{\rho\bar{c}}=\frac{\mu}{\rho}\sqrt{\frac{\pi}{2RT}}\,,
\end{equation}
where $\mu$ is the viscosity, $\rho$ is the mass density and $\bar{c}$ is the mean molecular speed.
The boundary condition for the translational-rotational temperature jump is similarly simplified 
as follows~\cite{Maccormack1987}
\begin{equation}\label{T_{tr}^j}
T_{tr}^j=T_{tr}-T_{tr}^w=\frac{2-\alpha}{\alpha}\frac{2\gamma}{(\gamma+1)\Pr}\lambda
\left.\frac{\partial T_{tr}}{\partial n}\right|_{n=0},
\end{equation}
where $T_{tr}^w$ is the wall translational-rotational temperature, $T_{tr}$ is the translational-rotational 
temperature of the gas at the wall (and where $T_{tr}-T_{tr}^w$ is the temperature jump), $\alpha$ is the 
thermal accommodation coefficient, 
Pr is the Prandtl number, $\gamma$ is the speciflc heat ratio.
Similarly, the vibrational temperature jump condition is given as follows:
\begin{equation}
	T_{v}^j=T_{v}-T_{v}^w=\frac{2-\alpha}{\alpha}\frac{2\gamma}{(\gamma+1)\Pr}Z_v \lambda 
	\left.\frac{\partial T_v}{\partial n}\right|_{n=0},
\end{equation}
where $T_{v}^w$ is the wall vibrational temperature, $T_{v}$ is the vibrational temperature of the gas 
at the wall. In this study, it is assumed that a fully diffuse wall ($\sigma = 1$) that is 
also thermally accommodating ($\alpha = 1$).

In this study, ghost cells are employed to enforce boundary conditions. The process of calculating the 
physical quantities in the ghost cell is as follows. 

\noindent 
1) The fluid properties at the wall are determined using velocity slip and temperature jump 
conditions. Taking velocity slip as an example, the velocity gradient on the right-hand side of the 
slip equation can be expressed as follows:
\begin{equation}
	U_s=U_0=A\left(\frac{2-\sigma}{\sigma}\right)
	\lambda\left.\frac{\partial u_x}{\partial n}\right|_{n=0}
	=A\left(\frac{2-\sigma}{\sigma}\right)
	\lambda \frac{U_1-U_0}{\Delta x}.
\end{equation}
Here, $U_1$ denotes the tangential velocity in the first fluid cell adjacent to the wall, and 
$\Delta x$ represents the normal distance from this cell to the wall.
From this equation, $U_0$ (the fluid velocity at the wall) can be solved. Similarly, the wall temperature 
can be obtained using the temperature jump condition in the same manner. 

\noindent
2) Once the wall fluid velocity and temperature are determined, the physical quantities in the ghost cell 
can be computed based on the following symmetry relation, assuming that the pressure gradient at the wall 
is zero, and thus the pressure on both sides of the wall is assumed to be equal. 
\begin{equation}
\begin{aligned}
U_{-1}&=-U_1, \\ V_{-1}&=2V_0-V_1, \\ P_{-1}&=P_1, \\ T_{tr}^{-1}&=2T_{tr}^{0}-T_{tr}^{1}, \\
T_{v}^{-1}&=2T_{v}^{0}-T_{v}^{1}.
\end{aligned}
\end{equation}

\section{Numerical examples}
In the following, the proposed method is referred to as TR-V 2T GKS. The numerical examples include 
one-dimensional (1D) and two-dimensional (2D) test cases, and the abbreviations 1D and 2D will be used hereafter for brevity.

\subsection{1D shock structure}
Accurately computing the inner structure of normal shock waves is crucial for many
hypersonic applications. This section computes the flow of one-dimensional nitrogen in vibrationally 
non-equilibrium across a planar shock wave. Reference numerical results were obtained 
by the present authors using the one-temperature GKS and by Cai \textit{et al.}~\cite{Cai2008-3T} 
employing the translational-rotational-vibrational multi-temperature GKS approach.
In the following, 3T GKS refers to the multi-temperature GKS proposed by Cai \textit{et al.}~\cite{Cai2008-3T}, 
while 1T GKS denotes the one-temperature gas-kinetic scheme.
The freestream gas is nitrogen, and the initial conditions are specified as:
\begin{equation}
M_{\infty}=5, \quad T_{\infty}=226.149\,\mathrm{K}, \quad \rho_{\infty}=1.7413 
\times 10^{-2}\,\mathrm{kg/m^3}. 
\end{equation}
Since the initial temperature is far below the vibrational excitation temperature of nitrogen, 
it can be assumed that the initial vibrational temperature is equal to the translational-rotational 
temperature, i.e., $T_v=T_{tr}=T_{\infty}$. 
This assumption is also applied in the subsequent test cases presented in this study.
Accurately prescribing the post-shock equilibrium state as an appropriate initial and 
downstream boundary condition has a significant impact on the simulation. Since vibrational 
excitation is considered, the specific heat ratio $\gamma$ is no longer constant, 
and the classical Rankine--Hugoniot relations fail to provide the correct post-shock equilibrium state. 
Instead, the generalized Rankine--Hugoniot relations must be employed, as given below.
\begin{equation}
	\begin{gathered}
\frac{{T}_{2}}{{T}_{1}} = \frac{\left\lbrack  {{\gamma }_{1}/\left( {{\gamma }_{1} - 1}\right) }
\right\rbrack   + \left( {{\gamma }_{1}/2}\right) {M}_{1}^{2}}{\left\lbrack  {{\gamma }_{2}/
\left( {{\gamma }_{2} - 1}\right) }\right\rbrack   + \left( {{\gamma }_{2}/2}\right) {M}_{2}^{2}}, \\
\frac{{u}_{2}}{{u}_{1}} = \sqrt{\frac{{\gamma }_{2}}{{\gamma }_{1}}}\frac{{M}_{2}}{{M}_{1}}
\sqrt{\frac{\left\lbrack  {{\gamma }_{1}/\left( {{\gamma }_{1} - 1}\right) }\right\rbrack   
+ \left( {{\gamma }_{1}/2}\right) }{\left\lbrack  {{\gamma }_{2}/\left( {{\gamma }_{1} - 1}\right) }
\right\rbrack   + \left( {{\gamma }_{2}/2}\right) }}, \\
\frac{{p}_{2}}{{p}_{1}} = \frac{1 + {\gamma }_{1}{M}_{1}^{2}}{1 + {\gamma }_{2}{M}_{2}^{2}}, \\
\frac{\left(1+\gamma_1M_1^2\right)^2}{\left\{[\gamma_1/(\gamma_1-1)]+(\gamma_1/2)M_1^2\right\}
\gamma_1M_1^2}\\
=\frac{\left(1+\gamma_2M_2^2\right)^2}{\left\{[\gamma_2/(\gamma_2-1)]
+(\gamma_2/2)M_2^2\right\}\gamma_2M_2^2}.
	\end{gathered}
\end{equation}
The derivation process is similar to that for the classical Rankine--Hugoniot relation, 
which can be found in many textbooks, such as those in Refs.~\cite{Saad1985,Liepmann2001}.
The above equations provide an appropriate approach for determining the downstream boundary conditions. 
To compute the downstream boundary conditions, first, with a specific $\gamma_2$, use $M_1$ and $\gamma_1$ 
to compute an intermediate post-shock Mach number $M_2$. Second, use this Mach number $M_2$ and the 
specific heat ratio $\gamma_2$ to determine a post-shock temperature $T_2$. Third, use Eq.~\eqref{K_v} 
and Eq.~\eqref{gamma} to determine a new specific heat ratio $\gamma_2$. Repeat the above three steps 
until convergence is achieved within an appropriate tolerance.

To enable direct comparison with the reference results, the vibrational collision number $Z_v$ is set 
to 100, consistent with the original setup.
The dynamic viscosity coefficient is defined following the same formulation:
\begin{equation}
\mu  = {1.656} \times {10}^{5} {\left( \frac{T}{273}\right) }^{0.74}.
\end{equation} 
The numerical viscosity coefficient $C$ in Eq.~\eqref{tau} is set to 1, 
in accordance with the reference method.
The computational domain spans a total length of 80 mean free paths and is discretized 
into 300 uniform spaced cells.
\begin{figure*}[htp]	
	\centering
	\includegraphics[width=0.4\textwidth]
	{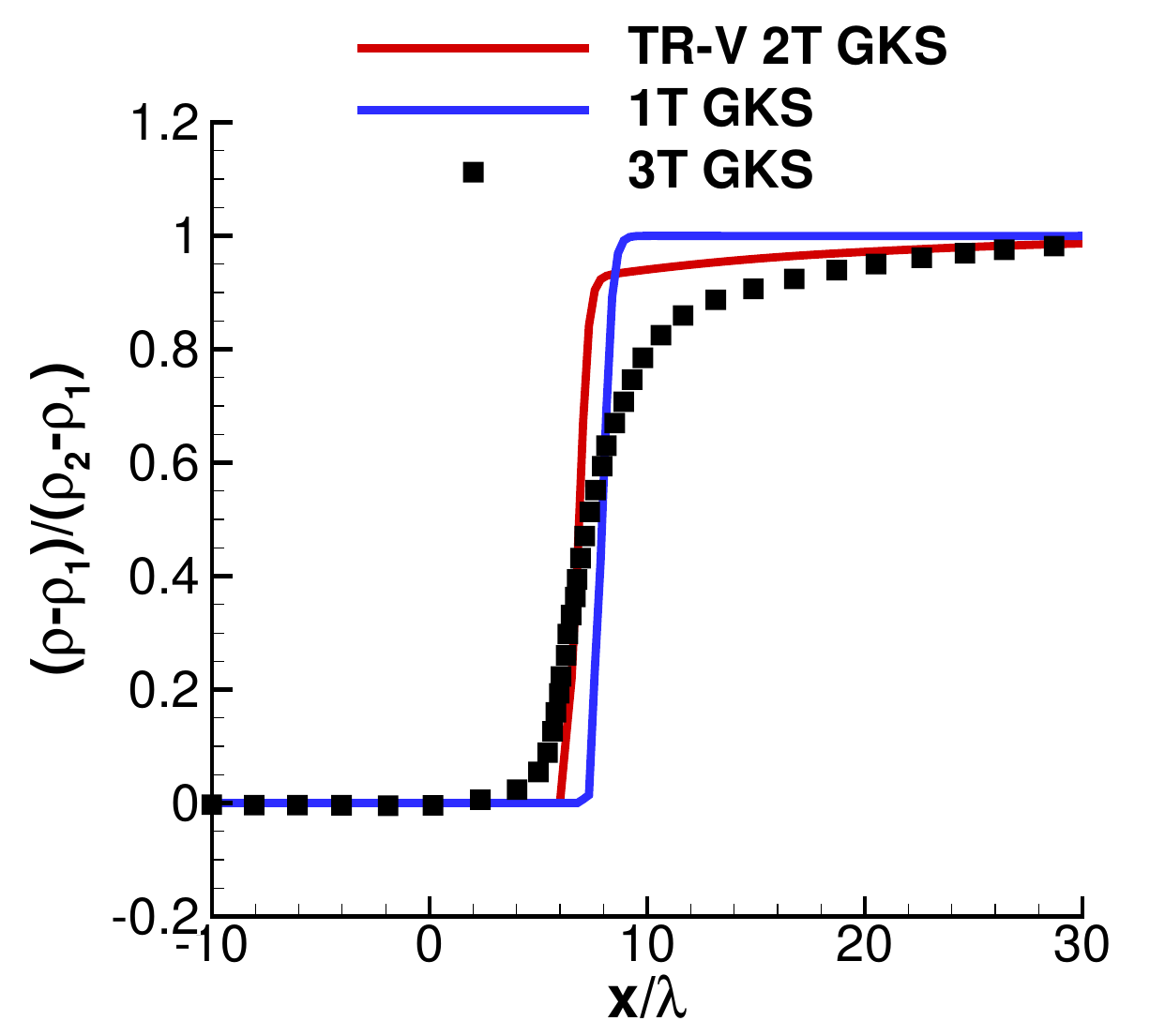}
	\includegraphics[width=0.4\textwidth]
	{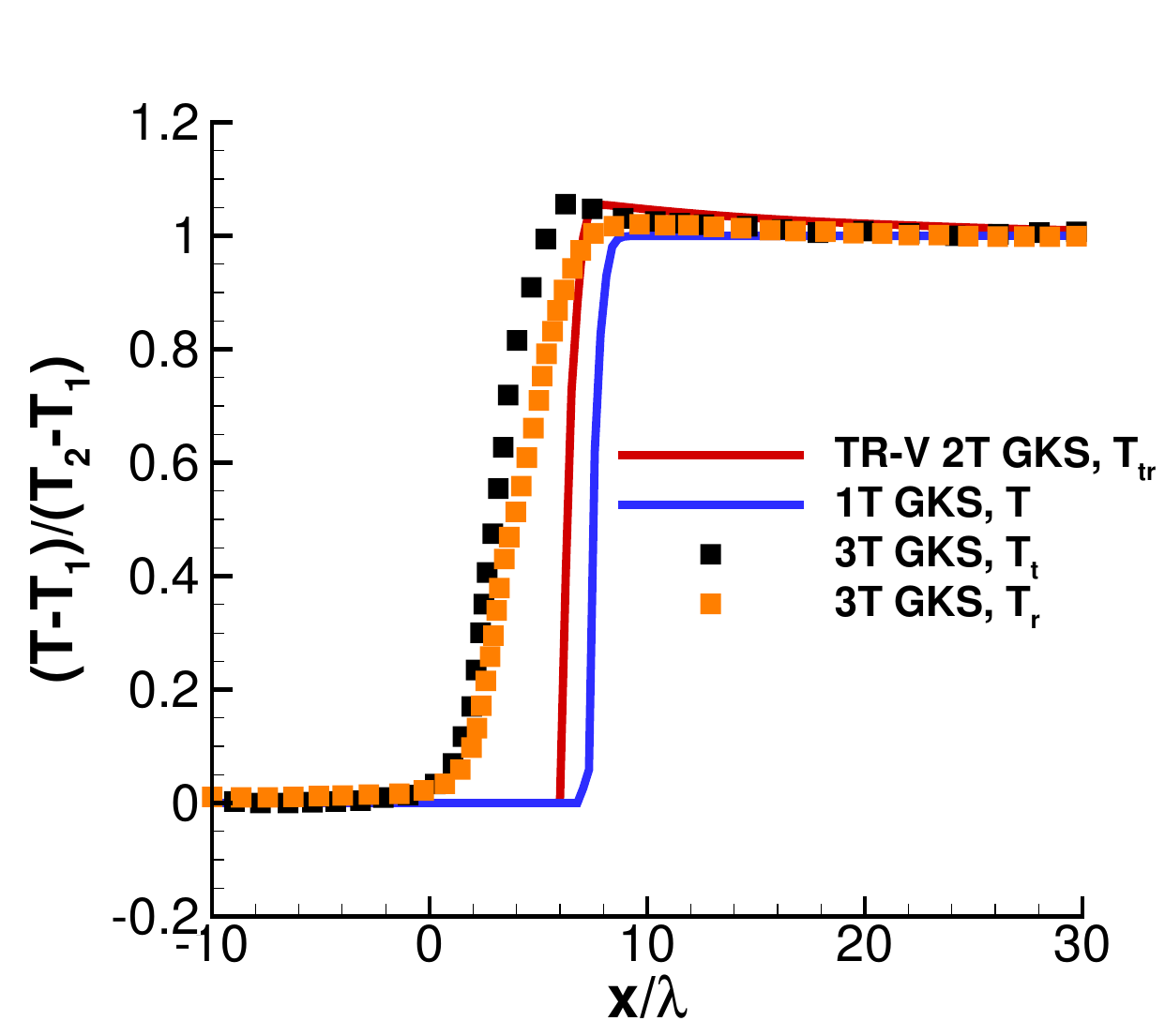}
	\vspace{-4mm} \caption{
	Density and translational-rotational temperature distributions. 
	Left: Density distribution. 
	Right: Translational-rotational temperature distribution.}
	\label{shock_structure_1}
\end{figure*}

\begin{figure*}[htp]	
	\centering
	\includegraphics[width=0.4\textwidth]
	{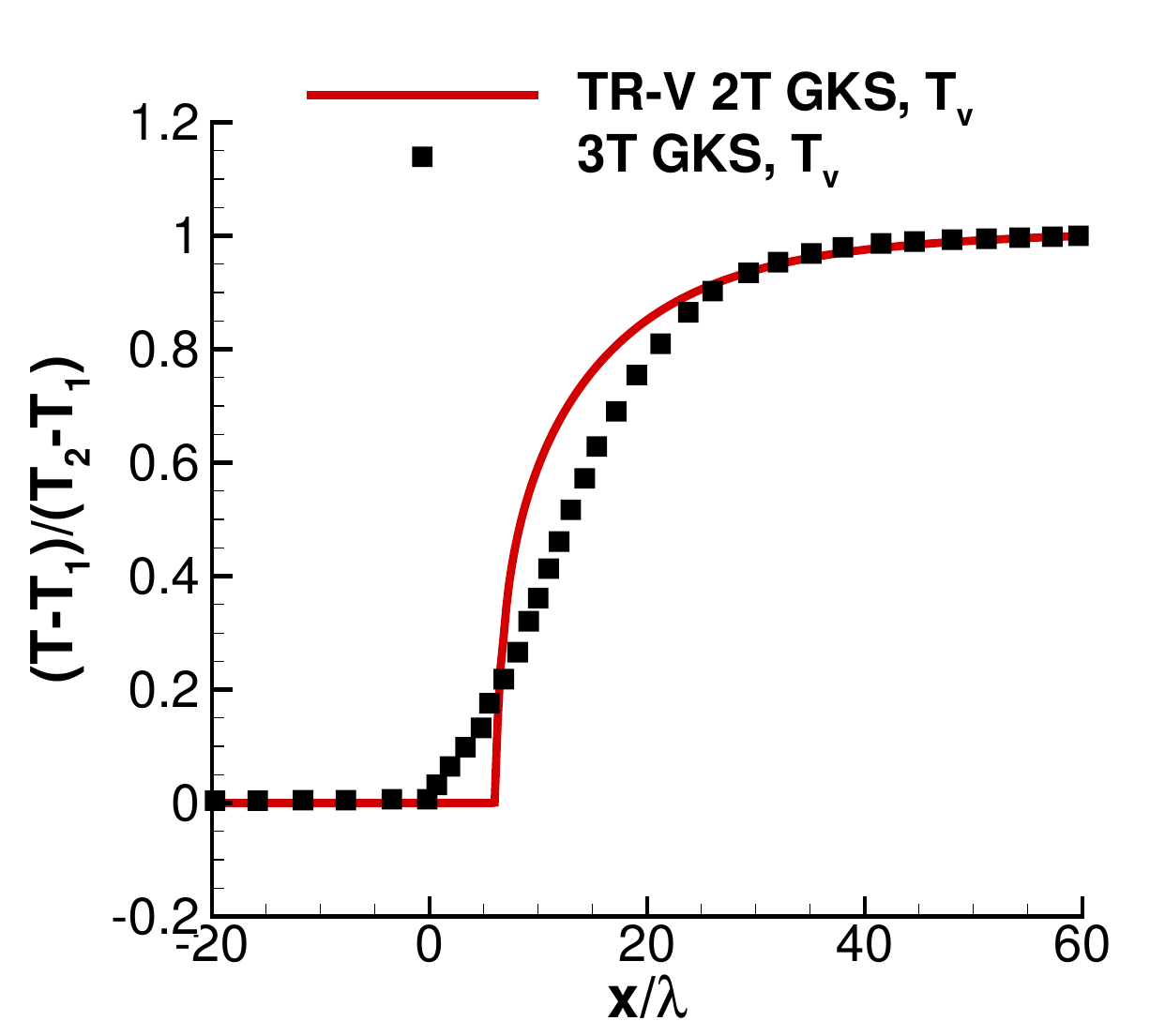}
	\includegraphics[width=0.4\textwidth]
	{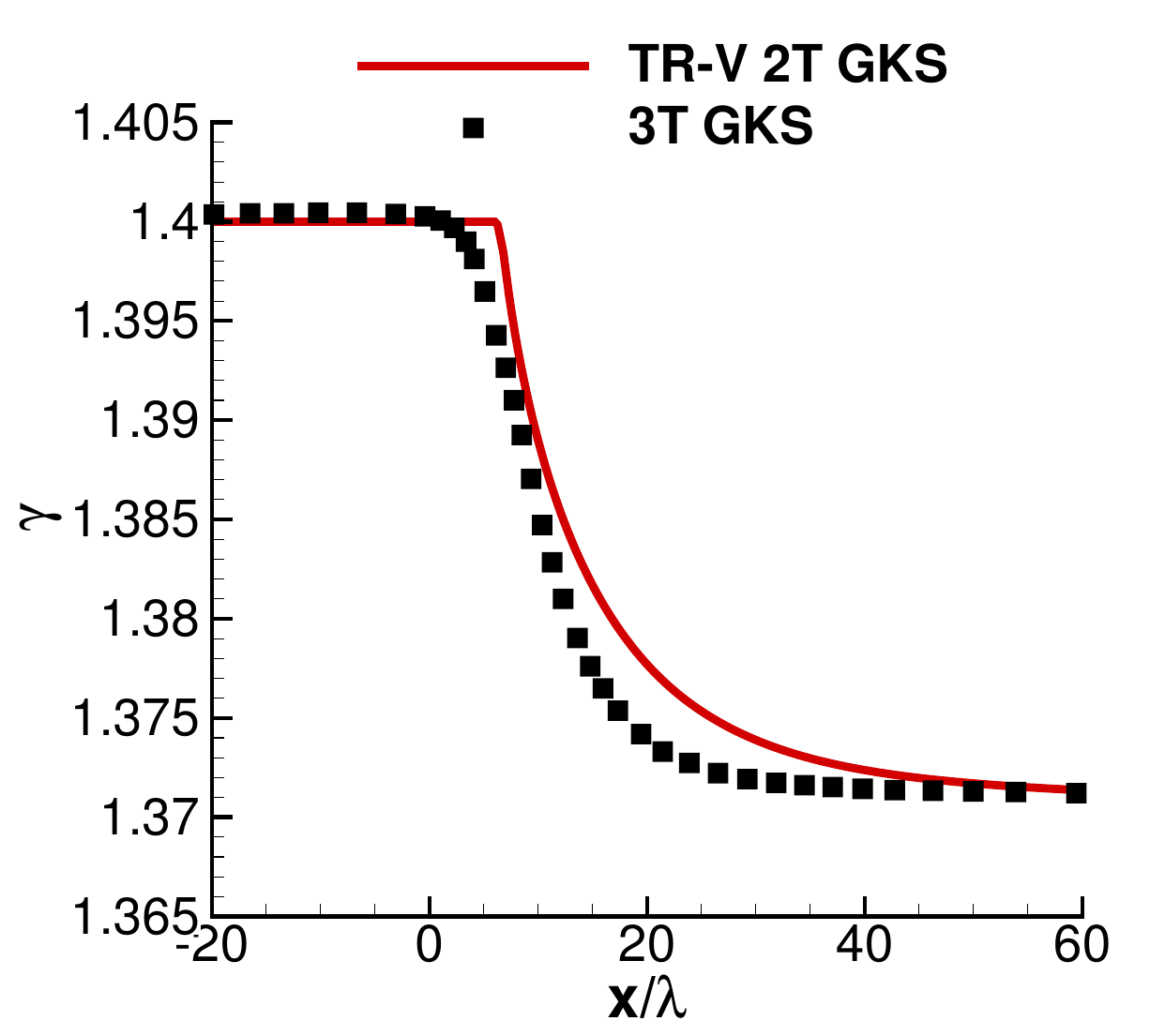}
	\vspace{-4mm} \caption{
	Vibrational temperature and specific heat ratio $\gamma$ distributions. 
	Left: Vibrational temperature distribution. 
	Right: Specific heat ratio $\gamma$ distribution.}
	\label{shock_structure_2}
\end{figure*}

Figs.~\ref{shock_structure_1}–\ref{shock_structure_2} present the simulation results for the shock structure 
using the TR-V 2T GKS, 1T GKS, and 3T GKS models.
Fig.~\ref{shock_structure_1} displays the density and translational-rotational temperature distributions, 
while Fig.~\ref{shock_structure_2} illustrates the vibrational temperature and specific heat ratio $\gamma$.
The density and temperature profiles are normalized using the values at the two ends of the shock.
For example, $\rho' = (\rho - \rho_1)/(\rho_2 - \rho_1)$. The x-axis is normalized by the mean free path 
$\lambda$.
The results clearly capture the thermal non-equilibrium effects and associated relaxation processes.
Given the vibrational collision number $Z_v = 100$, the vibrational relaxation occurs approximately 100 
times more slowly than translational-rotational relaxation.
This is evident in the slower evolution of the vibrational temperature across the shock.
Consequently, the post-shock translational-rotational temperature does not immediately equilibrate to its 
downstream value, as in one-temperature model.
Instead, it gradually decreases from a higher peak value, as vibrational energy continues to rise and absorbs 
part of the translational-rotational energy in the downstream region.

A noticeable difference is observed in the shock thickness between the present and reference methods. 
Specifically, the predicted shock thickness in this study is approximately 3 mean free paths, whereas the reference 
method gives a value of about 10 mean free paths. This discrepancy primarily stems from the different approaches 
used to compute the relaxation time. For a freestream Mach number of 5, the predicted shock thickness in this 
study (\textasciitilde 3 mean free paths) still falls within the typical range reported in previous DSMC simulations, 
theoretical estimates, 
and available experimental data (3\textasciitilde 10 mean free paths)~\cite{Bird1994, Vincenti1966, Guo2007}. 
Although a definitive experimental measurement for this specific case is not available, both the present 
and reference results lie within the physically reasonable range, indicating that the observed differences 
are consistent with expected physical trends.

\subsection{2D hypersonic flow over a wedge}\label{subsec:wedge}
Hypersonic vehicles are generally categorized as either blunt-body or sharp-
leading-edge vehicles. The flows around each of these two types of vehicles are 
significantly different and emphasize unique physical phenomena. This subsection considers 
a hypersonic flow over a sharp-leading-edge vehicle.

\begin{figure}[htp]	
	\centering
	\includegraphics[width=0.5\textwidth]
	{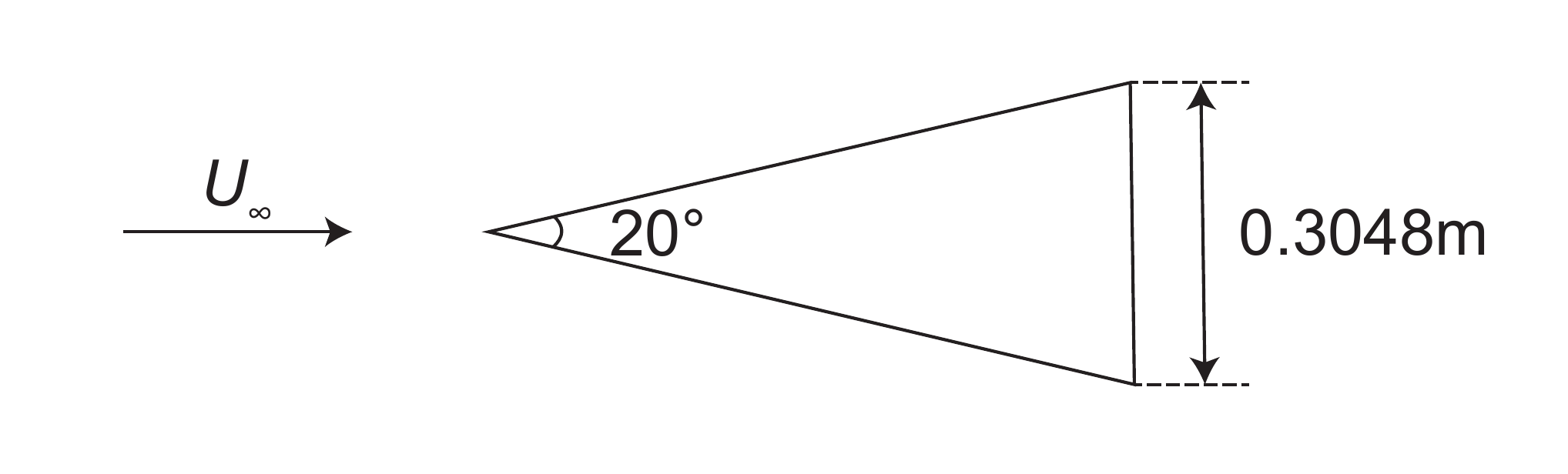}
	\vspace{0mm} 
	\caption{
	2D wedge geometry definition.}
	\label{wedge_geometry}
\end{figure}
The 2D wedge considered here has a 10-degree half-angle and a base height of 12 inches, as illustrated in 
Fig.~\ref{wedge_geometry}. The reference numerical simulations were carried out by Lofthouse~\cite{Lofthouse2008} 
using the Direct Simulation Monte Carlo (DSMC) method and the Michigan Aerothermodynamic Navier--Stokes (LeMANS) 
code. LeMANS is a finite-volume CFD solver that incorporates a two-temperature model.
The freestream gas is nitrogen, and the initial conditions are specified as:
\begin{equation}
\begin{gathered}
M_{\infty} = 10, \quad T_{\infty} = 200\,\mathrm{K}, \\
\quad \rho_{\infty} = 9.872 \times 10^{-5}\,\mathrm{kg/m^3}, 
\quad Re_{\infty} = 8000.
\end{gathered}
\end{equation}
A slip boundary condition is applied at the wall, with the wall temperature set as $T_{\text{wall}} = 500\,\mathrm{K}$.
A mesh-independence study is performed to determine the final mesh resolution. Due to the large gradients near the 
wedge's leading edge, refinement in the wall-parallel direction significantly influences the accuracy of surface 
property predictions. The wall-normal spacing is also critical for capturing surface characteristics accurately. 
To ensure both directional resolutions, the number of grid nodes near the leading edge is progressively doubled, 
alongside wall-normal refinement.
Given the symmetry of the flow field, only half of the computational domain is simulated.
\begin{figure*}[htp]	
	\centering
	\includegraphics[width=0.4\textwidth]
	{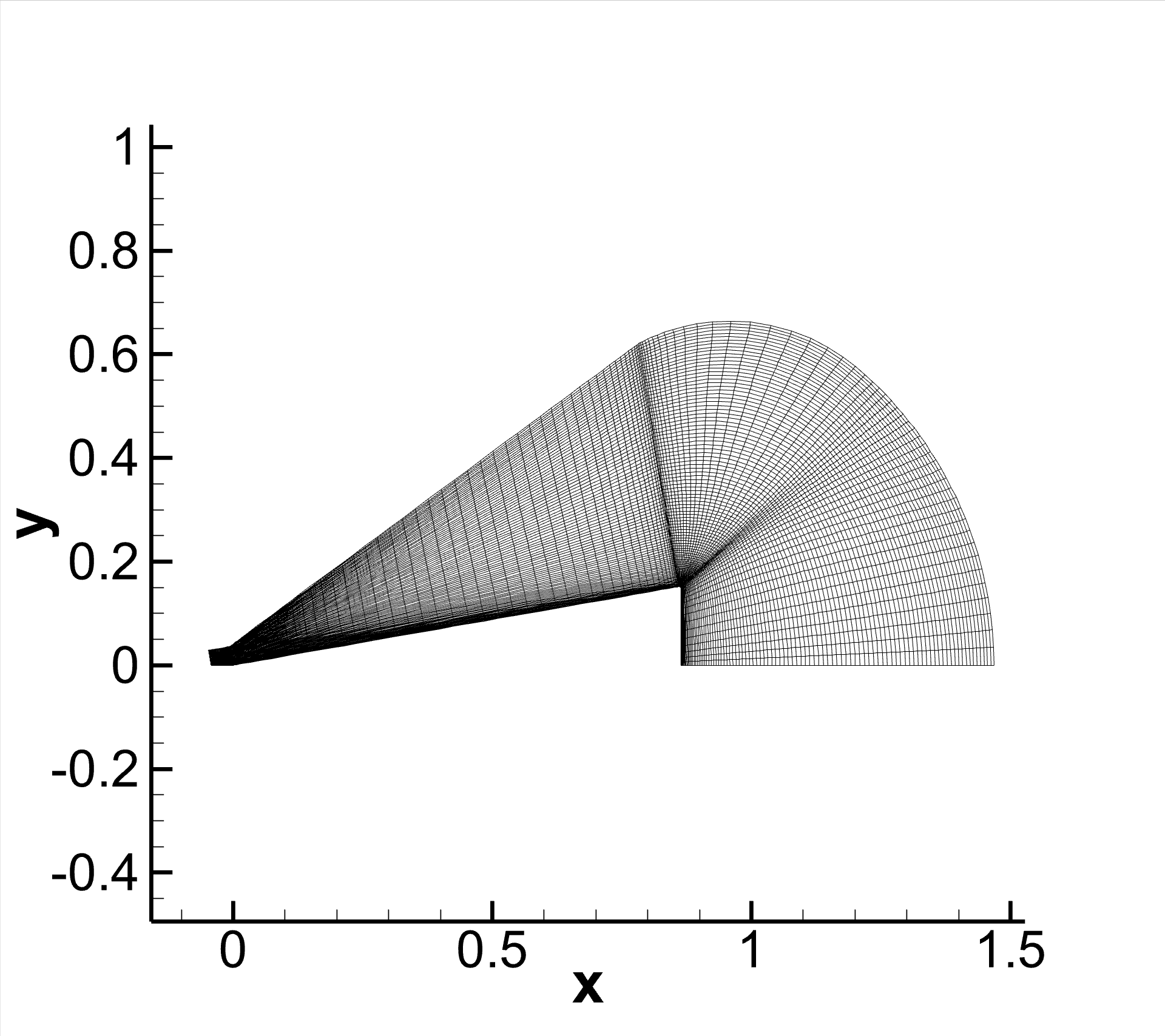}
	\includegraphics[width=0.4\textwidth]
	{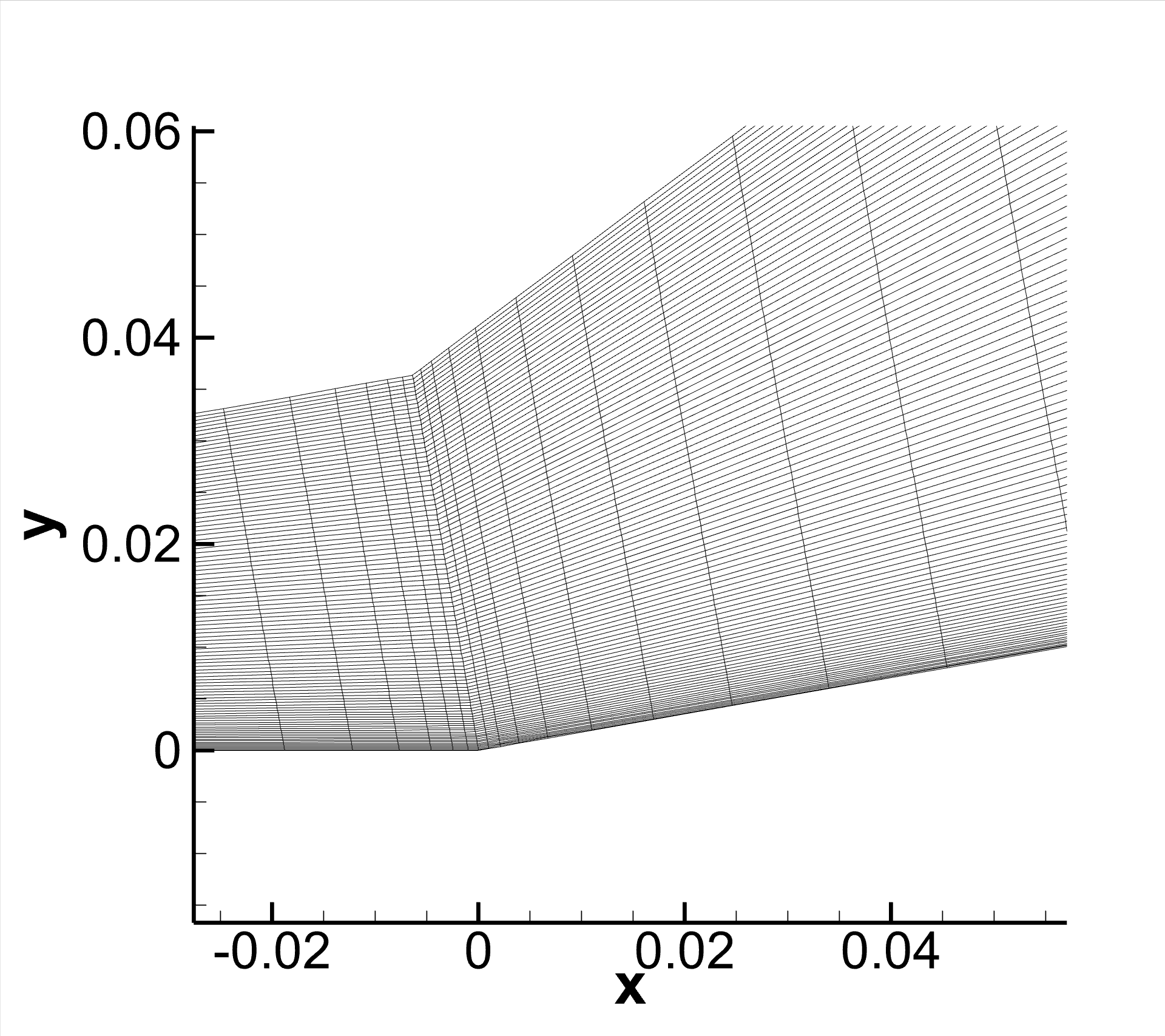}
	\vspace{-4mm} \caption{
	global view for the whole grid (left) and local view near the leading edge wall surface (right).}
	\label{wedge_grid}
\end{figure*}
The simulations are performed using a structured grid and a structured solver.

\subsubsection{Grid Independence Study}
\begin{figure*}[htp]	
	\centering
	\includegraphics[width=0.4\textwidth]
	{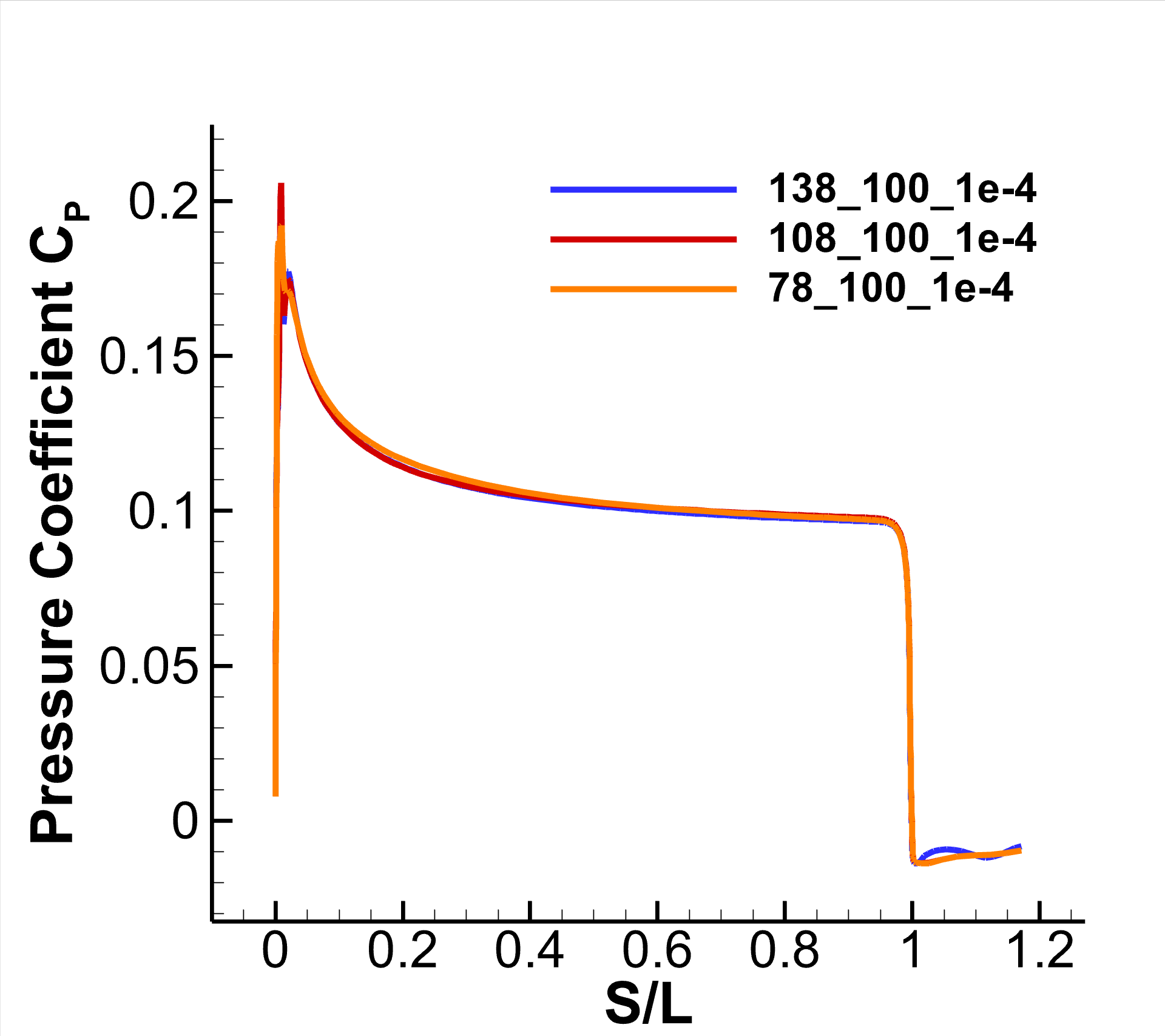}
	\includegraphics[width=0.4\textwidth]
	{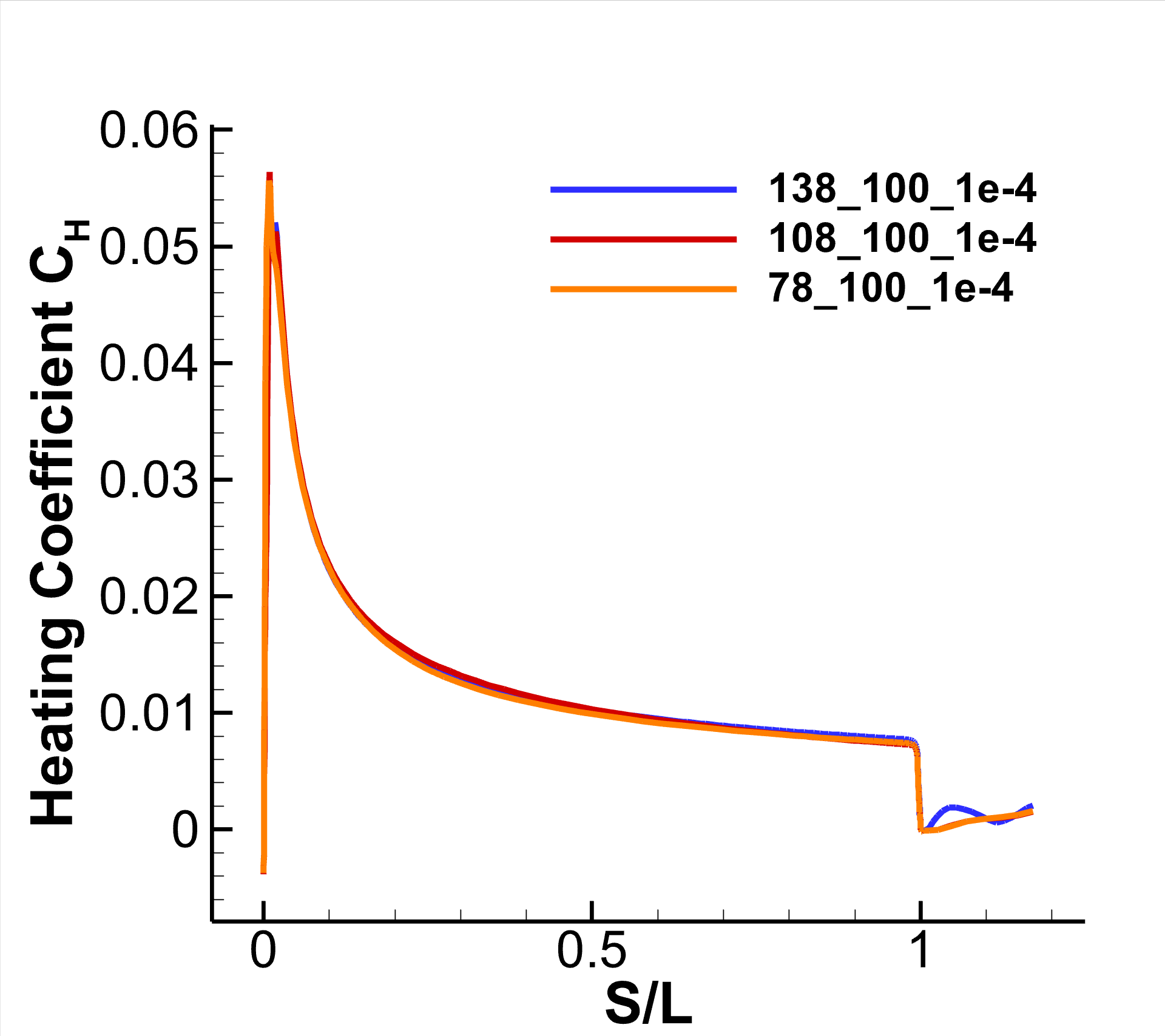}
	\vspace{-4mm} \caption{
	Effect of tangential grid resolution on surface pressure (left) and heat flux (right) for wedge.}
	\label{grid_study_wedge_surface1}
\end{figure*}

\begin{figure*}[htp]	
	\centering
	\includegraphics[width=0.4\textwidth]
	{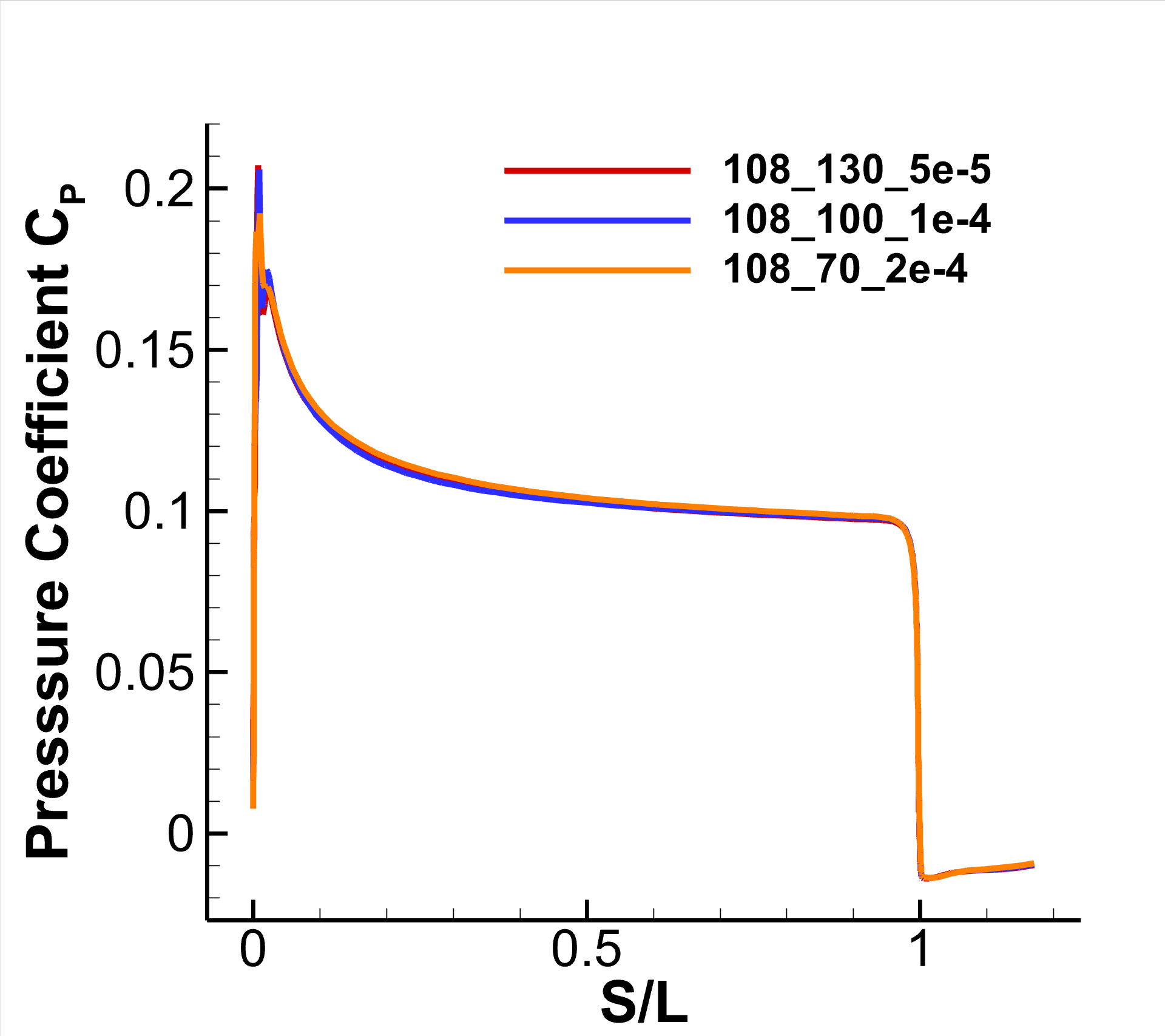}
	\includegraphics[width=0.4\textwidth]
	{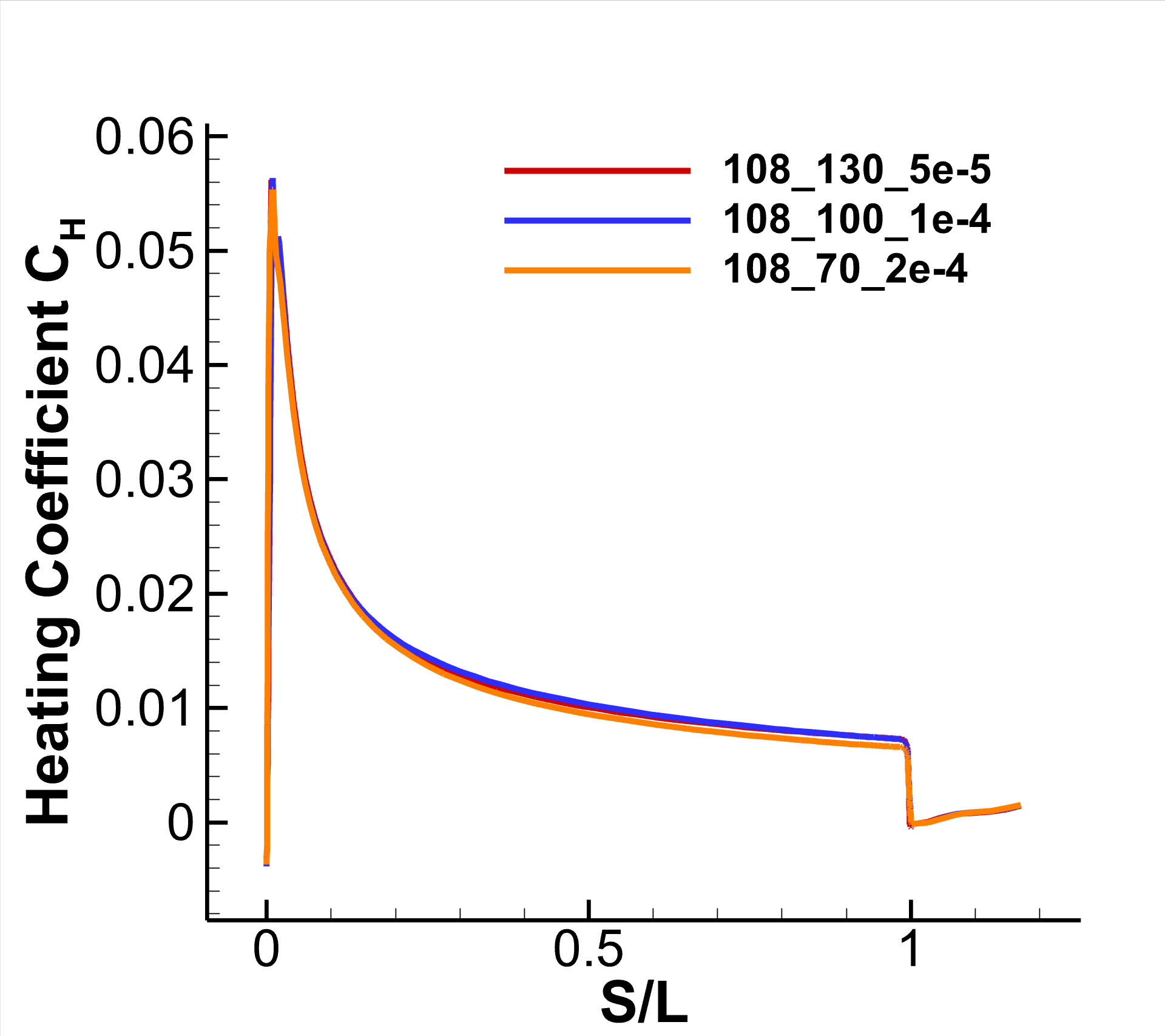}
	\vspace{-4mm} \caption{
	Efect of normal grid resolution and first-layer cell height on surface pressure (left) and heat flux (right) for wedge.}
	\label{grid_study_wedge_surface2}
\end{figure*}

The final mesh, verified for grid independence, is shown in Fig.~\ref{wedge_grid}. It consists of 108 cells in 
the tangential direction and 100 cells in the normal direction, with the first-layer cell height set to 
$1 \times 10^{-4},\mathrm{m}$, corresponding to a cell Reynolds number of $Re_{\text{cell}}=2.6455$. Grid 
independence was assessed by examining the effects of the normal grid resolution and first-layer cell height 
near the wall, as well as the tangential grid resolution. The meshes used for the verification are summarized 
in Table~\ref{grid_study_wedge}, and the corresponding results are shown in Fig.~\ref{grid_study_wedge_surface1} and  
Fig.~\ref{grid_study_wedge_surface2}. It can be seen that further refinement in both 
directions has no significant impact on the wall pressure and heat flux, indicating that the mesh with 
$108 \times 100$ cells and a first-layer height of $1 \times 10^{-4},\mathrm{m}$ is sufficiently fine.

\begin{table*}[htbp]
\centering
\caption{Grid configurations used for grid independence study for the wedge case.}
\label{grid_study_wedge}

\begin{tabular}{ccc}  
\hline\hline
\textbf{Tangential cells} & \textbf{Normal cells} & \textbf{First-layer height [m]} \\
\hline
\multicolumn{3}{c}{\textit{Part 1: Tangential grid refinement}} \\
78  & 100 & $1 \times 10^{-4}$ \\
108 & 100 & $1 \times 10^{-4}$ \\
138 & 100 & $1 \times 10^{-4}$ \\
\hline
\multicolumn{3}{c}{\textit{Part 2: Normal grid / first-layer height refinement}} \\
108 & 70  & $2 \times 10^{-4}$ \\
108 & 100 & $1 \times 10^{-4}$ \\
108 & 130 & $5 \times 10^{-5}$ \\
\hline\hline
\end{tabular}

\end{table*}

\subsubsection{Flow field properties}
\begin{figure*}[htp]	
	\centering
	\includegraphics[width=0.45\textwidth]
	{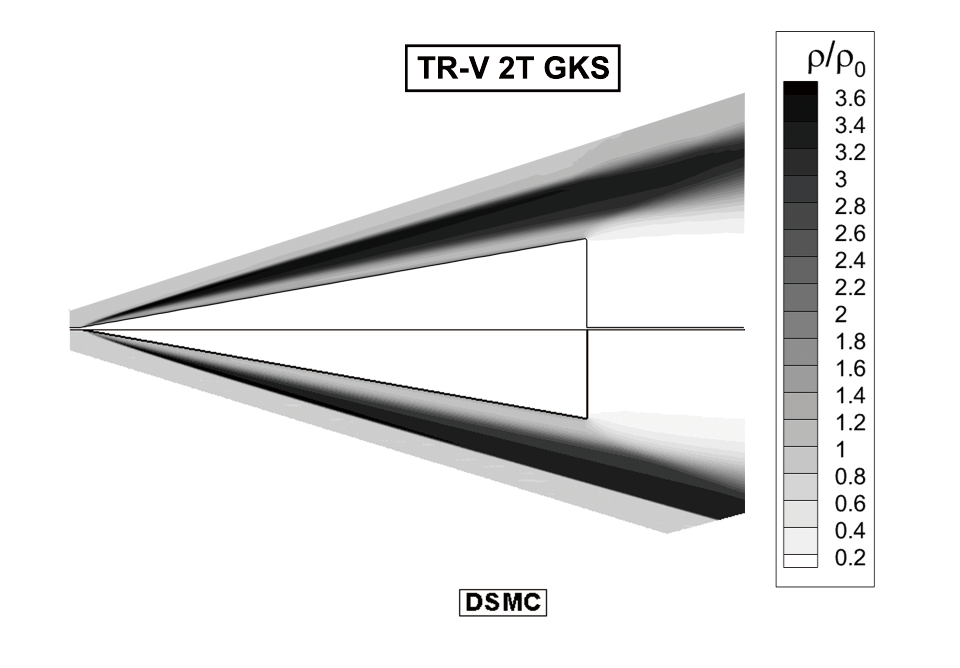}
	\includegraphics[width=0.45\textwidth]
	{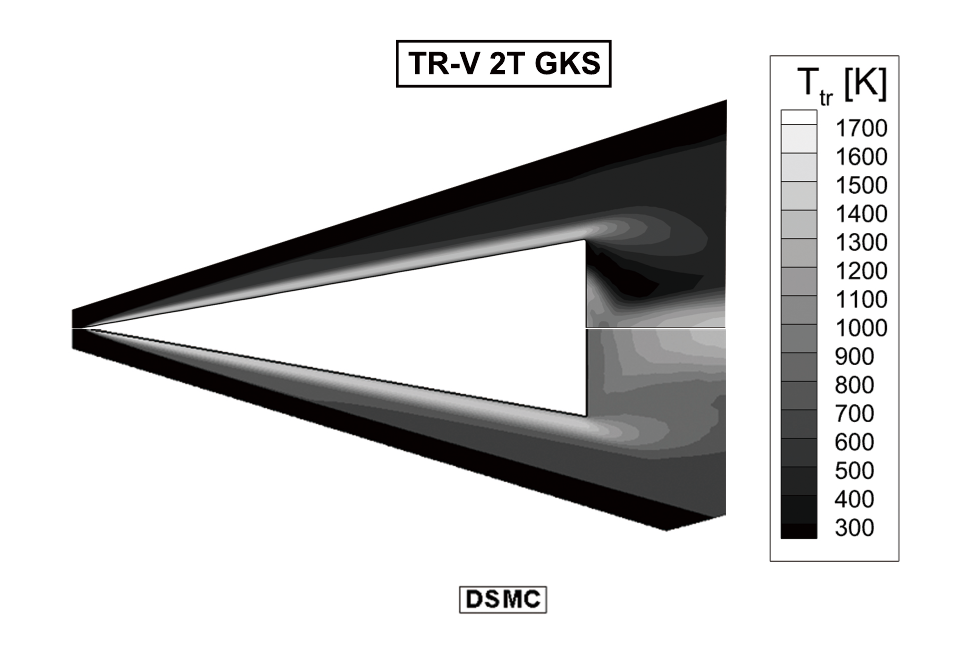}
	\vspace{-4mm} 
	\caption{
	Density ratio field (left) and translational-rotational temperature field (right).
	top: TR-V 2T GKS. bottom: DSMC.}
	\label{wedge_contour1}
\end{figure*}
\begin{figure*}[htp]	
	\centering
	\includegraphics[width=0.5\textwidth]
	{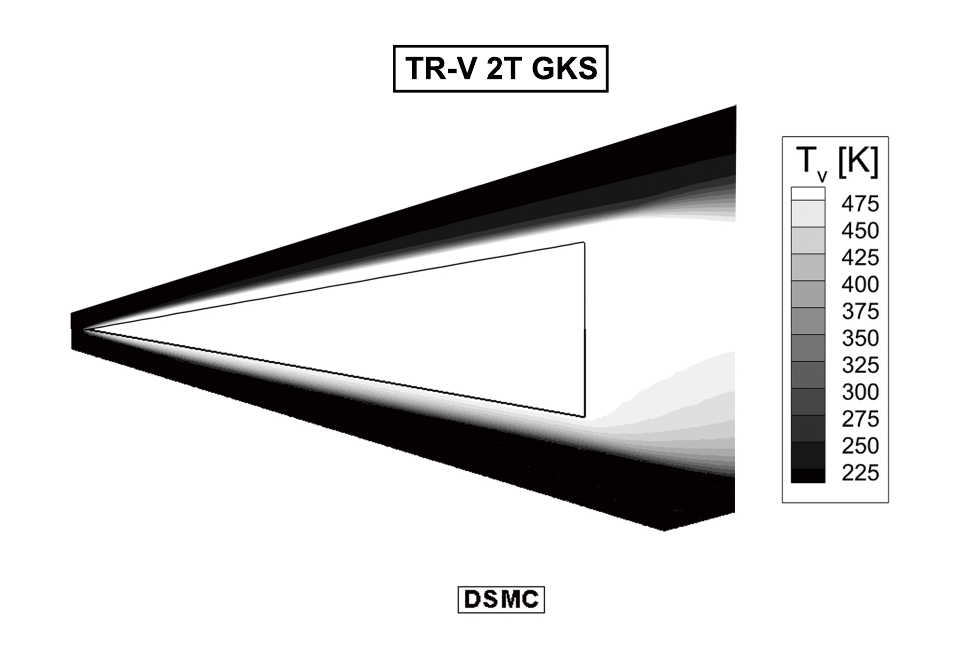}
	\vspace{-4mm} 
	\caption{
	Vibrational temperature field.
	top: TR-V 2T GKS. bottom: DSMC.}
	\label{wedge_contour2}
\end{figure*}
The fields of density, translational-rotational temperature, and vibrational temperature are shown in 
Fig.~\ref{wedge_contour1} and Fig.~\ref{wedge_contour2}, where the density is normalized by the freestream 
density. The density ratio distributions are similar in both models, with the maximum density ratio appearing 
immediately behind the leading-edge shock, reaching a value of 3.80891.
The region of greatest interest lies near the leading edge, where DSMC predicts significantly higher 
temperatures than the TR-V 2T GKS model. Specifically, DSMC estimates a peak temperature of approximately 
$1800\,\mathrm{K}$, while TR-V 2T GKS predicts a lower peak of about $1400\,\mathrm{K}$.
Due to the relatively low translational-rotational temperature, vibrational excitation remains weak. The 
GKS model predicts a higher vibrational temperature than DSMC, with the peak value reaching approximately 
$630\,\mathrm{K}$ in GKS and about $500\,\mathrm{K}$ in DSMC.
It should be emphasized that the prediction of vibrational temperature strongly depends on the choice of 
the vibrational collision number $Z_v$. Since the empirical formula for $Z_v$ lacks experimental verification, 
different studies may adopt different values, leading to variations in the predicted vibrational temperature. 
Nonetheless, such differences in vibrational temperature have minimal impact on the surface properties.

\subsubsection{Surface properties}

The design of hypersonic vehicles requires accurate prediction of the surface properties 
while in flight. These quantities are typically the heat flux, pressure, and shear
stress, from which the aerodynamic forces and moments can be calculated. These
variables govern not only the aerodynamic performance of the vehicle, but also determine 
the selection and sizing of the thermal protection system (TPS), which protects
the vehicle from the extreme temperatures encountered at hypersonic velocities.

In the results that follow, the surface properties are presented in terms of 
non-dimensional coefficients, 
\begin{equation}\label{nondim_surface}
	\begin{gathered}
{C}_{P} =  \frac{p - {p}_{\infty }}{\frac{1}{2}{\rho }_{\infty }{U}_{\infty }^{2}}, \\
{C}_{F} = \frac{\tau}  {\frac{1}{2}{\rho }_{\infty }{U}_{\infty }^{2}}, \\
{C}_{H} =  \frac{q} {\frac{1}{2}{\rho }_{\infty }{U}_{\infty }^{3}}.
	\end{gathered}
\end{equation}
where $p$ is the pressure, $\tau$ is the shear stress, $q$ is the heat transfer rate, 
$p_{\infty}$ is the freestream pressure, $\rho_{\infty}$ is the freestream density, 
and $U_{\infty}$ is the freestream velocity. 
The surface properties in each case are plotted as a function of the distance $S$ 
along the wedge surface, normalized by the length $L$ of the lateral faces. Thus, 
$S/L = 1$ corresponds to the wedge shoulder, which marks the beginning of the wake region. 

\begin{figure*}[htp]	
	\centering
	\includegraphics[width=0.4\textwidth]{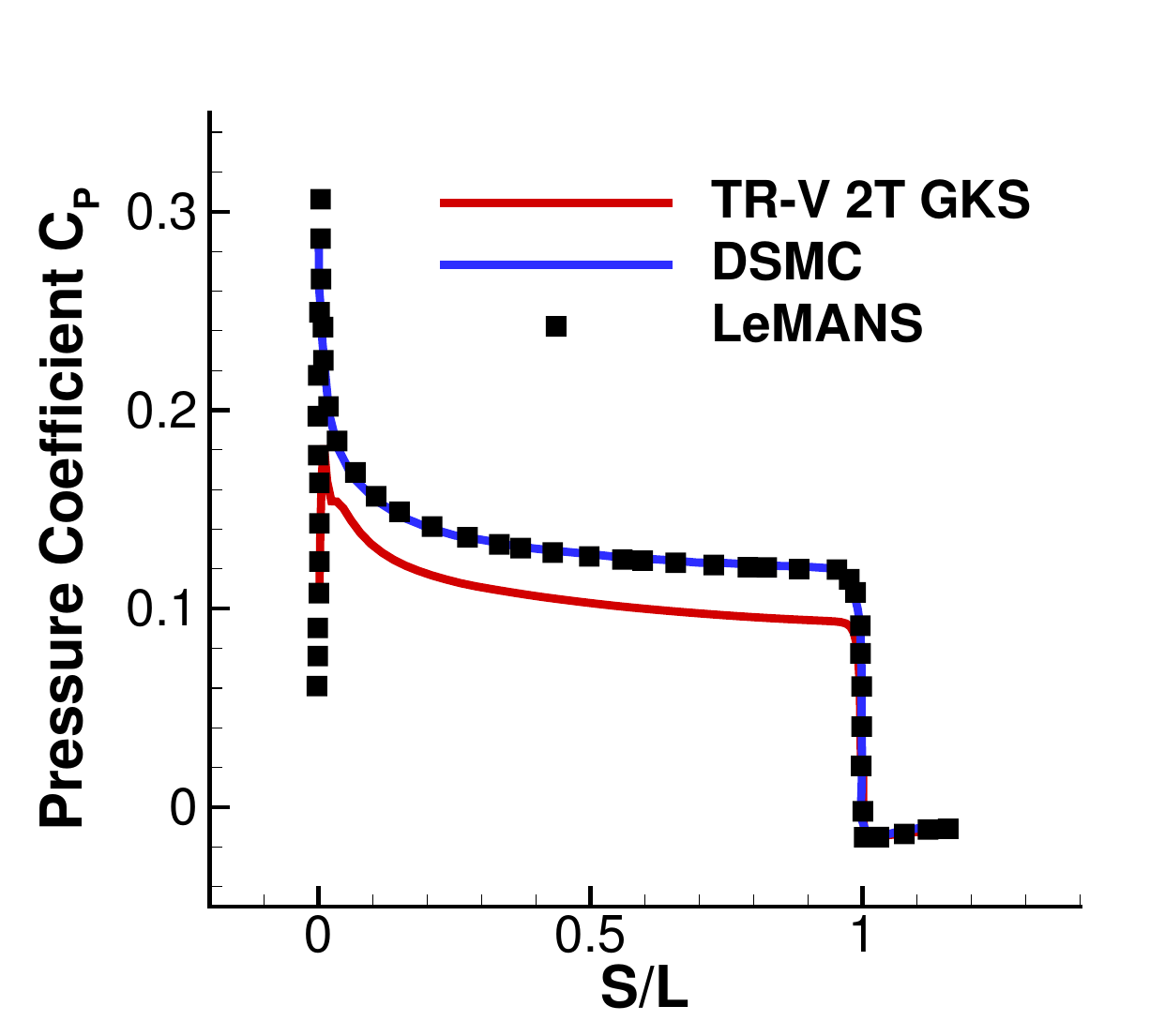}
	\includegraphics[width=0.4\textwidth]{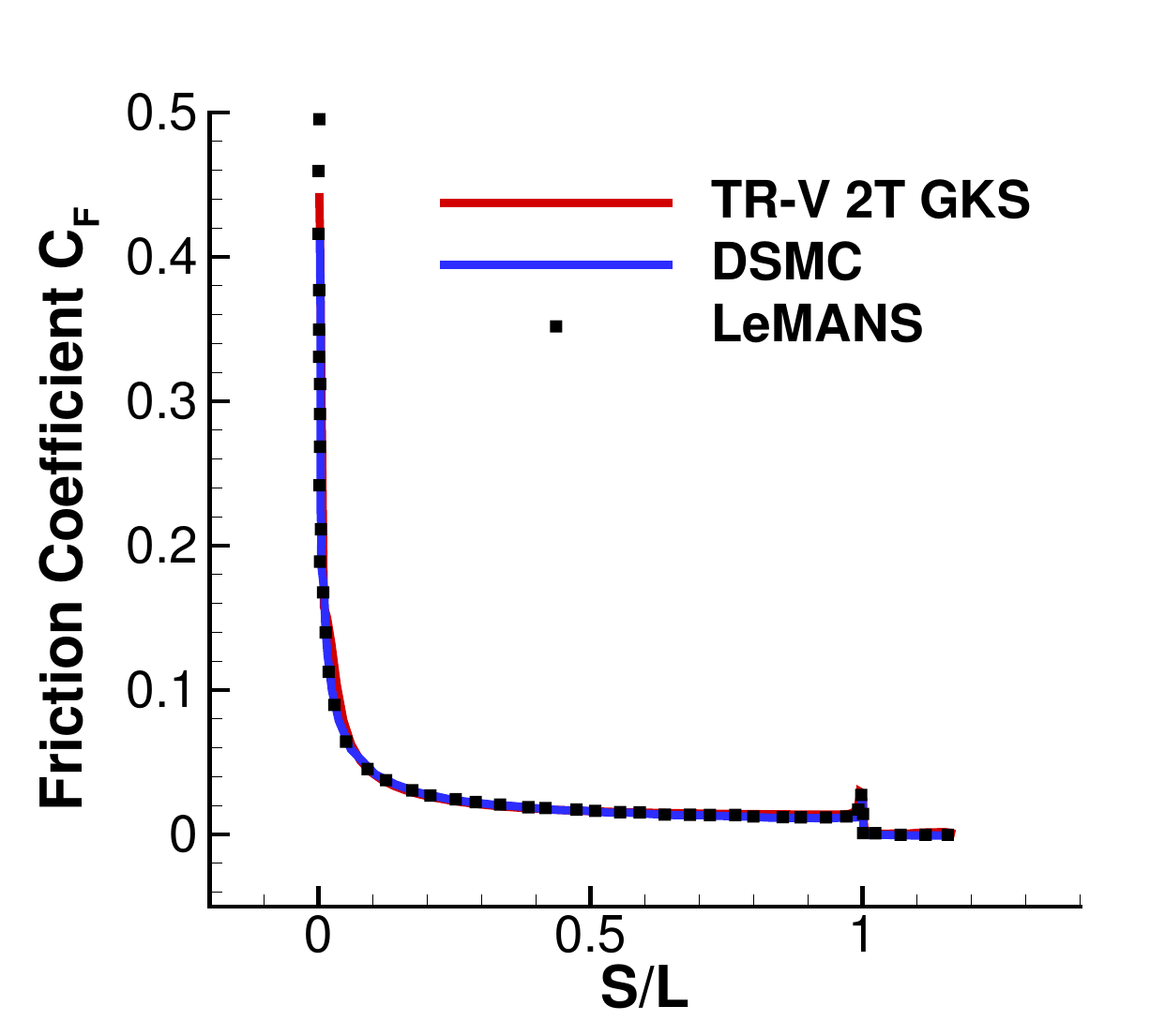}
	\vspace{-4mm} 
	\caption{
	The surface pressure coefficient (left), friction coefficient (right).}
	\label{wedge_surface}
\end{figure*}

\begin{figure*}[htp]	
	\centering
	\includegraphics[width=0.4\textwidth]{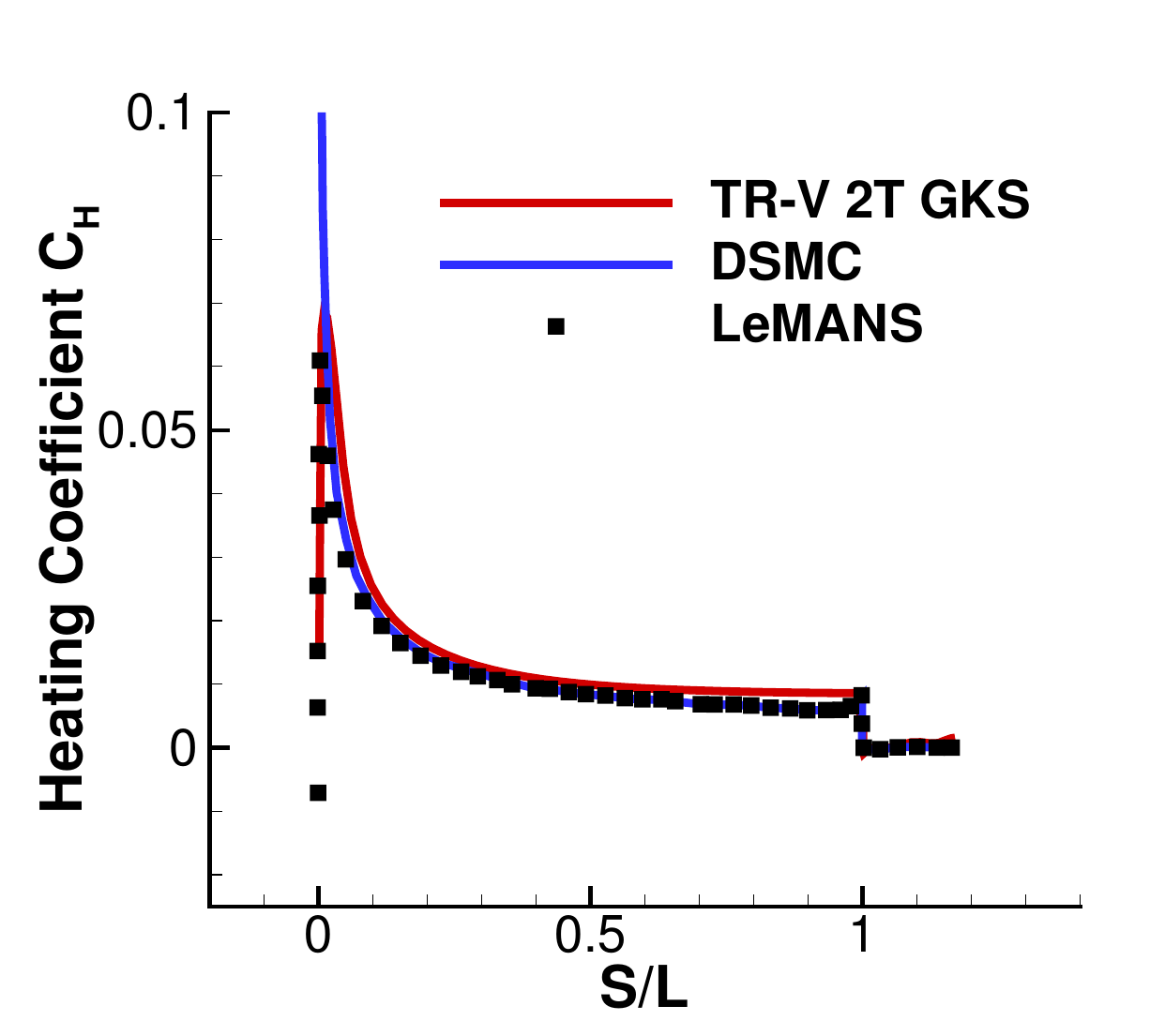}
	\vspace{-4mm} 
	\caption{
	The surface heating coefficient.}
	\label{wedge_surface}
\end{figure*}

The surface pressure coefficient, friction coefficient, and heating coefficient are shown in 
Fig.~\ref{wedge_surface}. The TR-V 2T GKS model predicts 
a pressure distribution that is qualitatively consistent with DSMC and LeMANS, although the magnitude 
is about 20\% lower. 
To analyze the cause of this underprediction, the gradient-length local Knudsen number 
($\mathrm{Kn}_{\mathrm{GLL}}$) is examined~\cite{Boyd1995,Lofthouse2007}. $\mathrm{Kn}_{\mathrm{GLL}}$ is defined as
\[
\mathrm{Kn}_{\mathrm{GLL}} = \frac{\lambda}{Q}\frac{dQ}{dl},
\]
where $\lambda$ is the molecular mean free path and $Q$ represents a representative macroscopic 
quantity, such as density, pressure, or temperature. Unlike the global Knudsen number, 
$\mathrm{Kn}_{\mathrm{GLL}}$ characterizes the local strength of non-equilibrium and provides a criterion for 
assessing the validity of continuum-based models. Empirical guidelines suggest that the continuum 
assumption with no-slip conditions is valid for $\mathrm{Kn}_{\mathrm{GLL}} < 0.01$, continuum with slip 
conditions is applicable for $0.01 \le \mathrm{Kn}_{\mathrm{GLL}} \le 0.1$, and regions with 
$\mathrm{Kn}_{\mathrm{GLL}} > 0.1$ require rarefied treatment. 
\begin{figure}[htp]	
	\centering
	\includegraphics[width=0.4\textwidth]{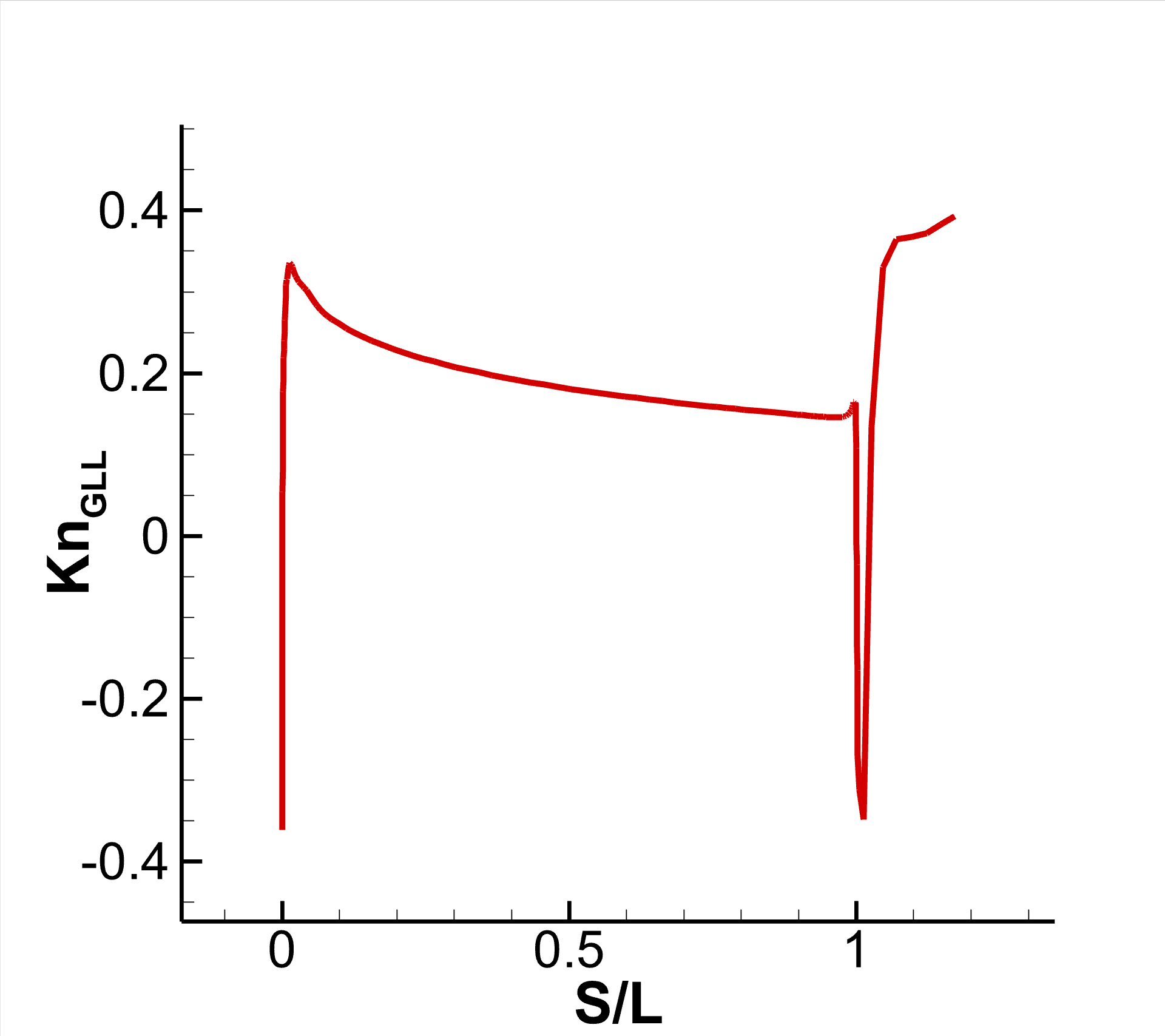}
	\vspace{-4mm} 
	\caption{
	Distribution of $\mathrm{Kn}_{\mathrm{GLL}}$ on the surface.}
	\label{wedge_surface_KnGLL}
\end{figure}
Figure~\ref{wedge_surface_KnGLL} presents the surface distribution of $\mathrm{Kn}_{\mathrm{GLL}}$ 
on the wedge, where translational-rotational temperature is taken as the representative quantity $Q$. 
It can be observed that $\mathrm{Kn}_{\mathrm{GLL}}$ values in the shoulder region behind the leading 
edge exceed 0.2, which is clearly beyond the applicability range of the continuum model with slip boundary conditions. 
This local breakdown of the continuum assumption explains the underprediction of surface pressure.

The shear stress predicted by TR-V 2T GKS shows excellent agreement with results from both DSMC and LeMANS. 
As for the heat transfer rate distribution, from the temperature contours shown in Fig.~\ref{wedge_contour1}, 
it is evident that DSMC predicts significantly higher temperatures at the leading edge. 
Consequently, the heat flux predicted by DSMC at the leading edge is also much higher. 
However, the predictions by TR-V 2T GKS are in close agreement with LeMANS in this region.

\subsubsection{Sensitivity Analysis of the Vibrational Collision Number $Z_v$}
Vibrational collision number $Z_v$ plays an important role in predicting vibrational non-equilibrium in high-temperature 
hypersonic flows. In this work, the ($c_1, c_2, \omega$) values from Liu \textit{et al.}~\cite{Liu2021-3T} are adopted, based 
on the corrected Millikan-White 
formulation that accounts for the distribution of energy among translational, rotational, and vibrational modes. This ensures 
that the vibrational relaxation predicted by our TR-V 2T GKS is physically consistent with prior DSMC and theoretical results.

To quantify the effect of parameter selection, a sensitivity study was conducted using this wedge flow case. Three 
sets of parameters ($c_1, c_2, \omega$) were tested, corresponding to Bird~\cite{Bird1994}, Wang~\cite{Wang2017-UGKS}, and the 
present work. The resulting vibrational temperature fields, as well as surface pressure and heat flux coefficients, are shown 
in the Figures below.
\begin{figure*}[htp]	
	\centering
	\includegraphics[width=0.4\textwidth]{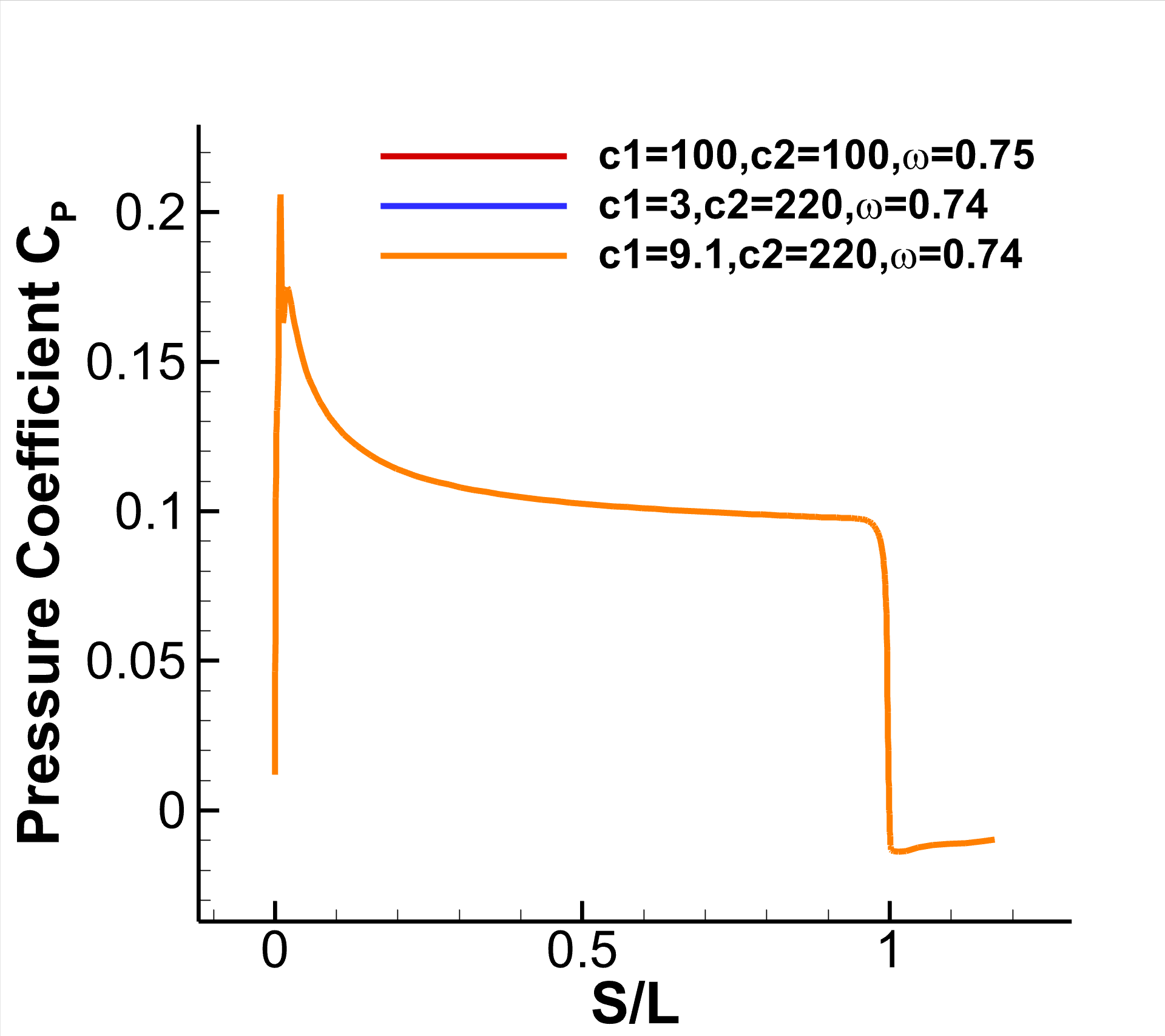}
	\includegraphics[width=0.4\textwidth]{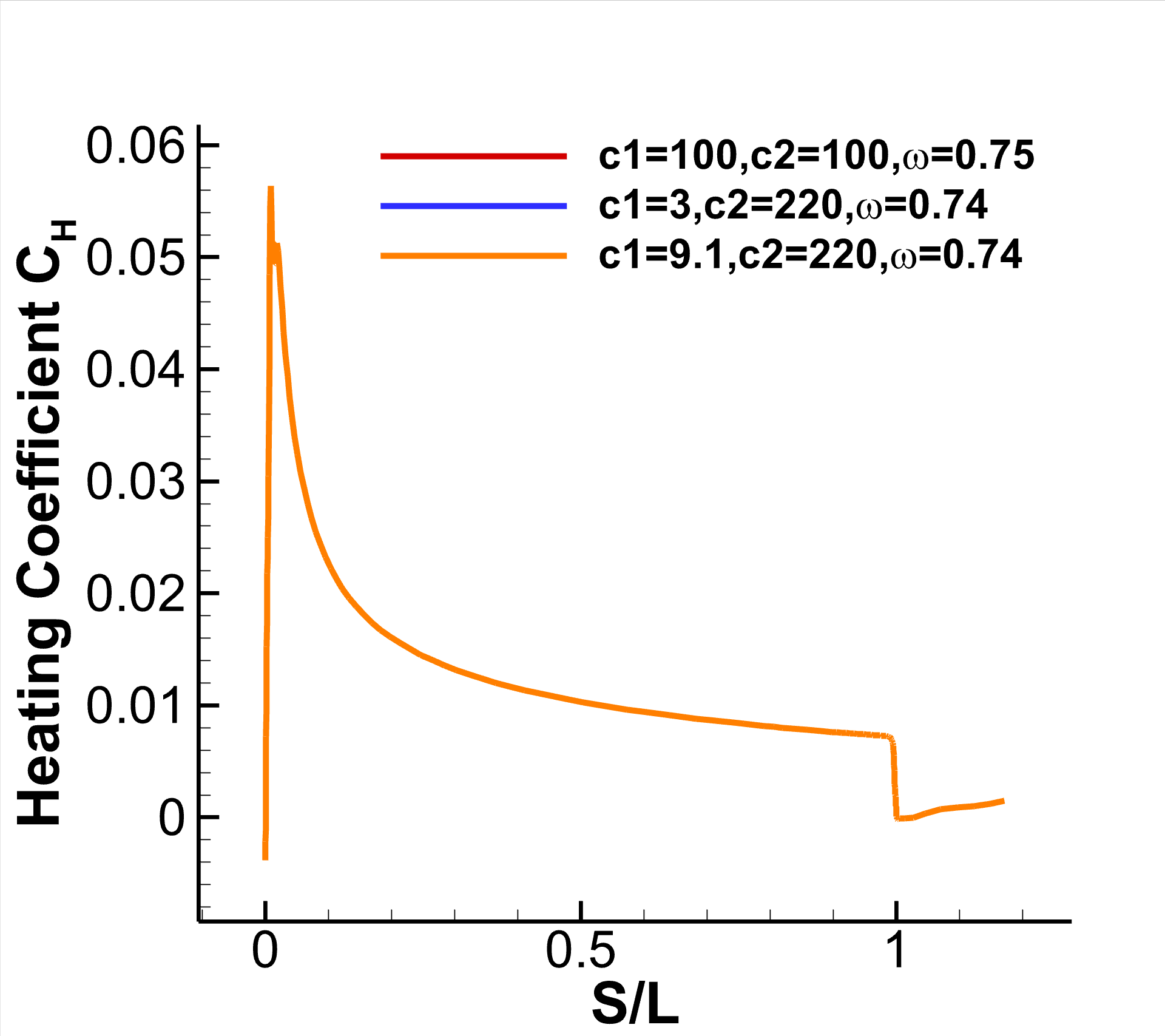}
	\vspace{-4mm} 
	\caption{
	Surface pressure (left) and heating (right) coefficient curves corresponding to the three sets of parameters.}
	\label{wedge_sens_surf}
\end{figure*}

\begin{figure*}[htp]	
	\centering
	\includegraphics[width=0.4\textwidth]{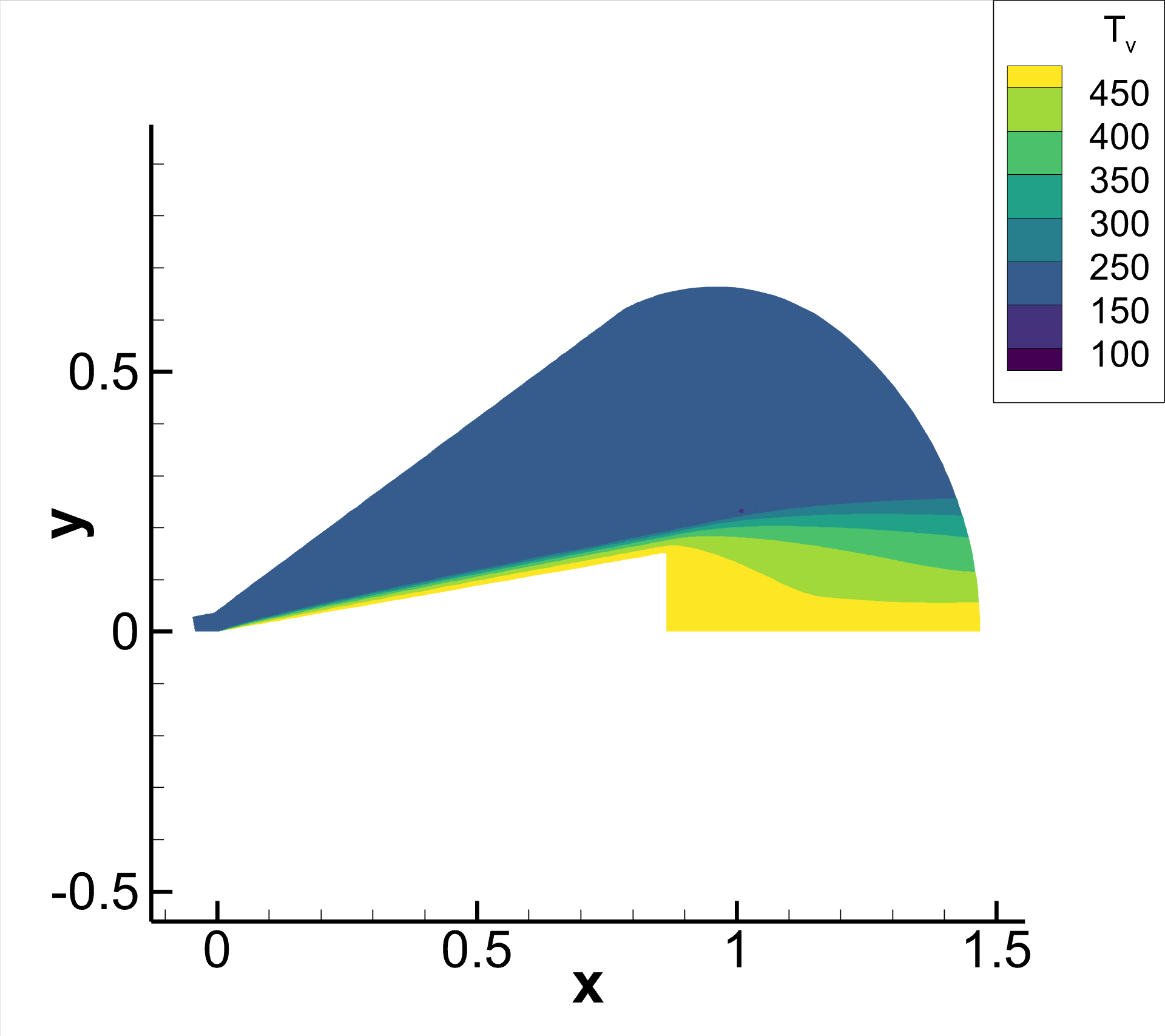}
	\includegraphics[width=0.4\textwidth]{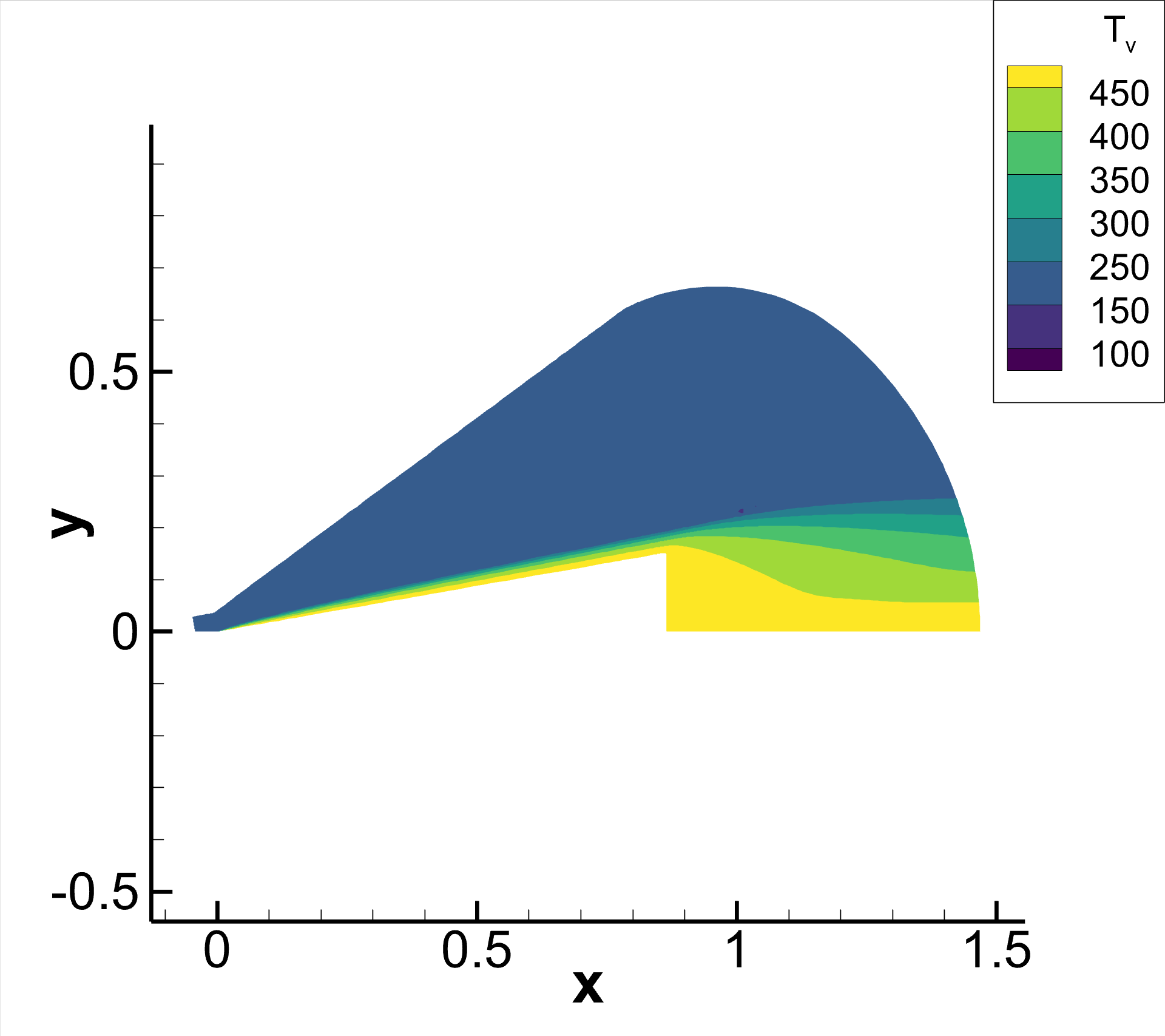}
	\vspace{-4mm} 
	\caption{
	Vibrational temperature contours corresponding to Bird~\cite{Bird1994} (left) and Wang~\cite{Wang2017-UGKS} (right) parameters}
	\label{wedge_sens_Tv1}
\end{figure*}

\begin{figure*}[htp]	
	\centering
	\includegraphics[width=0.4\textwidth]{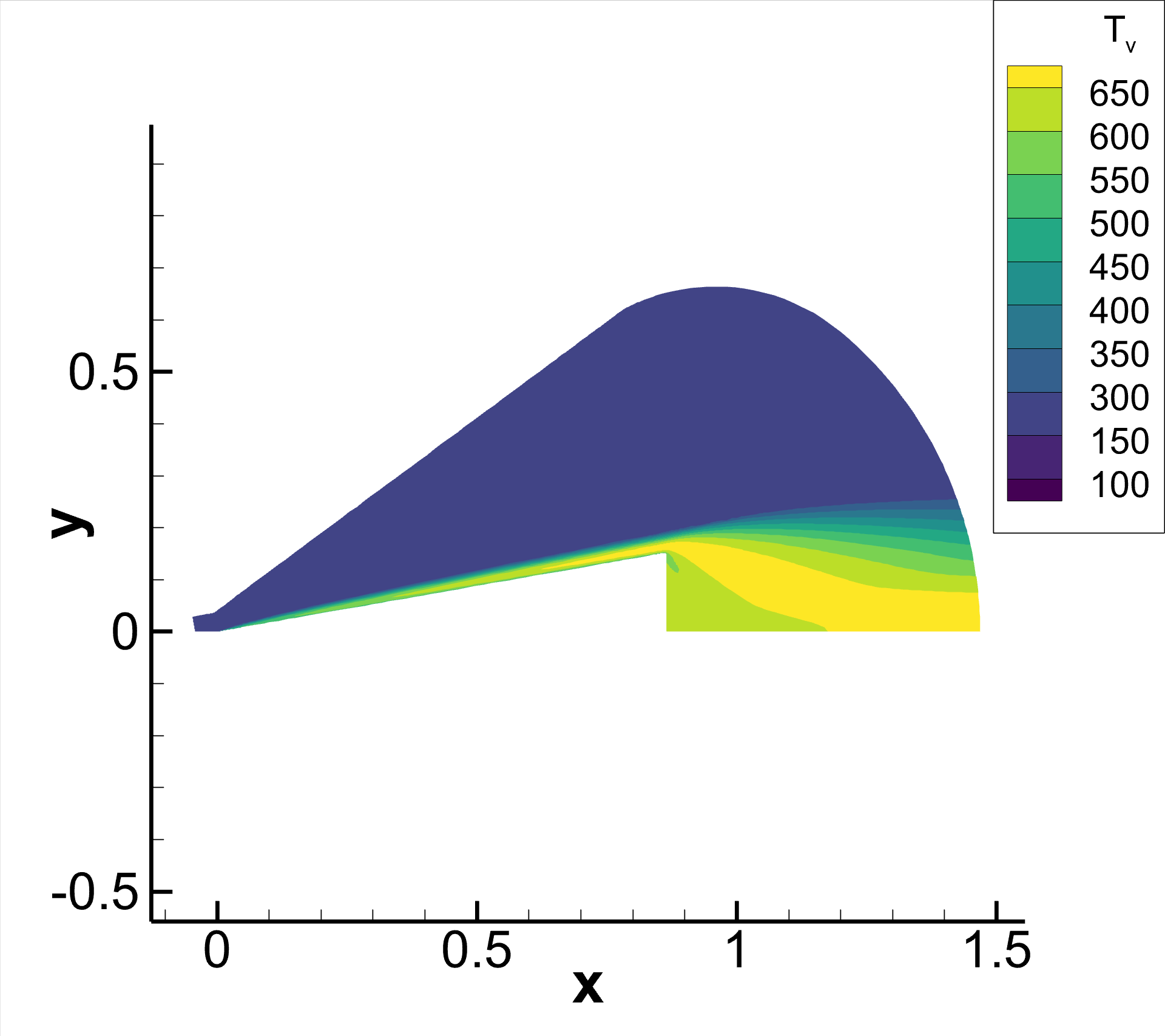}
	\vspace{-4mm} 
	\caption{
	Vibrational temperature contours corresponding to the parameters used in this study.}
	\label{wedge_sens_Tv1}
\end{figure*}

\begin{figure}[htp]	
	\centering
	\includegraphics[width=0.4\textwidth]{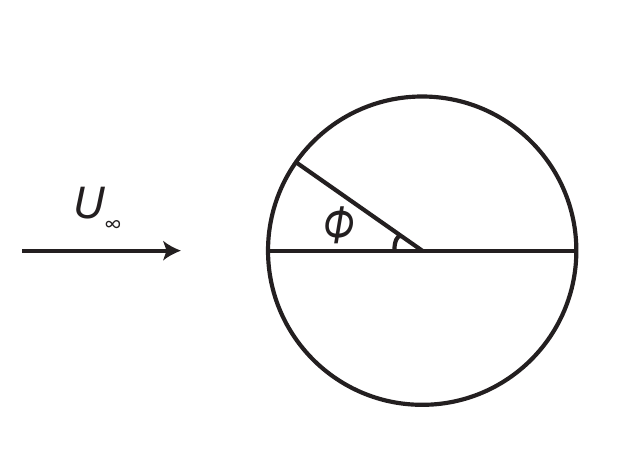}
	\vspace{0mm} 
	\caption{2D cylinder geometry definition.}
	\label{cylinder_geometry}
\end{figure}

The results indicate that surface pressure and heat flux remain nearly identical across all three parameter sets, while 
the vibrational temperature exhibits noticeable differences, with peak values of 499 K, 497 K, and 680 K, respectively. 
This demonstrates that reasonable variations in the vibrational collision parameters have minimal impact on overall flow 
predictions, mainly affecting the vibrational temperature distribution.
\begin{color}{blue}
\subsection{2D Cylinder Flow at 4000 K (Chemically Frozen)}
\end{color}
Besides sharp-body vehicles, another typical aerodynamic configuration is the blunt-body vehicle.
This subsection considers a hypersonic flow over a blunt-body.  
The 2D cylinder considered here has a diameter of 12 inches, as illustrated in Fig.~\ref{cylinder_geometry}.

The reference numerical simulations were carried out by Lofthouse~\cite{Lofthouse2008} using the DSMC method.
The freestream gas is nitrogen, and the initial conditions are specified as:
\begin{equation}
	\begin{gathered}
	M_{\infty} = 10, \quad T_{\infty} = 200\,\mathrm{K}, \\
	\quad \rho_{\infty} = 9.872 \times 10^{-5}\,\mathrm{kg/m^3}, 
	\quad Re_{\infty} = 8000.
	\end{gathered}
\end{equation}
A slip boundary condition is applied at the wall, with the wall temperature set as $T_{\text{wall}} = 500\,\mathrm{K}$.
Due to the unidirectional propagation of information in hypersonic flow, 
the leeward side is not considered, and only the windward flow field is computed. 
Additionally, since the flow field is axisymmetric about the cylinder centerline, 
only the upper half is simulated. This case is computed using a unstructured solver. 

\subsubsection{Grid Independence Study}
\begin{figure*}[htp]	
	\centering
	\includegraphics[width=0.4\textwidth]
	{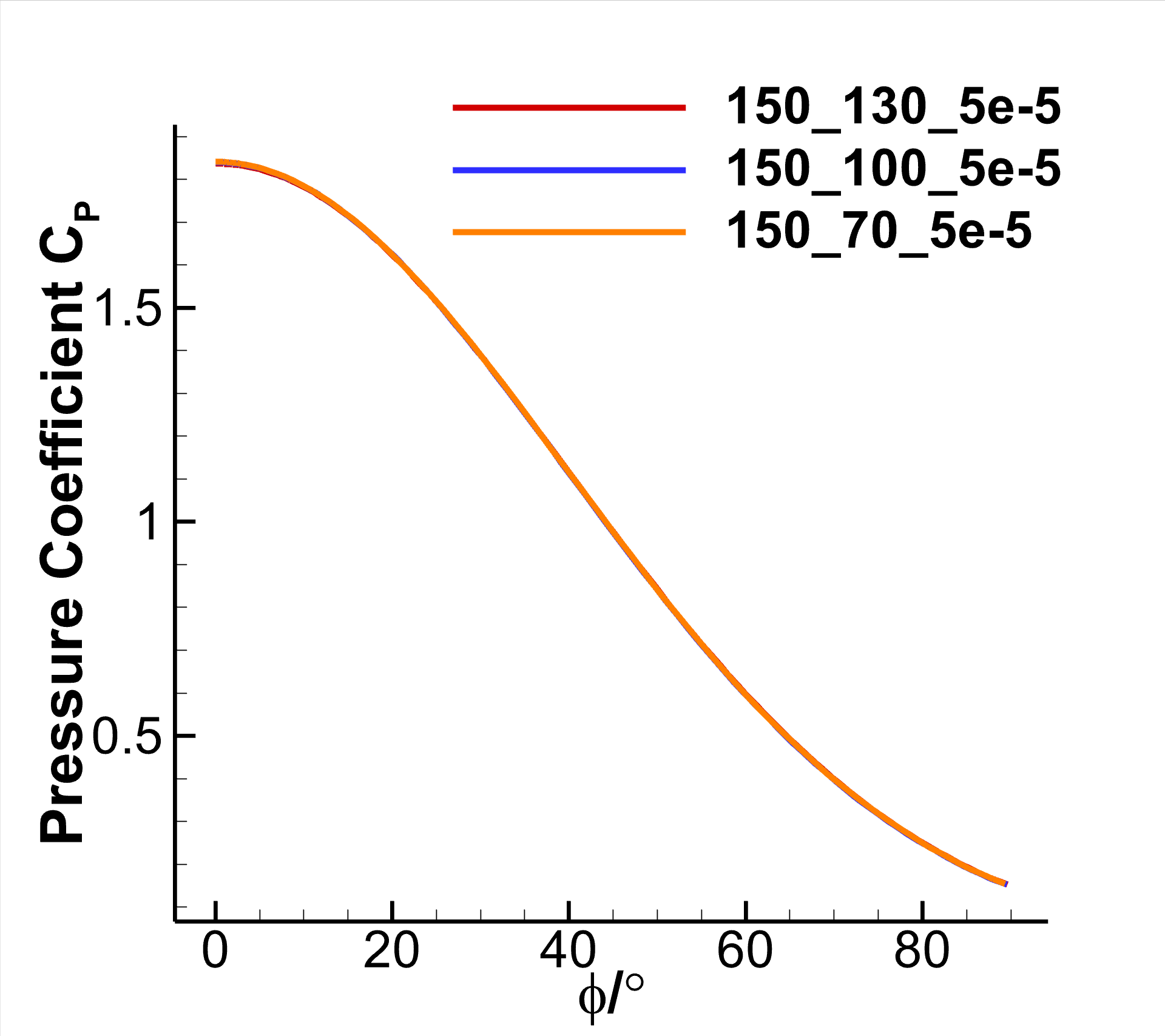}
	\includegraphics[width=0.4\textwidth]
	{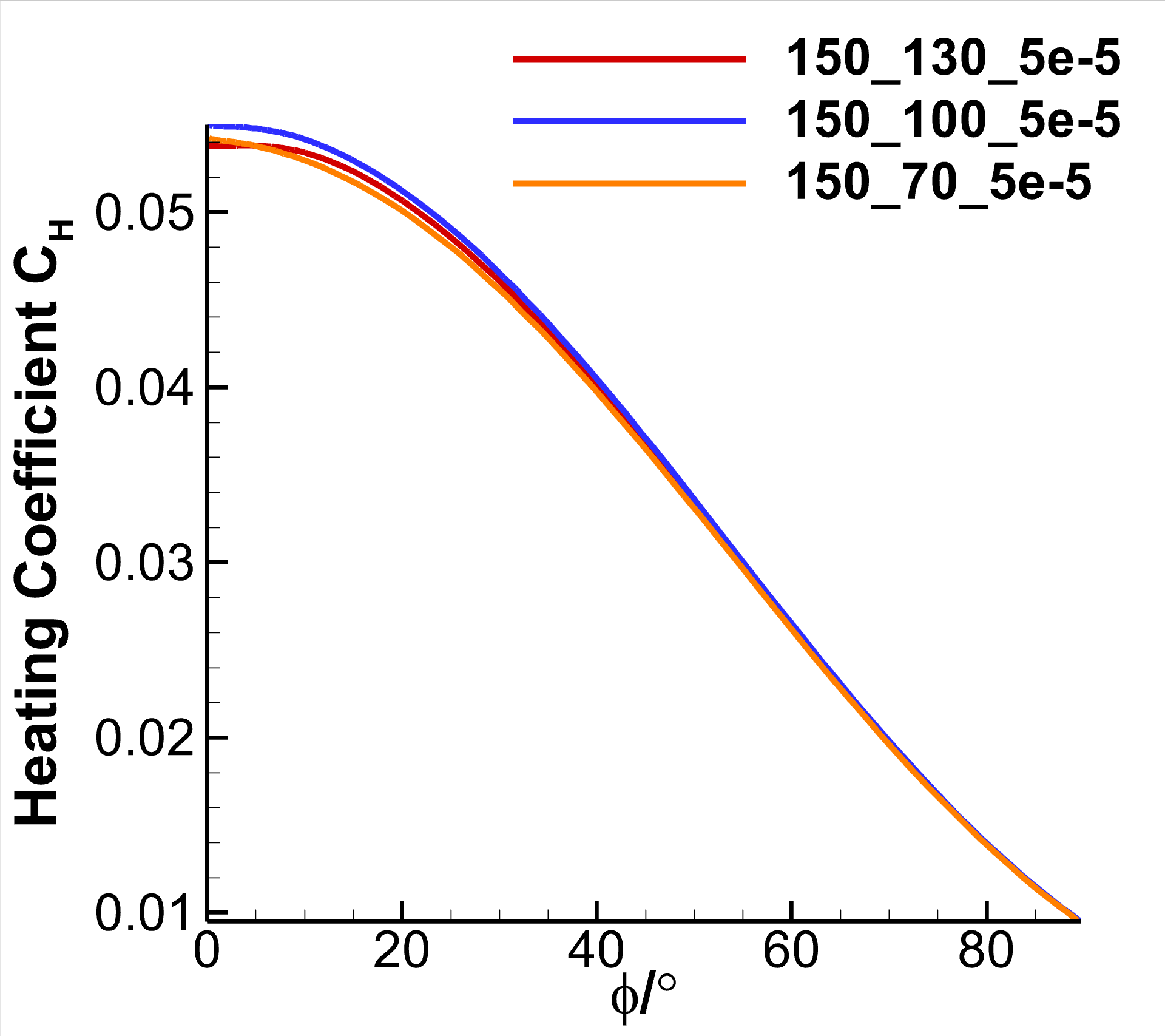}
	\vspace{-4mm} \caption{
	Effect of tangential grid resolution on surface pressure (left) and heat flux (right) for 4000K cylinder.}
	\label{grid_study_cylinder_surface1}
\end{figure*}

\begin{figure*}[htp]	
	\centering
	\includegraphics[width=0.4\textwidth]
	{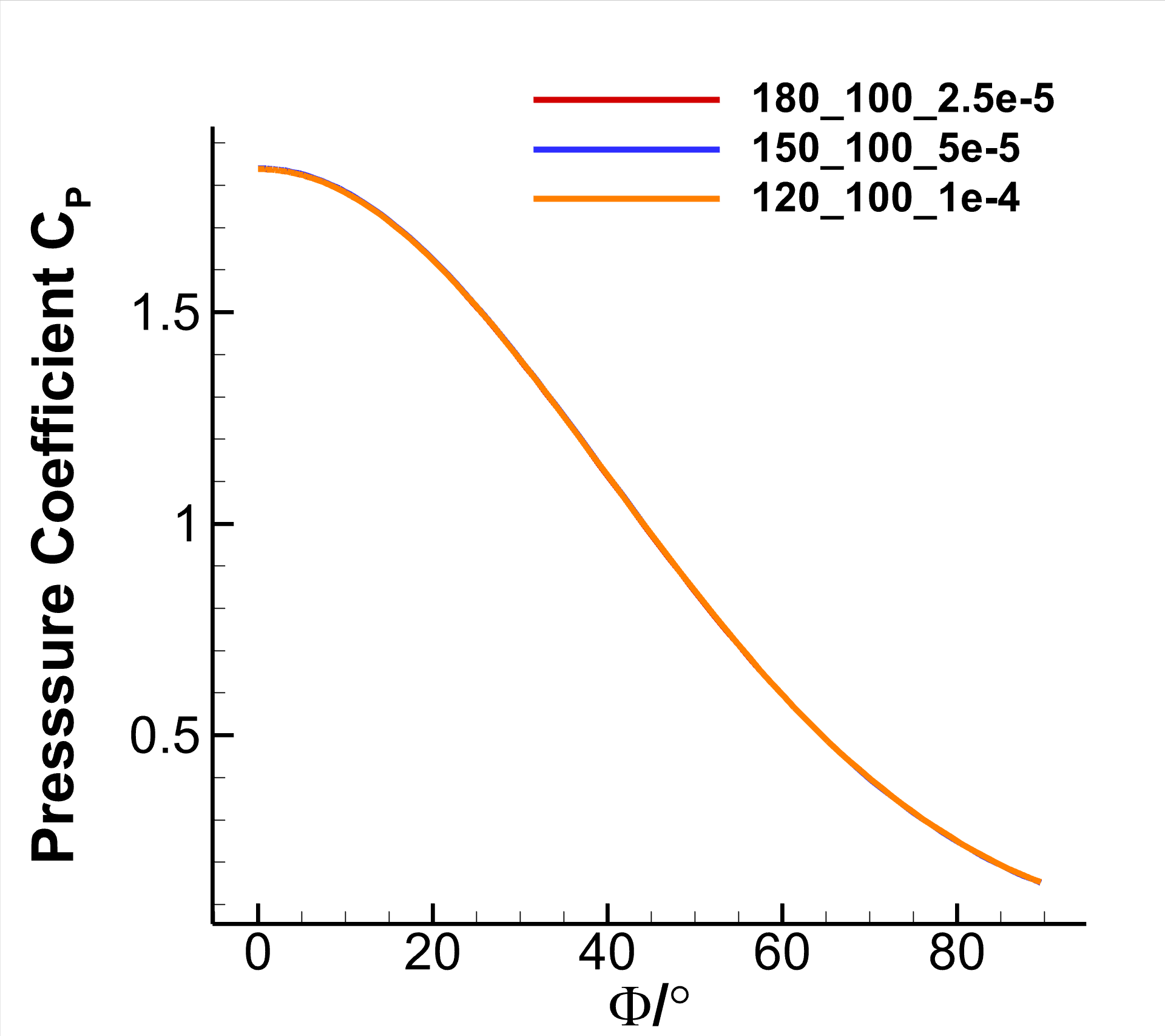}
	\includegraphics[width=0.4\textwidth]
	{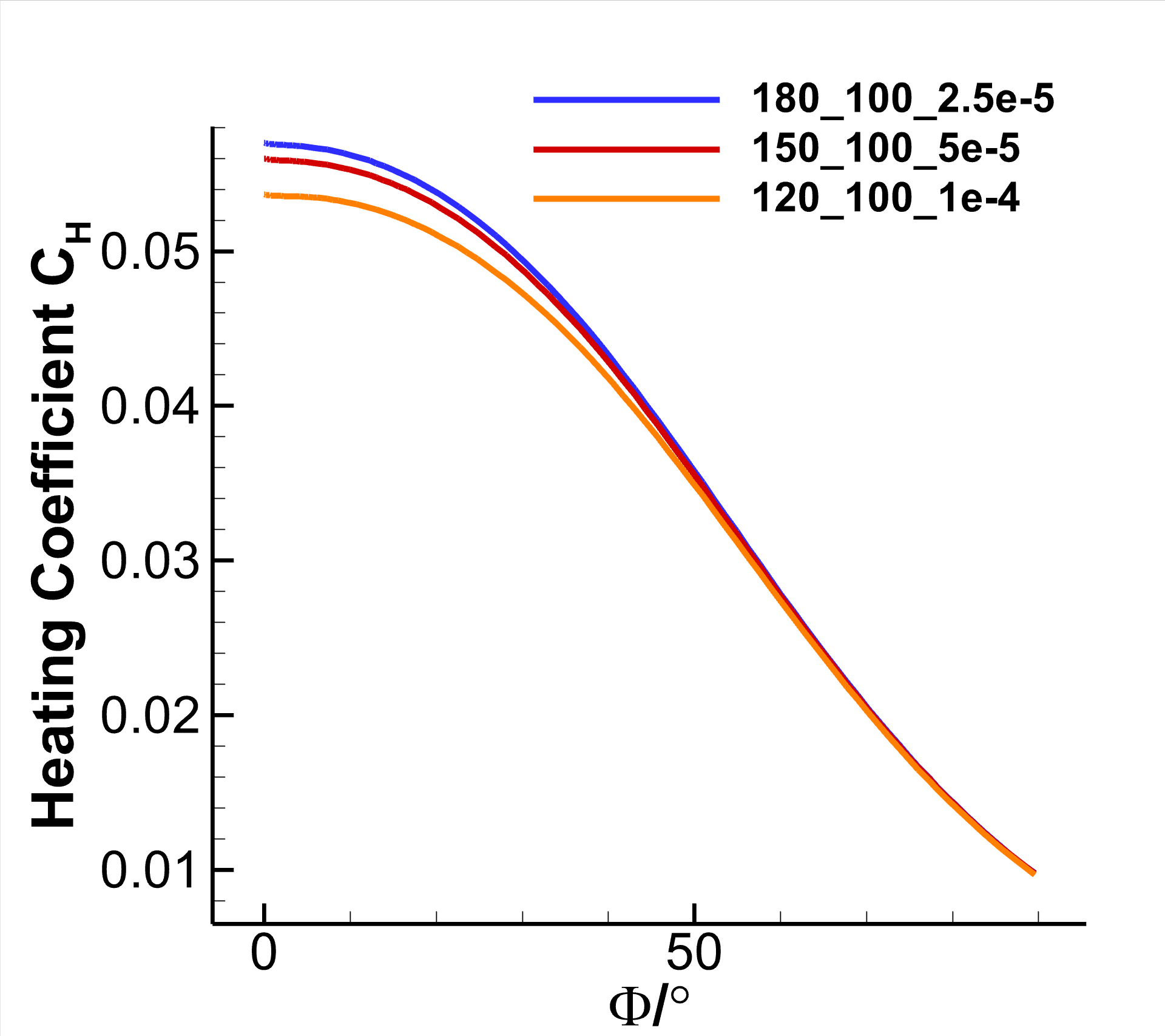}
	\vspace{-4mm} \caption{
	Efect of normal grid resolution and first-layer cell height on surface pressure (left) and heat flux (right) for 4000K cylinder.}
	\label{grid_study_cylinder_surface2}
\end{figure*}

Similar to the previous section, a grid independence study is conducted, 
confirming that the wall-normal spacing has the most significant impact on surface properties. 
The grid is also refined in the shock region to better capture the shock structure. 
The meshes used for the verification are summarized 
in Table~\ref{grid_study_cylinder}, and the corresponding results are shown in Fig.~\ref{grid_study_cylinder_surface1} and  
Fig.~\ref{grid_study_cylinder_surface2}. 

\begin{figure}[htp]	
	\centering
	\includegraphics[width=0.4\textwidth]{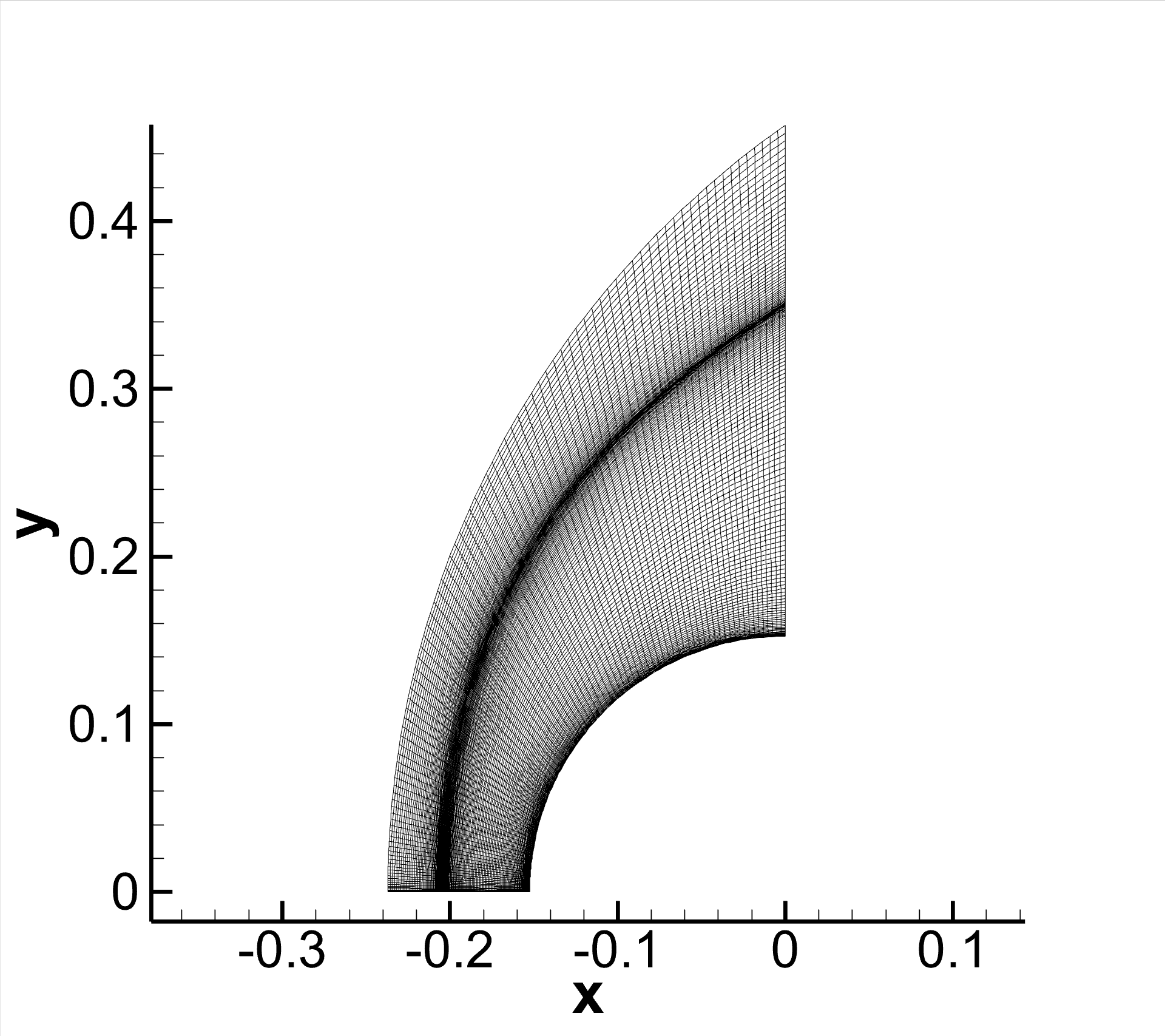}
	\vspace{-4mm} 
	\caption{Global view of the computational grid for 4000K cylinder.}
	\label{cylinderM_grid}
\end{figure}

The final grid-independent solution is shown in Fig.~\ref{cylinderM_grid}, 
consisting of 150 cells in the normal direction and 100 cells in the tangential direction, 
with the first-layer cell height set to $5\times10^{-5}$, 
corresponding to a cell Reynolds number $Re_{\text{cell}}=1.1359$.

It can be seen that, for the mesh with $150 \times 100$ cells and a first-layer height of $5 \times 10^{-5},
\mathrm{m}$, further refinement in both directions has no noticeable effect on the surface pressure and only 
results in less than 1\% change in the surface heat flux, indicating that the mesh is sufficiently resolved for 
accurate prediction of surface quantities.

\begin{table*}[htbp]
\centering
\caption{Grid configurations used for grid independence study for the moderate-temperature cylinder case.}
\label{grid_study_cylinder}

\begin{tabular}{ccc}
\hline
\textbf{Tangential cells} & \textbf{Normal cells} & \textbf{First-layer height [m]} \\
\hline
\multicolumn{3}{c}{\textit{Part 1: Tangential grid refinement}} \\
150  & 70  & $5 \times 10^{-5}$ \\
150  & 100 & $5 \times 10^{-5}$ \\
150  & 130 & $5 \times 10^{-5}$ \\
\hline
\multicolumn{3}{c}{\textit{Part 2: Normal grid / first-layer height refinement}} \\
120  & 100 & $1 \times 10^{-4}$ \\
150  & 100 & $5 \times 10^{-5}$ \\
180  & 100 & $2.5 \times 10^{-5}$ \\
\hline
\end{tabular}
\end{table*}

Similar to the previous section, a grid independence study is conducted, 
confirming that the wall-normal spacing has the most significant impact on surface properties. 
The grid is also refined in the shock region to better capture the shock structure. 
The final grid-independent solution is shown in Fig.~\ref{cylinderM_grid}, 
consisting of 150 cells in the normal direction and 100 cells in the tangential direction, 
with the first-layer cell height set to $5\times10^{-5}$, 
corresponding to a cell Reynolds number $Re_{\text{cell}}=1.1359$.

\subsubsection{Flow field properties}
\begin{figure*}[htp]	
	\centering
	\includegraphics[width=0.3\textwidth]
	{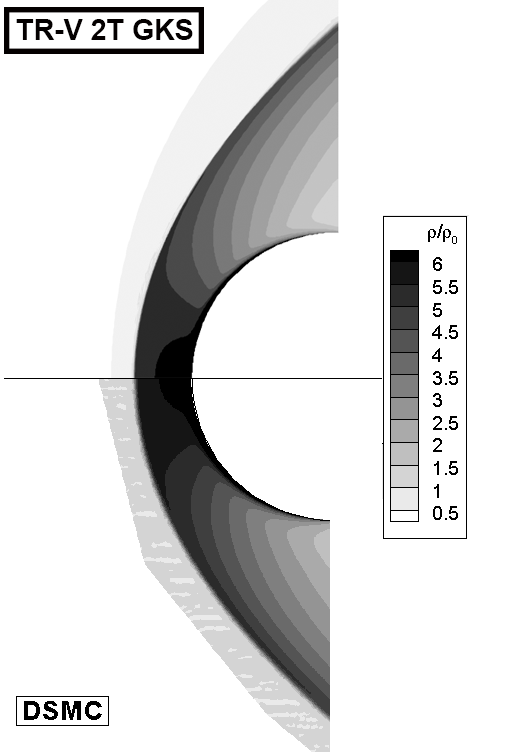}
	\includegraphics[width=0.32\textwidth]
	{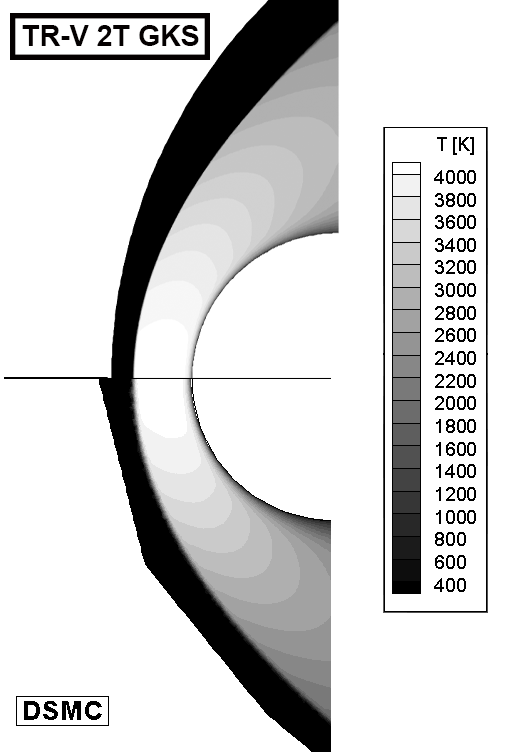}
	\includegraphics[width=0.3\textwidth]
	{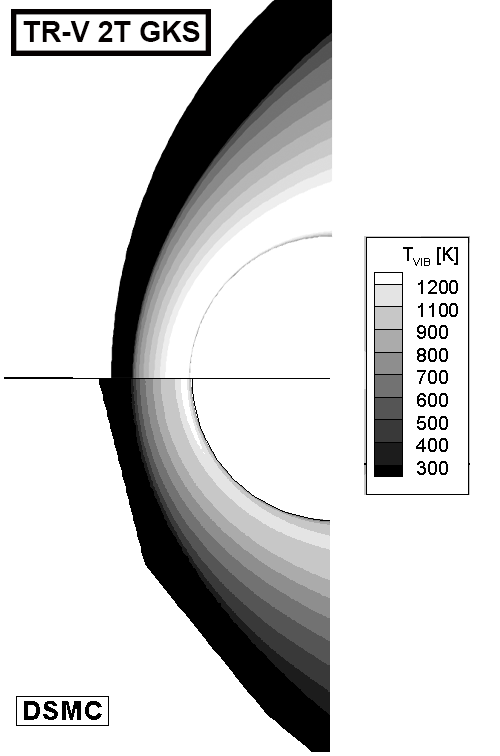}
	\vspace{-4mm} 
	\caption{
	Density ratio field (left), translational-rotational temperature field (middle)
	and Vibrational temperature field (right).
	top: TR-V 2T GKS. bottom: DSMC.}
	\label{cylinderM_contour}
\end{figure*}

The fields of density, translational-rotational temperature, and vibrational temperature are shown in 
Fig.~\ref{cylinderM_contour}, where the density is normalized by the freestream density. 
It is evident that in the post-shock region near the cylinder nose, a peak temperature of approximately 4000 K 
is reached. At this temperature, nitrogen is generally considered to undergo noticeable dissociation; hence, 
this case serves as a suitable benchmark to assess the applicability of the chemically frozen-flow assumption 
employed in this work.
The predicted shock standoff distance shows excellent agreement with the DSMC results,
while both the density and translational-rotational temperature distributions match closely,
and the vibrational temperature exhibits only a slight deviation, indicating that the present method 
is able to accurately predict the main flow features under such thermal conditions.

\subsubsection{Stagnation line properties}

In cylinder flow, the stagnation line is defined as the streamline that passes through the stagnation point, 
where the velocity of the flow drops to zero due to direct impingement on the surface. 
This line represents the axis of symmetry of the incoming flow and plays a critical role 
in determining the distributions of pressure, temperature, and density. 
In this section, stagnation line temperatures are compared with those obtained by 
Lofthouse~\cite{Lofthouse2008} using the DSMC method.

\begin{figure}[htp]	
	\centering
	\includegraphics[width=0.4\textwidth]{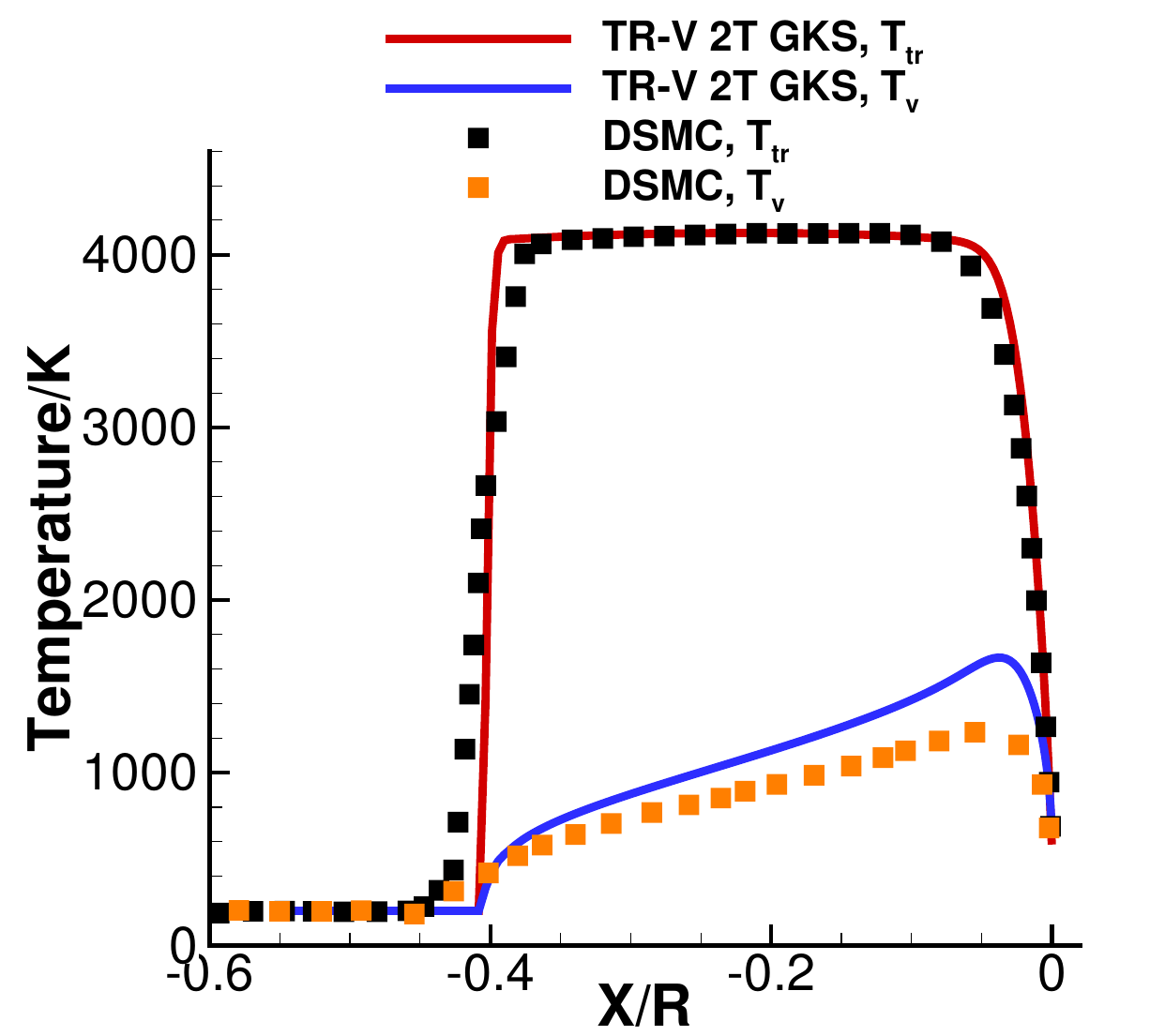}
	\vspace{-4mm} 
	\caption{Comparison of stagnation line temperatures between TR-V 2T GKS and the DSMC method.}
	\label{cylinderM_stagnation}
\end{figure}

Figure~\ref{cylinderM_stagnation} presents comparisons of stagnation line temperatures 
between TR-V 2T GKS and the DSMC method. Consistent with the observations above, the translational-rotational 
temperature shows good agreement, whereas the vibrational temperature is slightly overpredicted.

\subsubsection{Surface properties}
\begin{figure*}[htp]	
	\centering
	\includegraphics[width=0.3\textwidth]{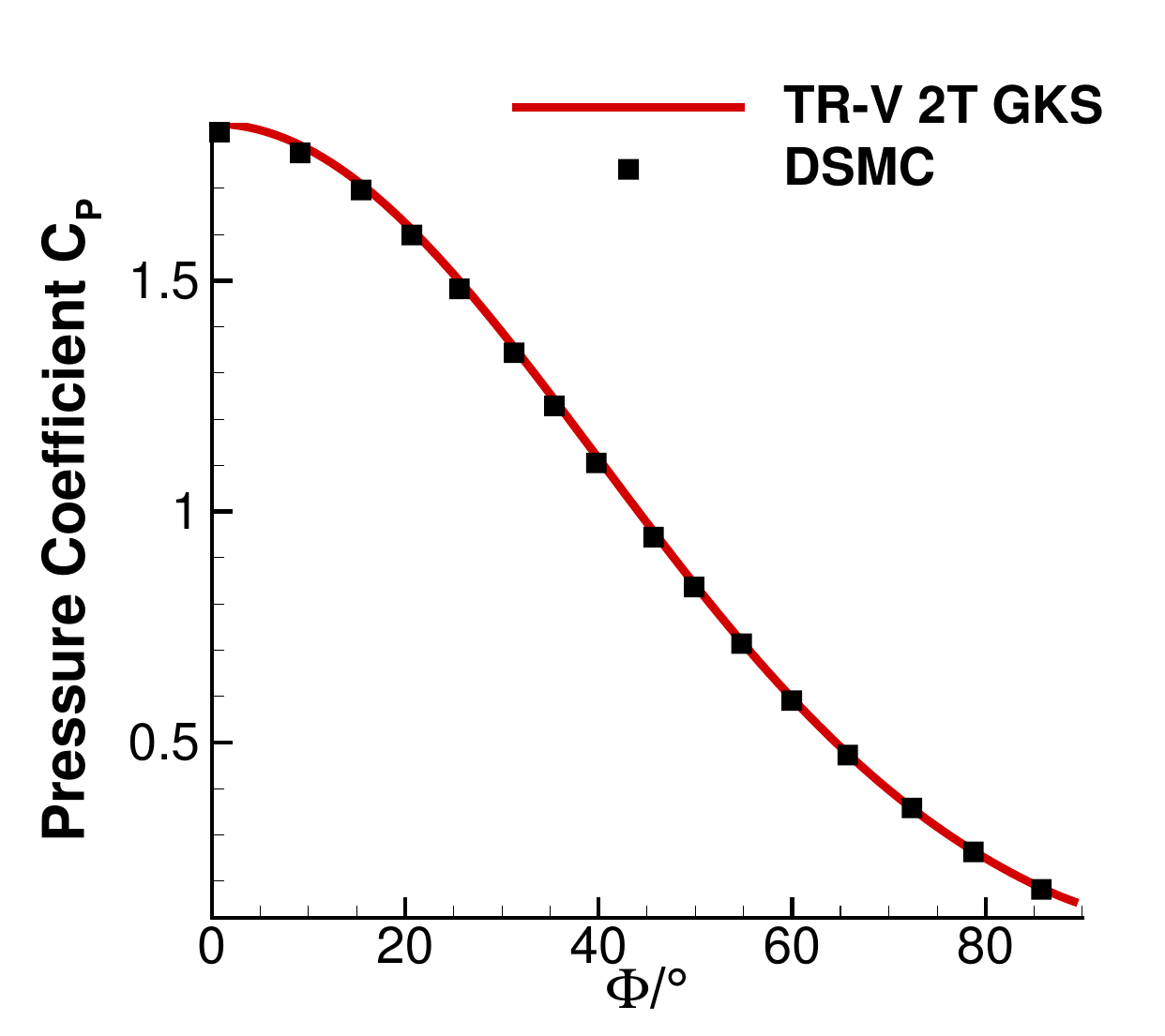}
	\includegraphics[width=0.3\textwidth]{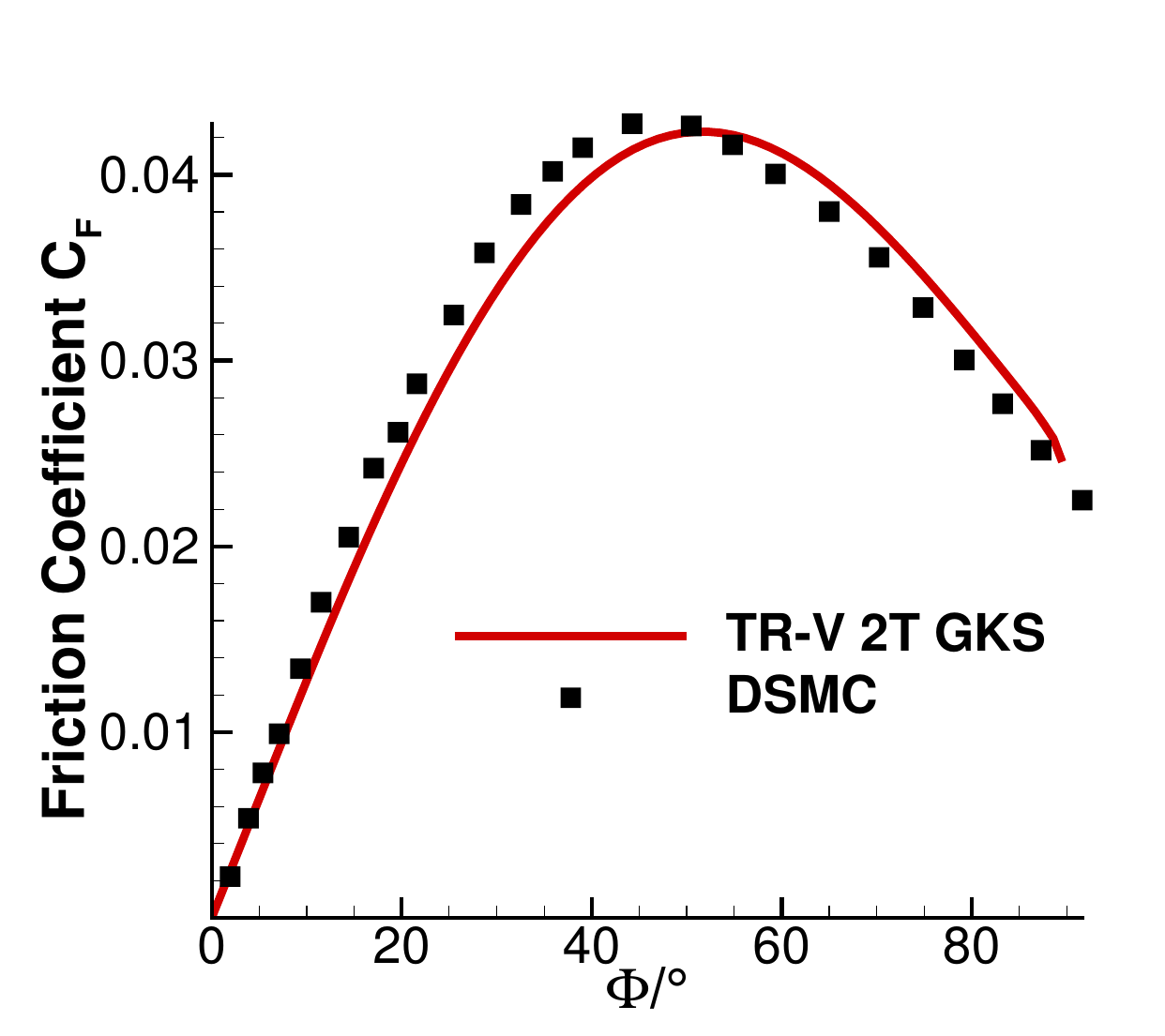}
	\includegraphics[width=0.3\textwidth]{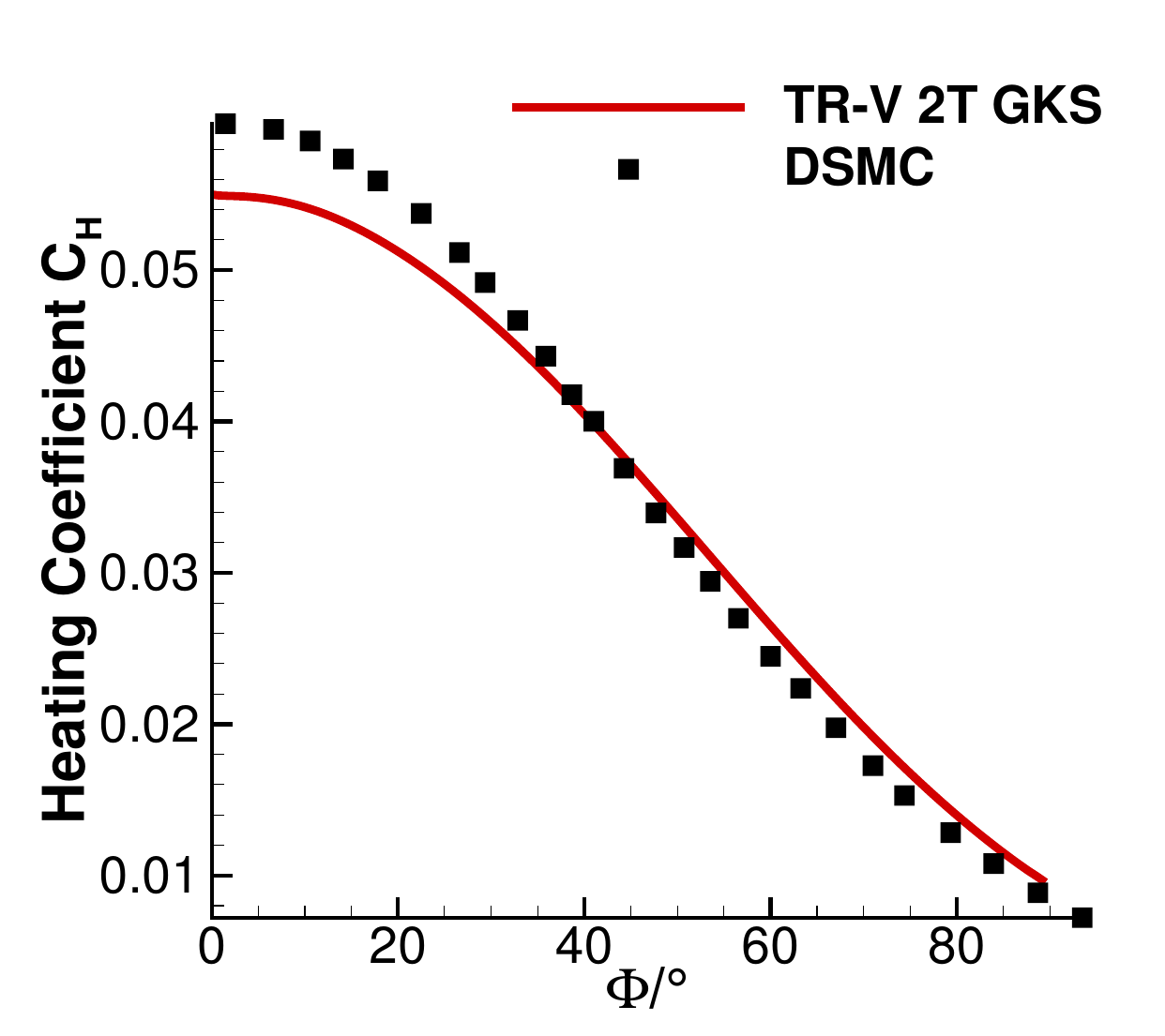}
	\vspace{-4mm} 
	\caption{
	The surface pressure coefficient (left), friction coefficient (center), and heating coefficient (right).}
	\label{cylinderM_surface}
\end{figure*}

Similar to the previous case, we compute the surface properties along the cylinder surface.  
Fig.~\ref{cylinderM_surface} shows a comparison between numerical results and experimental data  
for surface pressure, shear stress and heat flux along the cylinder, 
using the same nondimensional parameters as those defined in Eq.~\eqref{nondim_surface} of the previous section.
The x-axis represents the rotational angle measured from the stagnation point,  
as illustrated in Fig.~\ref{cylinder_geometry}. 

The predicted surface pressure shows good agreement with the experimental data.
It can be seen that the pressure coefficient agrees very well with the measurements, 
the friction coefficient exhibits only minor deviations, and the heating coefficient shows an 
underestimation of approximately 10\% near the cylinder nose, where the heat transfer peaks.

Overall, even at a peak temperature of approximately 4000 K, the chemically frozen-flow assumption is capable 
of accurately predicting the flow field and yielding reliable surface properties, thereby confirming its 
continued applicability under such thermal conditions.
\begin{color}{blue}
\subsection{2D Cylinder Flow at 9500 K (Limitation Case)}
\end{color}
To further assess the applicability range of the present method, a higher-temperature cylindrical case is 
considered in this section.
The experiments were performed in the High Enthalpy Shock Tunnel Göttingen (HEG) 
at the German Aerospace Center (DLR) 
\textcolor{blue}{under a freestream Mach number of $M_{\infty} = 8.78$}. 
\textcolor{blue}{This HEG test condition (DLR, HEG facility, $M = 8.78$) corresponds to the dataset reported 
in Ref.~\cite{Martinez2002}.}
The study measured the density field, as well as the surface pressure and heat flux.
Li \textit{et al.} conducted numerical simulations of this case using the AUSMPW+ scheme, 
coupled with a two-temperature model and a chemical reaction model~\cite{Li2013}. 
The AUSMPW+ scheme is an improved version of the AUSM-type schemes.

The cylinder, as shown in Fig.~\ref{cylinder_geometry}, has a diameter of $0.09\,\mathrm{m}$.  
The freestream gas is air, and the initial conditions are specified as:
\begin{equation}
\begin{gathered}
M_{\infty} = 8.78, \quad T_{\infty} = 694\,\mathrm{K},\\
\quad \rho_{\infty} = 3.26 \times 10^{-3}\,\mathrm{kg/m^3}, \quad Re_{\infty} = 42288.
\end{gathered}
\end{equation}

A slip boundary condition is applied at the wall. 
Since the experiment duration is approximately $2.5\,\mathrm{ms}$, 
the wall temperature is assumed to be the ambient temperature of $300\,\mathrm{K}$. 
The vibrational characteristic temperature of air is obtained by weighting the vibrational 
characteristic temperatures of nitrogen and oxygen according to their volume fractions. 

\begin{figure}[htp]	
	\centering
	\includegraphics[width=0.4\textwidth]{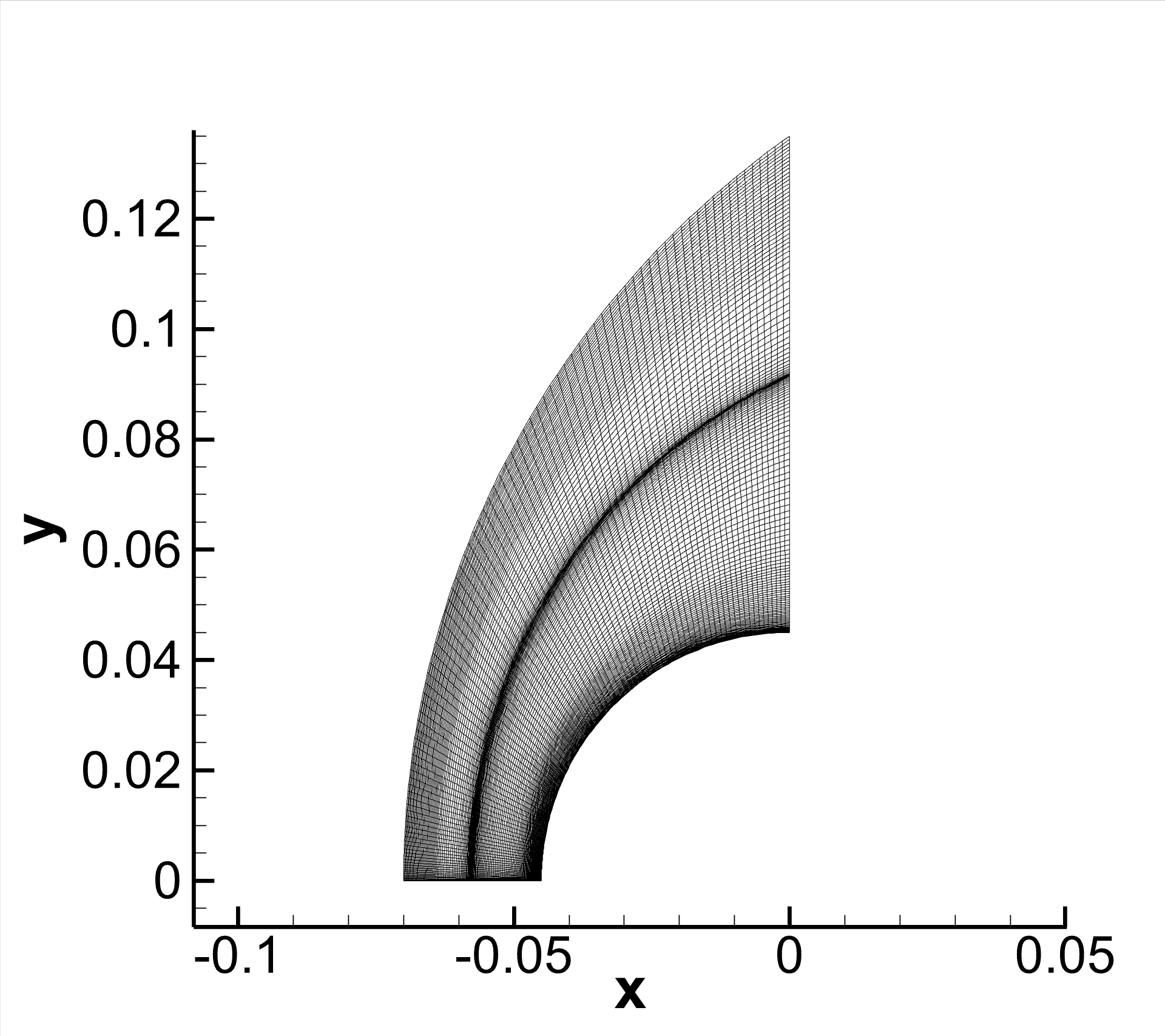}
	\vspace{-4mm} 
	\caption{Global view of the computational grid for 9500K cylinder.}
	\label{cylinder_grid}
\end{figure}

This case is computed using a structured grid and a structured solver. 
Similar to the previous section, a grid independence study is conducted, 
confirming that the wall-normal spacing has the most significant impact on surface properties. 
The grid is also refined in the shock region to better capture the shock structure. 
The final grid-independent solution is shown in Fig.~\ref{cylinder_grid}, 
consisting of 150 cells in the normal direction and 100 cells in the tangential direction, 
with the first-layer cell height set to $1.5\times10^{-6}$, 
corresponding to a cell Reynolds number $Re_{\text{cell}}=0.7048$.
Due to space limitations, the detailed results of the grid independence verification for this 
case are not presented.
\subsubsection{Flow field properties}

\begin{figure}[htp]	
	\centering
	\includegraphics[width=0.4\textwidth]{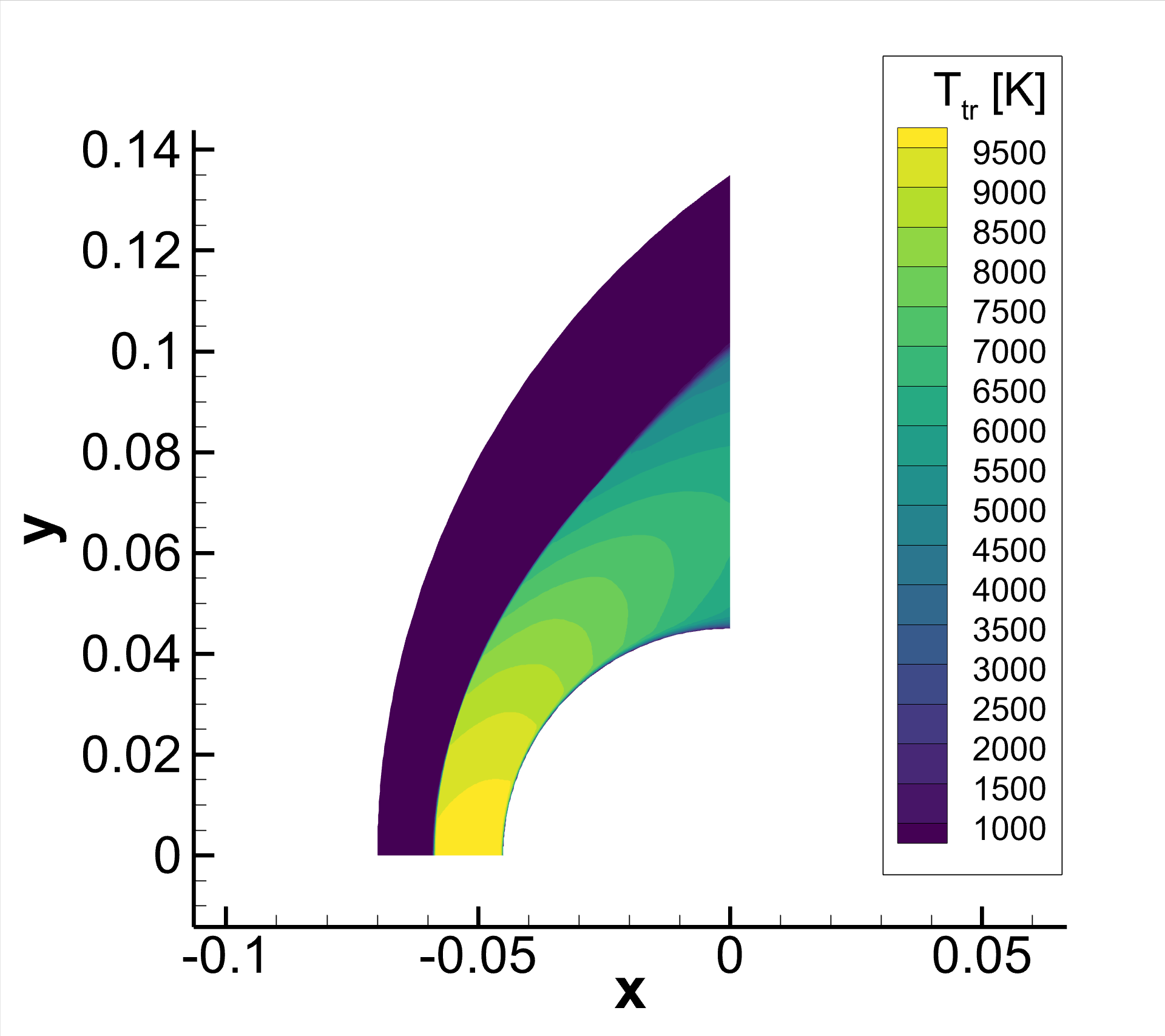}
	\vspace{-4mm} 
	\caption{Computed translational-rotational temperature field.}
	\label{cylinder_contour1}
\end{figure}
\begin{figure}[htp]	
	\centering
	\includegraphics[width=0.2\textwidth]{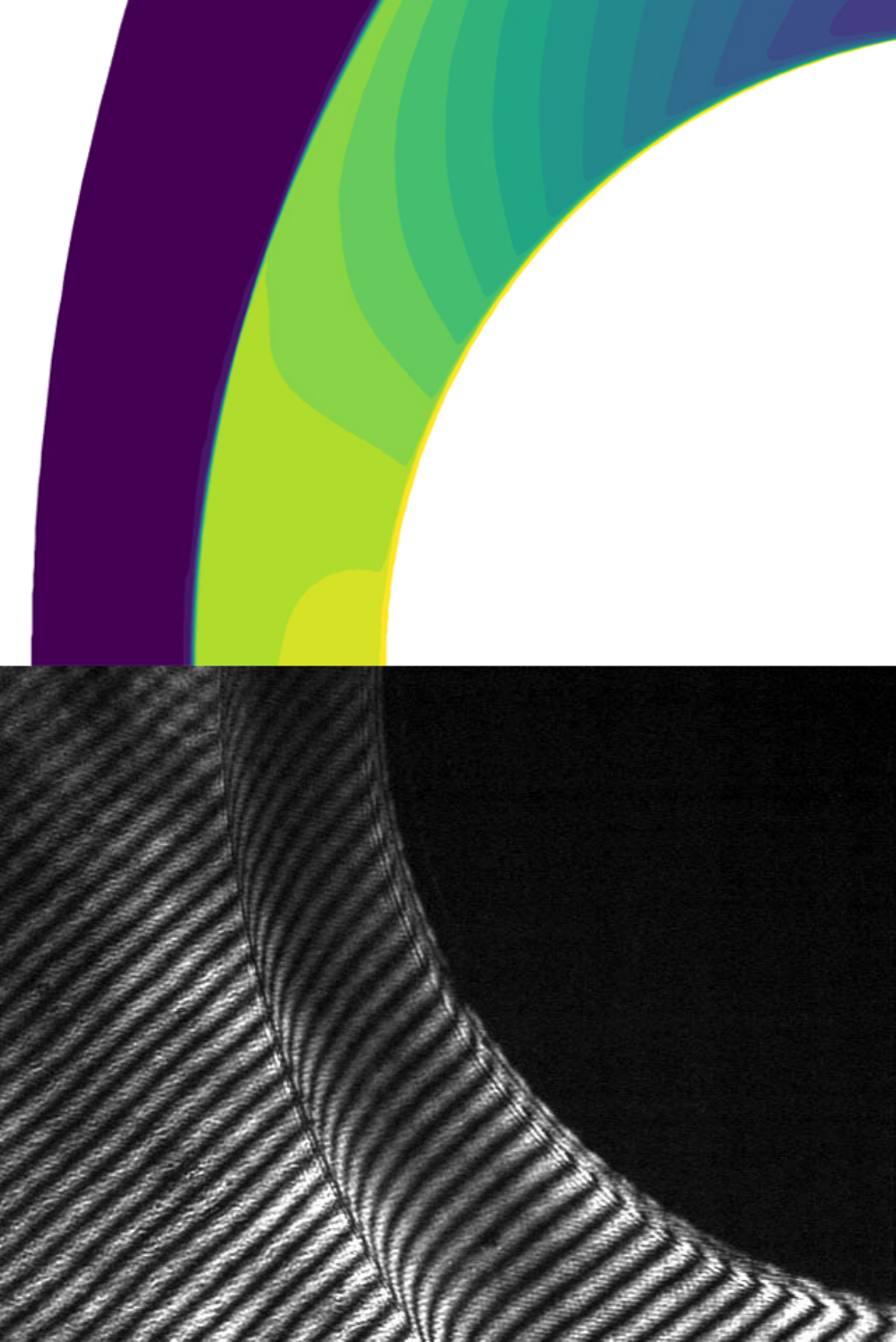}
	\caption{Density field comparison. Top: TR-V 2T GKS. 
	Bottom: HEG high-enthalpy wind tunnel experiment.}
	\label{cylinder_contour2}
\end{figure}

Fig.~\ref{cylinder_contour1} shows the computed translational-rotational 
temperature field. From Fig.~\ref{cylinder_contour1}, it can be observed that the temperature behind the shock 
at the cylinder's leading edge reaches 9500\,K, at which both oxygen and nitrogen molecules in air are expected 
to be highly dissociated, and noticeable ionization has already occurred.
Figure~\ref{cylinder_contour2} presents a comparison between the computed density field and the experimental 
results, where a noticeable deviation is observed in the predicted shock standoff distance using the present method.

\subsubsection{Stagnation line properties}

In this section, stagnation line properties are compared with those obtained by Li \textit{et al.}~\cite{Li2013}, 
who employed the calorically perfect gas model and the two-temperature model.

\begin{figure*}[htp]	
	\centering
	\includegraphics[width=0.4\textwidth]{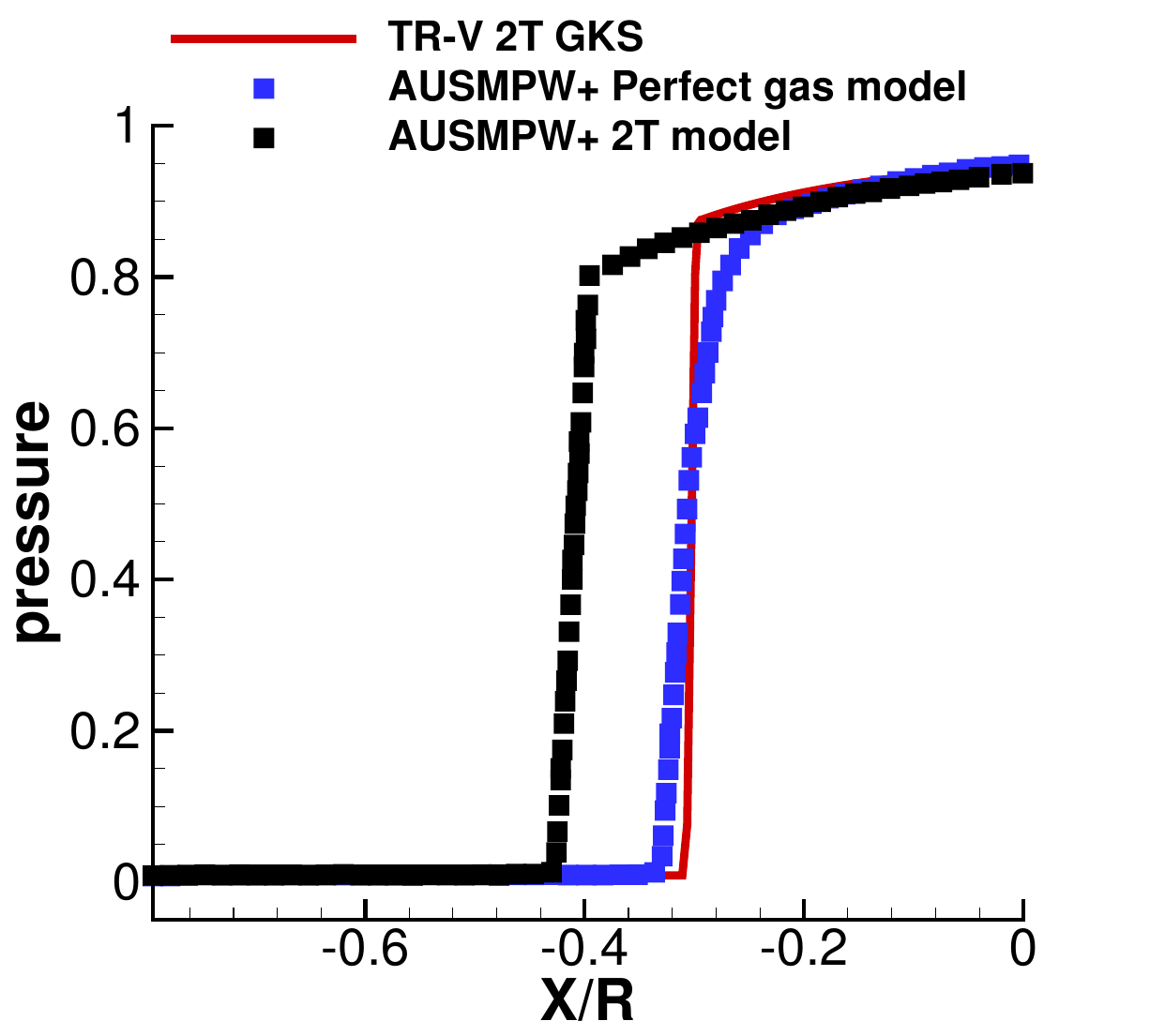}
	\includegraphics[width=0.4\textwidth]{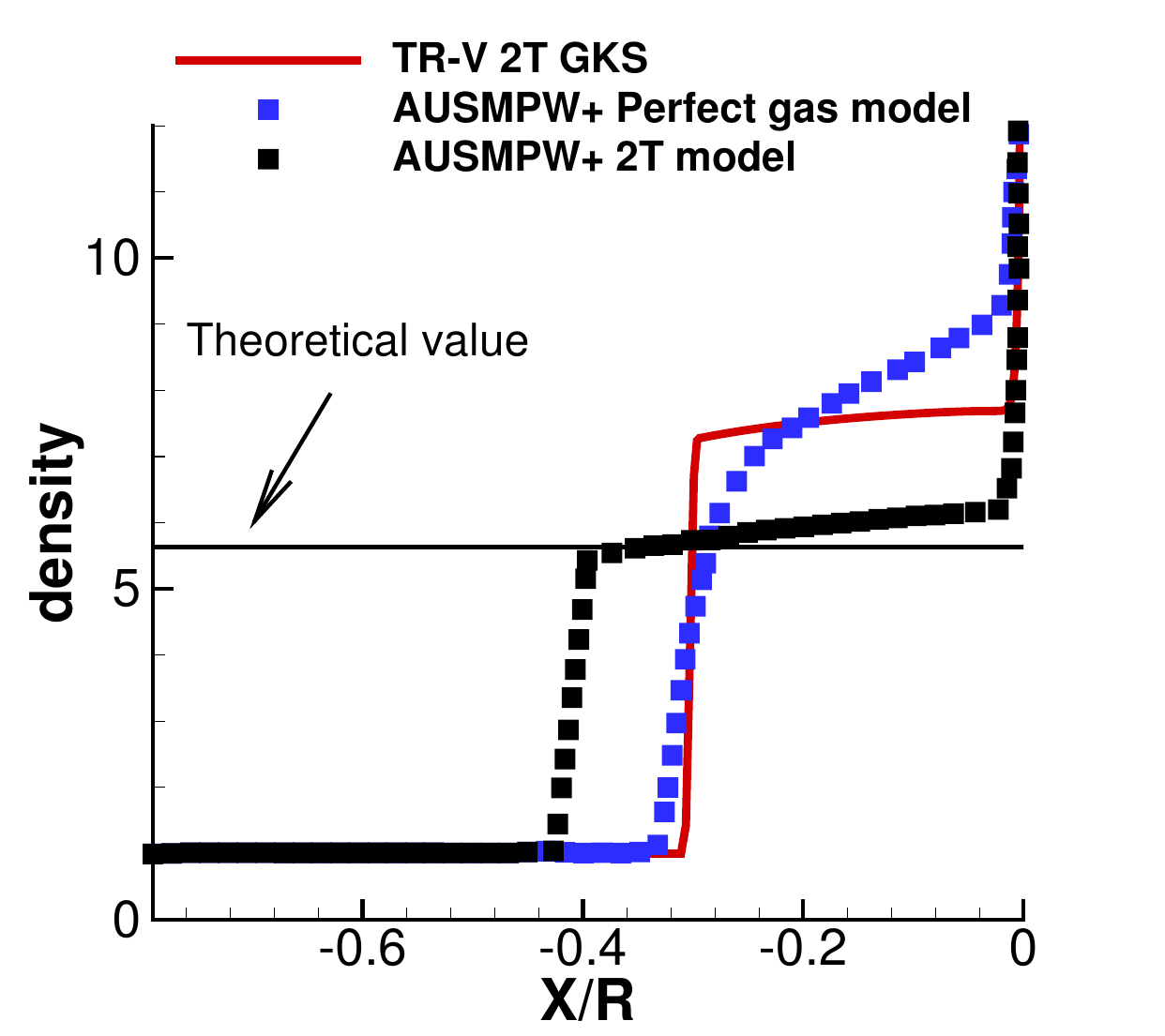}
	\vspace{-4mm} 
	\caption{Comparison of stagnation line properties between TR-V 2T GKS and the AUSMPW+ scheme 
	(using the calorically perfect gas model and the two-temperature model). 
	Left: Pressure; Right: Density.}
	\label{cylinder_stagnation1}
\end{figure*}

\begin{figure}[htp]	
	\centering
	\includegraphics[width=0.4\textwidth]{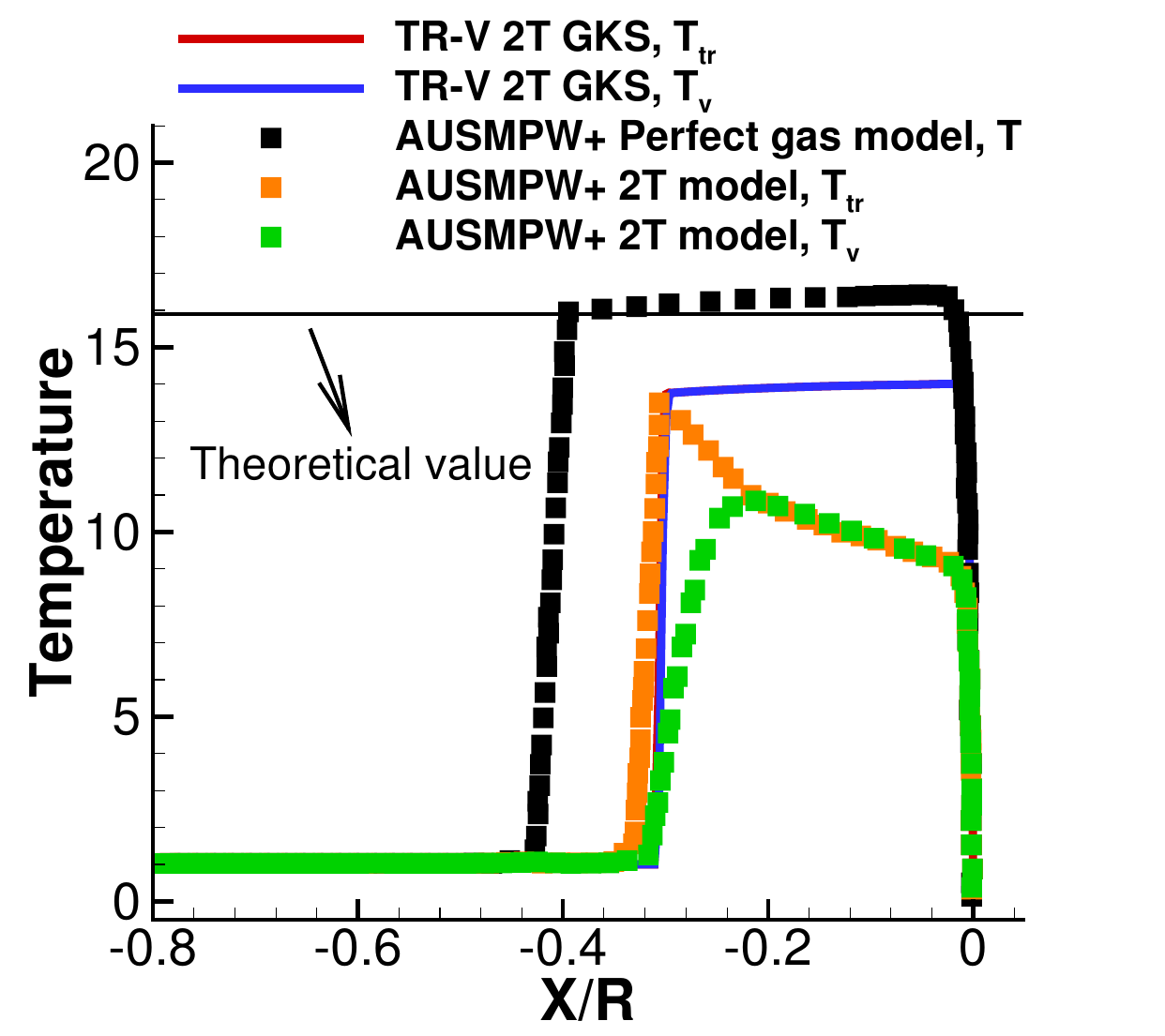}
	\vspace{-4mm} 
	\caption{Comparison of stagnation line temperature between TR-V 2T GKS and the AUSMPW+ scheme 
	(using the calorically perfect gas model and the two-temperature model).}
	\label{cylinder_stagnation2}
\end{figure}

Figure~\ref{cylinder_stagnation1} presents comparisons of stagnation line pressure and density 
between TR-V 2T GKS and the AUSMPW+ scheme using both the calorically perfect gas model and 
the two-temperature model. Figure~\ref{cylinder_stagnation2} compares the corresponding temperature profiles. 
The physical quantities on the $y$-axis are non-dimensionalized as follows:
\begin{equation}
	\begin{gathered}
		\bar{p} = \frac{p^*}{\rho_\infty V_\infty^2}, \\
		\bar{\rho} = \frac{\rho^*}{\rho_\infty}, \\
		\bar{T} = \frac{T^*}{T_\infty},
	\end{gathered}
\end{equation}
where the superscript “$\mspace{1mu}*\mspace{1mu}$” denotes dimensional quantities, “$\infty$” denotes 
freestream reference values, 
and the overbar “$\bar{\mspace{2mu}\cdot\mspace{2mu}}$” represents nondimensional quantities. 
The $x$-axis is normalized by the cylinder radius, with $x = 0$ corresponding to the stagnation point. 
In the density and temperature plots, theoretical values refer to those predicted by 
the calorically perfect gas normal shock relations.

As shown in Fig.~\ref{cylinder_stagnation1}, 
compared to the calorically perfect gas model (which further overpredicts the standoff distance), 
TR-V 2T GKS yields more accurate predictions by incorporating vibrational energy excitation. 
Moreover, the shock standoff distance predicted by TR-V 2T GKS closely matches the result of 
the two-temperature AUSMPW+ scheme coupled with chemical reactions. 
Both numerical models predict a higher post-shock density compared to the theoretical solution, 
which lacks vibrational effects; this leads to a reduced shock standoff distance.

However, it can also be observed that the post-shock density and pressure from the 
two-temperature AUSMPW+ scheme with chemistry vary more gradually from the shock 
to the wall. This behavior results from intense chemical non-equilibrium effects, 
where the relaxation time required for reactions to reach equilibrium is long, 
causing a slower evolution of flow properties behind the shock.

In the temperature distribution (Fig.~\ref{cylinder_stagnation2}), 
both TR-V 2T GKS and the AUSMPW+ scheme predict lower post-shock temperatures 
due to vibrational energy absorption. At this temperature level, 
vibrational relaxation occurs over a short time scale. 
Furthermore, the temperature computed by the AUSMPW+ scheme continues to decrease behind the shock, 
reflecting the dissociation of nitrogen and oxygen, which absorbs thermal energy and results in further cooling.

In summary, TR-V 2T GKS significantly improves predictions of shock thickness 
and post-shock properties compared to the AUSMPW+ scheme with a calorically perfect gas model. 
When compared to the AUSMPW+ scheme with the two-temperature and chemical reaction model, 
the predicted shock standoff distance is nearly identical, though some discrepancies remain 
in post-shock temperature and density distributions.

\subsubsection{Surface properties}
\begin{figure*}[htp]  
	\centering  
	\includegraphics[width=0.4\textwidth]{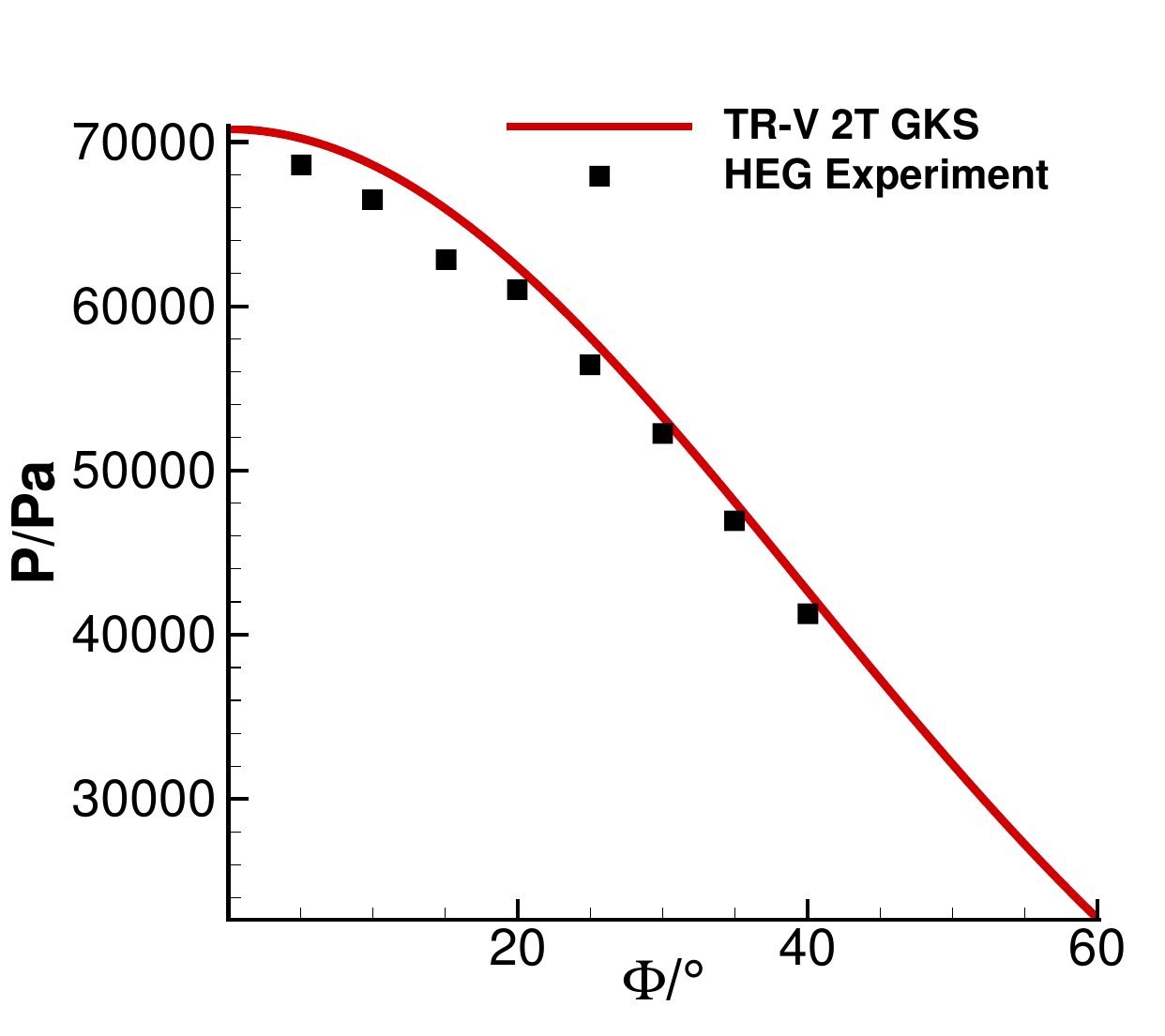}  
	\includegraphics[width=0.4\textwidth]{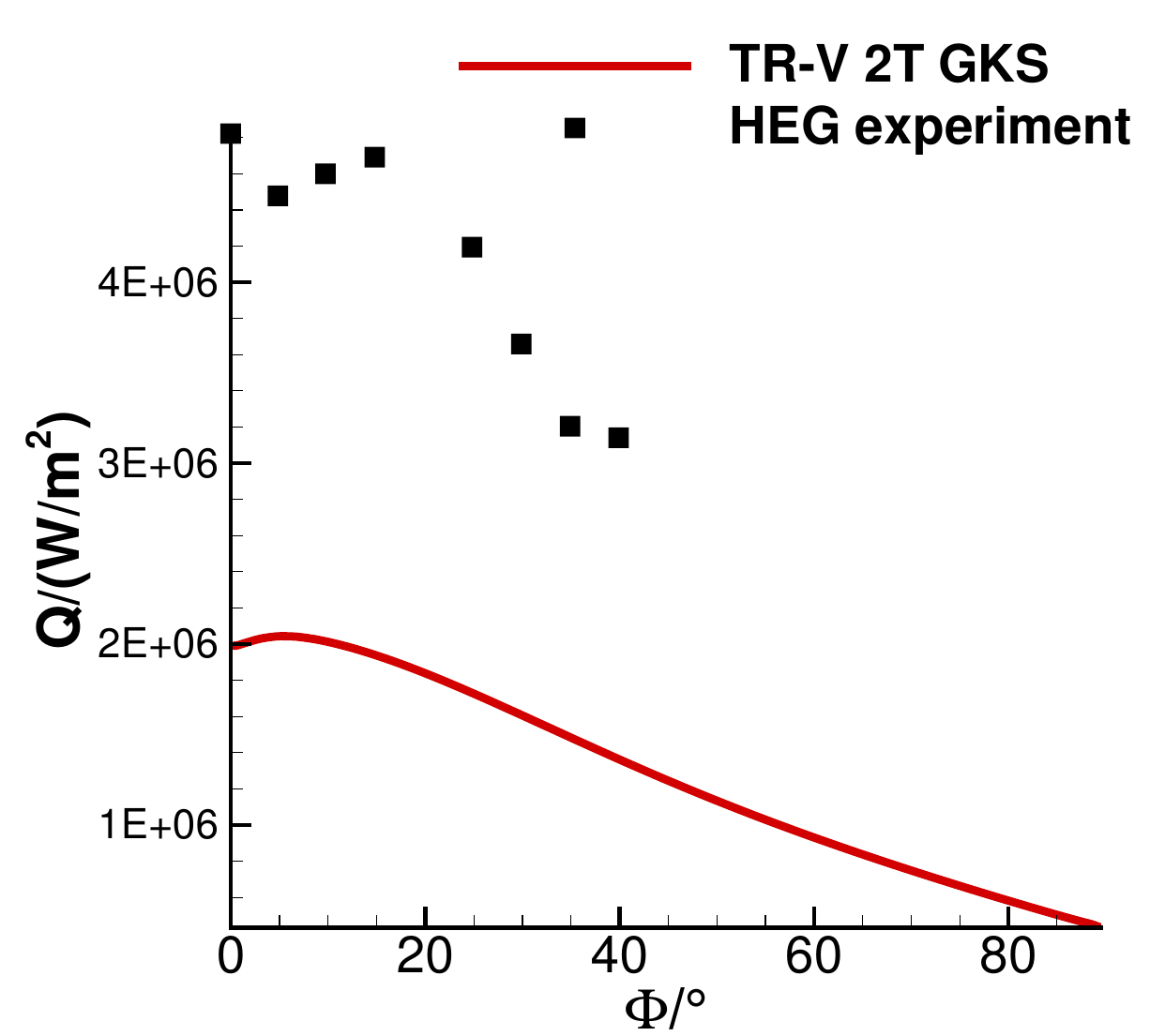} 
	\vspace{-4mm}  
	\caption{Comparison of surface pressure between TR-V 2T GKS and HEG experiments.}  
	\label{cylinder_surface}  
\end{figure*}

Similar to the previous case, we compute the pressure and heat flux along the cylinder surface.  
Fig.~\ref{cylinder_surface} shows a comparison between numerical results and experimental data  
for surface pressure along the cylinder.  
The x-axis represents the rotational angle measured from the stagnation point,  
as illustrated in Fig.~\ref{cylinder_geometry}.  

The predicted surface pressure agrees well with the experimental data; however, the surface heat-flux prediction 
shows a substantial discrepancy, with the peak value at the cylinder nose underestimated by a factor of 
approximately 2.5. This result indicates that, at such a peak temperature of 9500,K, neglecting chemical 
reactions renders the prediction of surface heat flux unreliable.

\subsection{2D Edney type IV shock/shock interaction}

Edney proposed a simplified canonical model for shock-shock interactions, in which a planar oblique shock  
impinges upon the detached bow shock formed ahead of a blunt cylindrical body~\cite{Edney1968}.  
Based on the relative position between the oblique shock and the bow shock, six representative types  
of shock interaction patterns were identified. Among these, the Type IV interaction has attracted particular  
attention due to the formation of a supersonic jet that impinges directly on the surface within the  
interaction region, leading to extremely high thermal and mechanical loads.  

This study focuses on the numerical simulation of this specific interaction type.  
The test was conducted by France's Office National d'Etudes et de Recherches Aerospatiales 
(ONERA)~\cite{Hedde1995,Thibert2000}. The experiments were performed in the ONERA R5CH wind tunnel.  
In addition, numerical simulation results obtained by Liu \textit{et al.}~\cite{Liu2021-3T} using the  
translational-rotational (T-R) and translational-rotational-vibrational (T-R-V) multi-temperature GKS  
are also included for comparison.  
In the following, T-R 2T GKS refers to the translational-rotational multi-temperature GKS method.
The free-stream air flow properties are  
\begin{equation}
	M_{\infty} = 9.95, \quad T_{\infty} = 52.5\,\mathrm{K}, \quad 
	p_{\infty} = 5.9\,\mathrm{Pa}, \quad Re_{\infty} = 2773.12.
\end{equation}

\begin{figure}[htp]	
	\centering
	\includegraphics[width=0.5\textwidth]{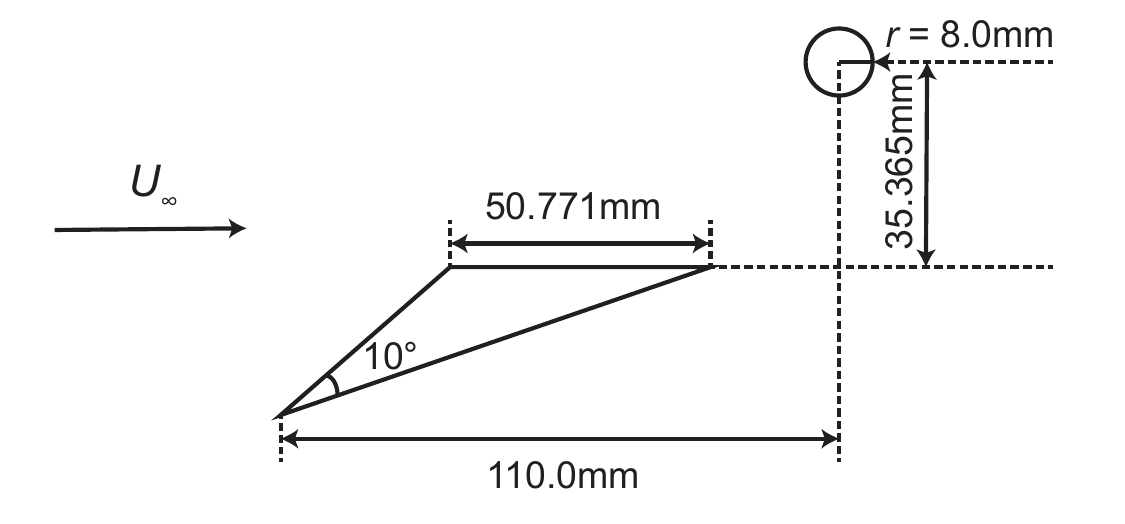}
	\vspace{0mm}
	\caption{
		Schematic for the configuration in ONERA test facility.}
	\label{shock_interaction_geometry}
\end{figure}

The schematic for the model configuration is shown in Fig.~\ref{shock_interaction_geometry}.  
An isosceles triangular wedge is employed as the shock generator to induce an oblique shock that  
interacts with the bow shock formed ahead of the cylindrical body.  
The spanwise width of the test facility was large enough to ensure that the case could be considered  
as a two-dimensional problem.  

\begin{figure*}[htp]	
	\centering
	\includegraphics[width=0.4\textwidth]{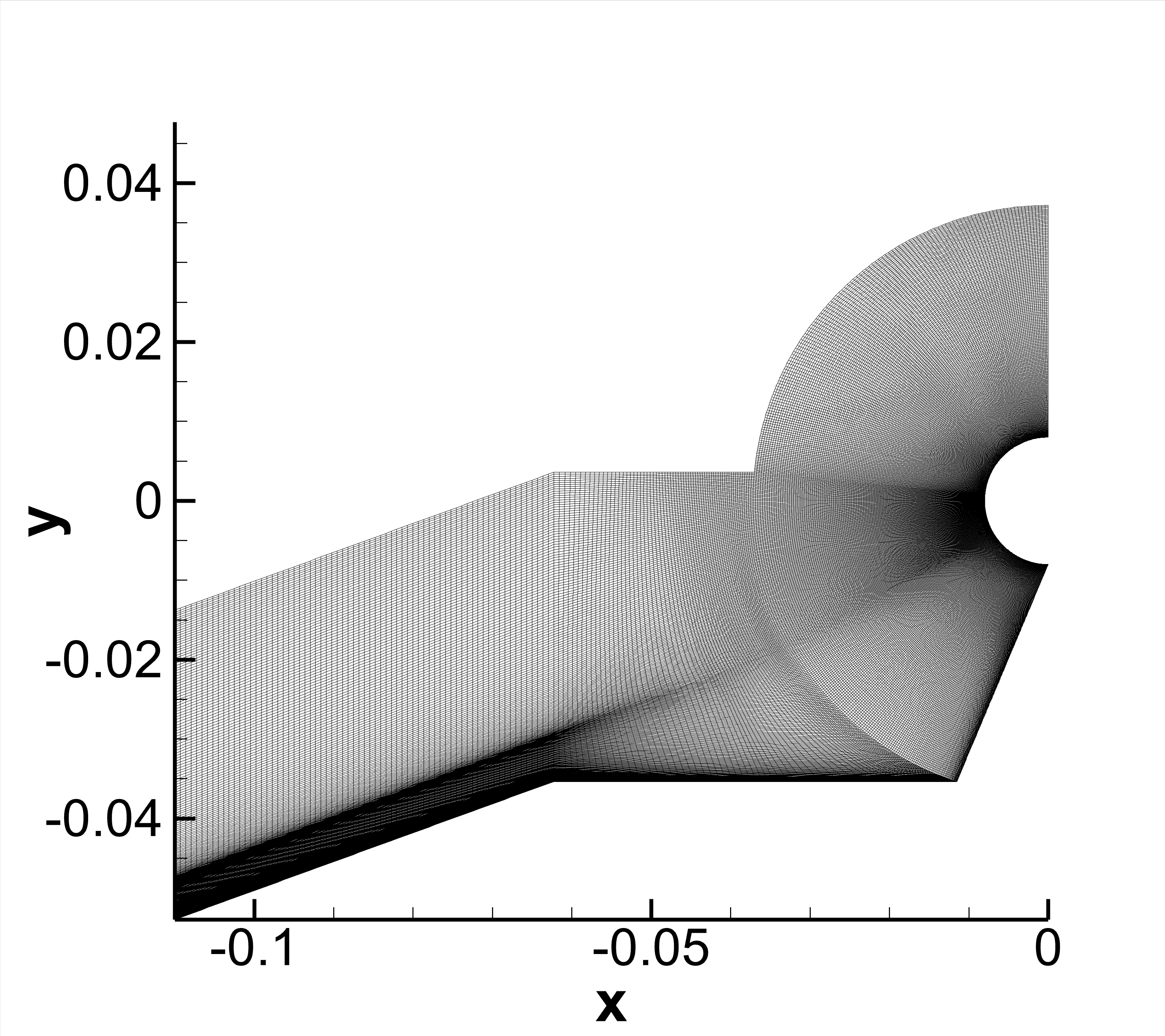}
	\includegraphics[width=0.4\textwidth]{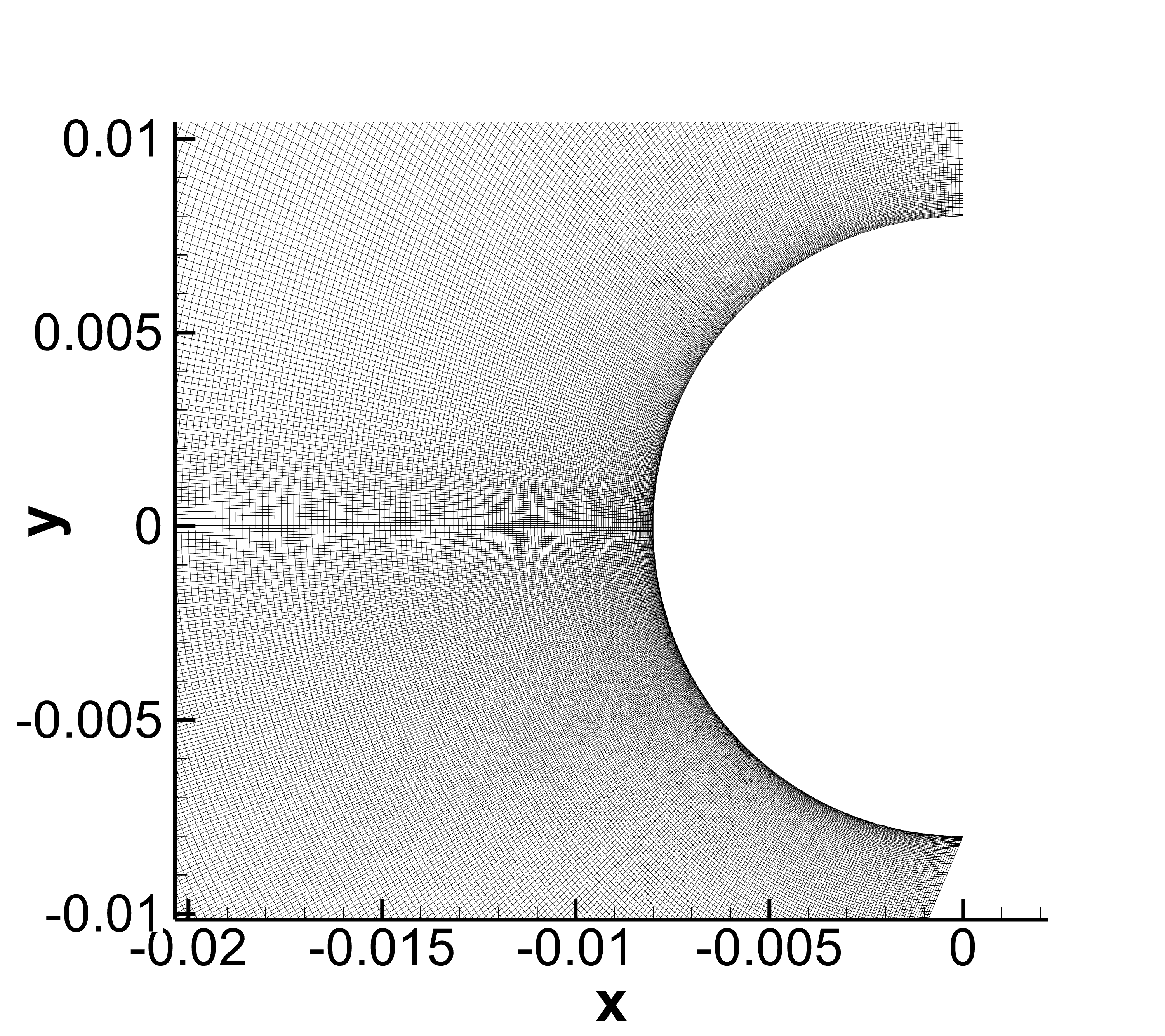}
	\vspace{-4mm}
	\caption{
		Schematic for grid: global view for the whole grid (left) and local view near the cylinder (right).}
	\label{shock_interaction_grid}
\end{figure*}

A quadrilateral unstructured grid and a corresponding solver are used for this case.  
A grid independence study is conducted, and the final grid-independent solution is shown in  
Fig.~\ref{shock_interaction_grid}.  
The cylindrical section consists of 200 cells in the normal direction and \textcolor{red}{300} 
cells in the tangential direction,  
with a total of 116{,}176 cells for the entire flow field.  
The first-layer cell height is set to $5\times10^{-6}\,\mathrm{m}$,  
corresponding to a cell Reynolds number $Re_{\text{cell}}=1.7332$.
Due to space limitations, the detailed results of the grid independence verification for this 
case are not presented.

\subsubsection{Flow field properties}
\begin{figure}[htp]	
	\centering
	\includegraphics[width=0.5\textwidth]
	{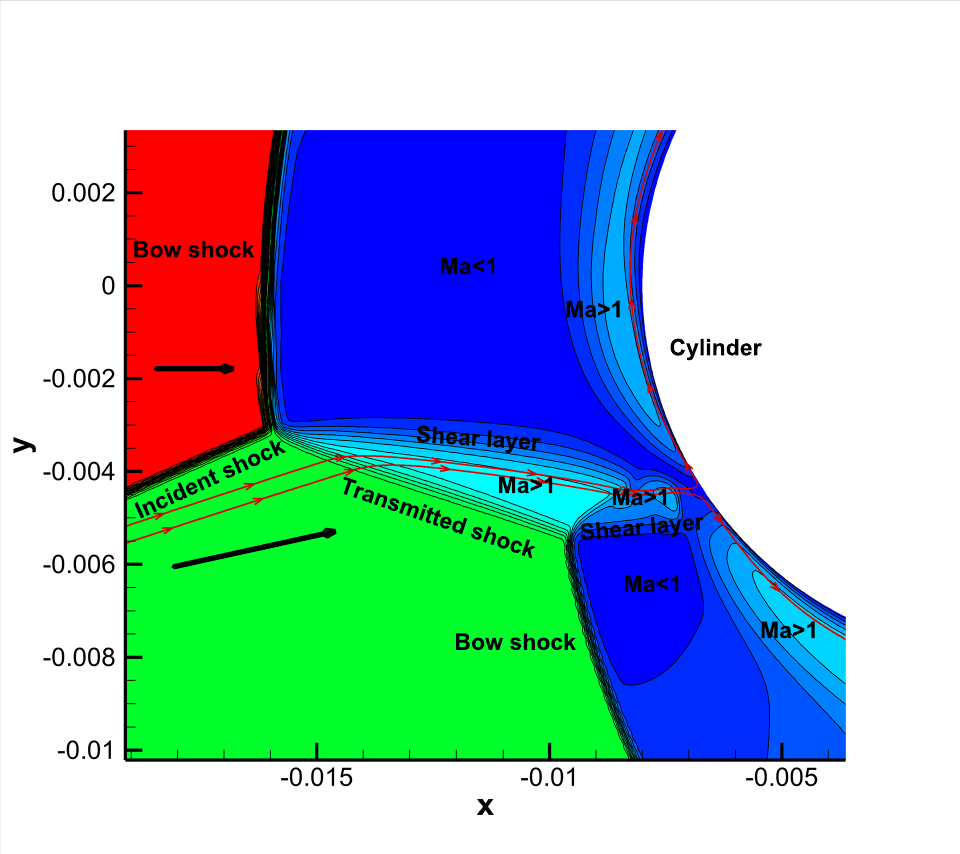}
	\vspace{0mm} 
	\caption{
	Schematic for flow field and streamlines.}
	\label{shock_interaction_flowfield}
\end{figure}

\begin{figure*}[htp]
    \centering
    \subfloat[\label{shock_interaction_upper_streamline}]{%
        \includegraphics[width=0.4\textwidth]{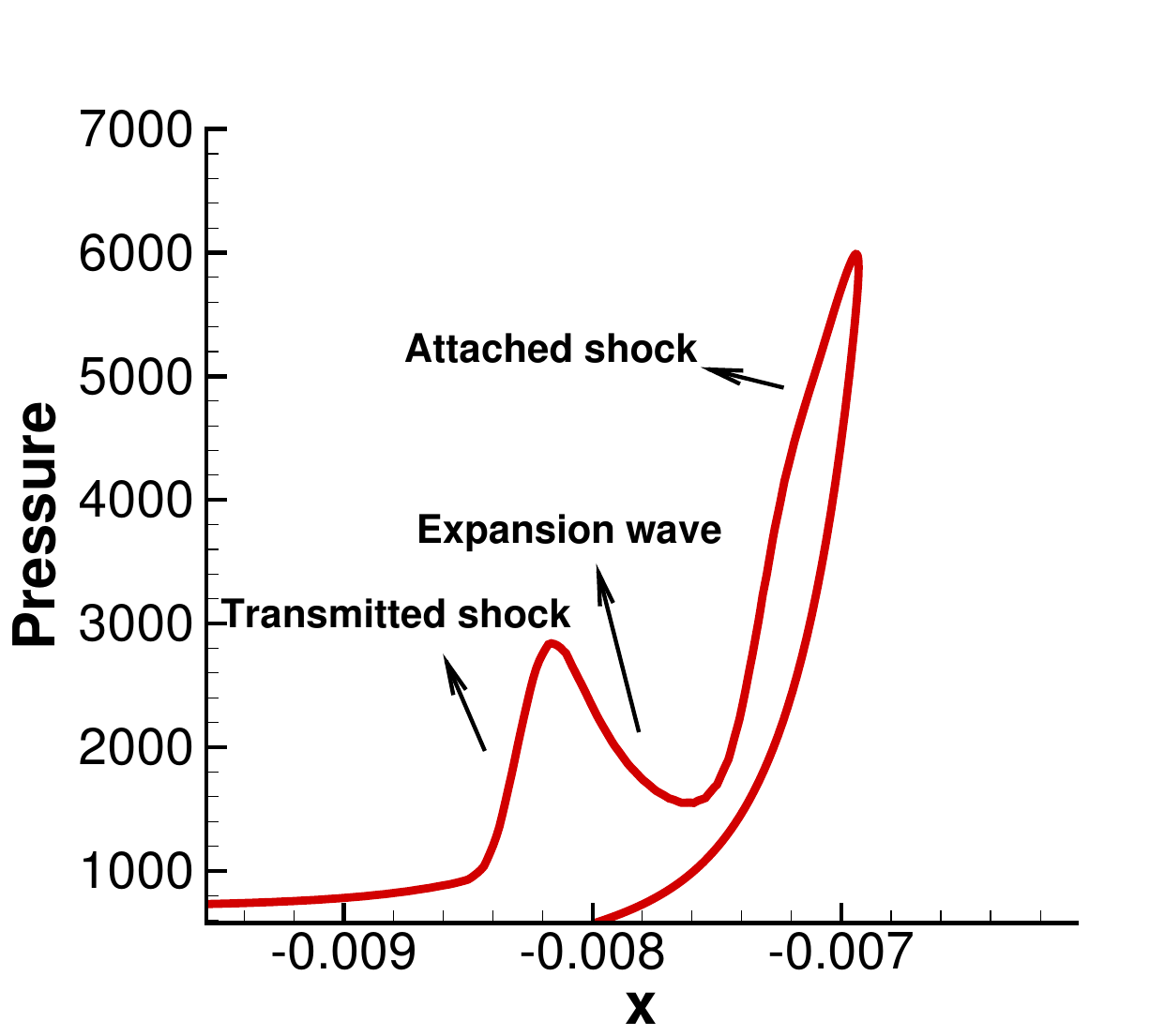}%
    }%
    \hspace{0.5cm} 
    \subfloat[\label{shock_interaction_lower_streamline}]{%
        \includegraphics[width=0.4\textwidth]{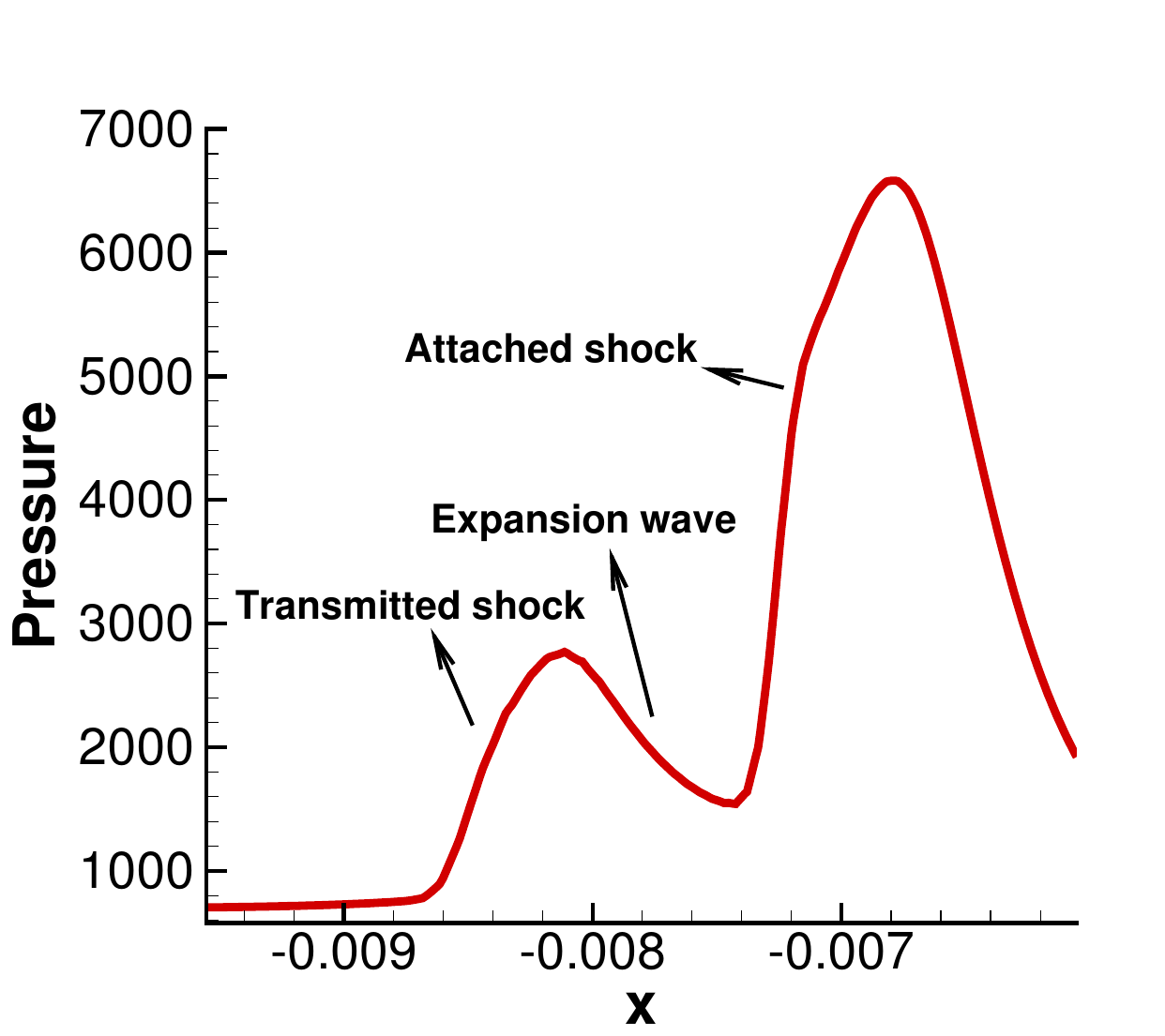}%
    }%
    
    \caption{Pressure variation along the upper streamline (a) and the lower streamline (b).}
    \label{shock_interaction_streamline}
\end{figure*}

Fig.~\ref{shock_interaction_flowfield} illustrates the characteristic flow structure of an Edney Type IV  
shock/shock interaction.  
The incident shock intersects the nearly normal portion of the bow shock,  
generating a localized region of supersonic jet flow.  
Strong shear layers form along the upper and lower boundaries of the jet,  
within which alternating compression and expansion waves develop.  
This supersonic jet terminates near the wall in an attached shock,  
downstream of which a small stagnation zone appears on the cylinder surface.  
In this region, pressure, temperature, and heat flux reach extremely high values.  
Therefore, accurately predicting both the location and magnitude of wall pressure and heat flux peaks  
relies critically on the precise resolution of the supersonic jet structure.

To further understand the wave system inside the jet,  
two streamlines are extracted and superimposed in Fig.~\ref{shock_interaction_flowfield}.  
The pressure distribution along these streamlines is shown in Fig.~\ref{shock_interaction_streamline}.  
For the upper streamline (Fig.~\ref{shock_interaction_upper_streamline}),  
the first pressure rise corresponds to the second transmitted shock within the jet channel.  
This is followed by a pressure drop caused by an expansion wave,  
and finally a pressure rise due to the attached shock near the cylinder wall.  
A similar sequence of pressure variations is observed along the lower streamline,  
as shown in Fig.~\ref{shock_interaction_lower_streamline}.
\begin{figure*}[htp]	
	\centering
	
	\subfloat[Mach number]{%
		\includegraphics[width=0.4\textwidth]{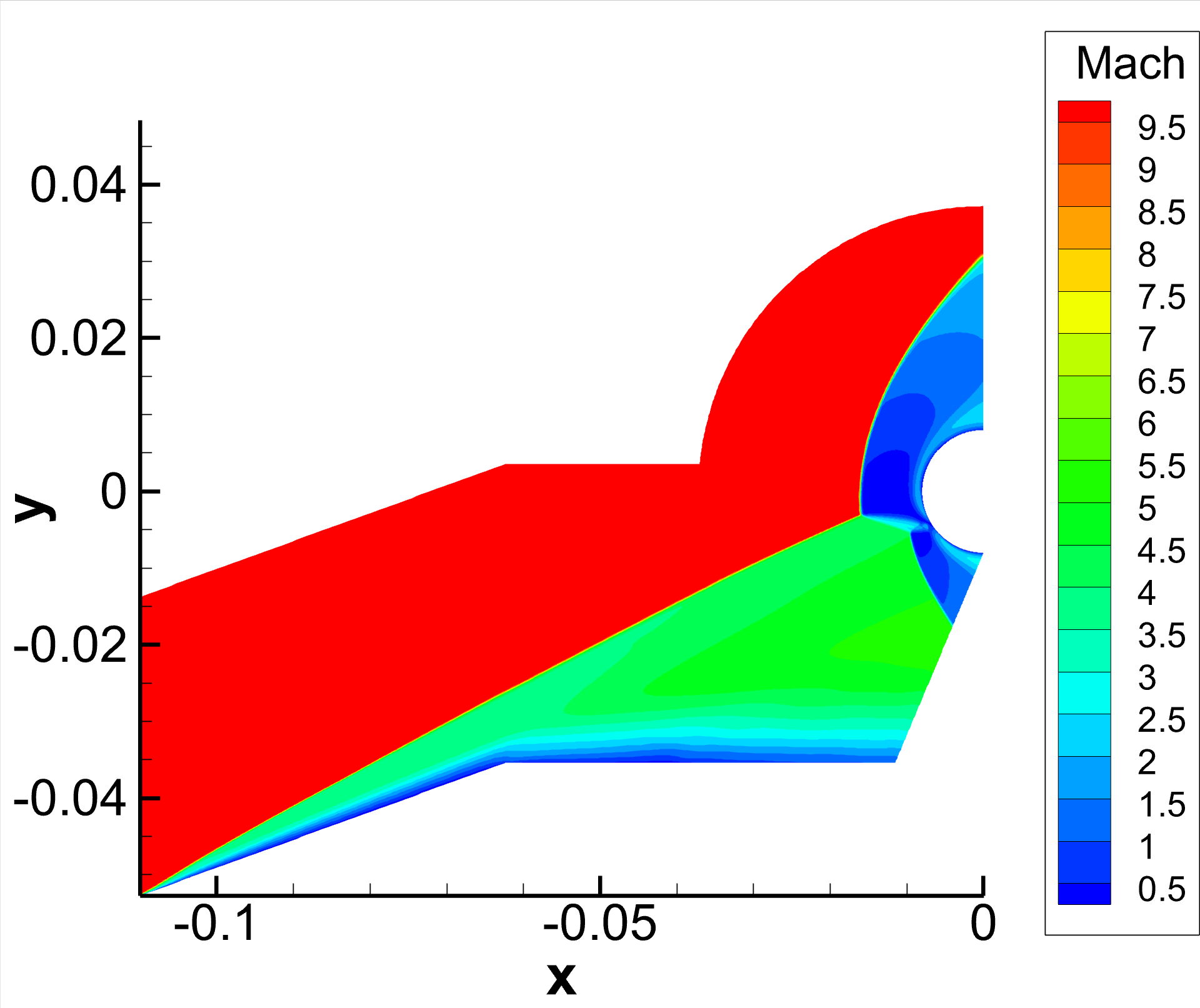}%
	}%
	\hspace{0.05\textwidth} 
	\subfloat[Pressure]{%
		\includegraphics[width=0.4\textwidth]{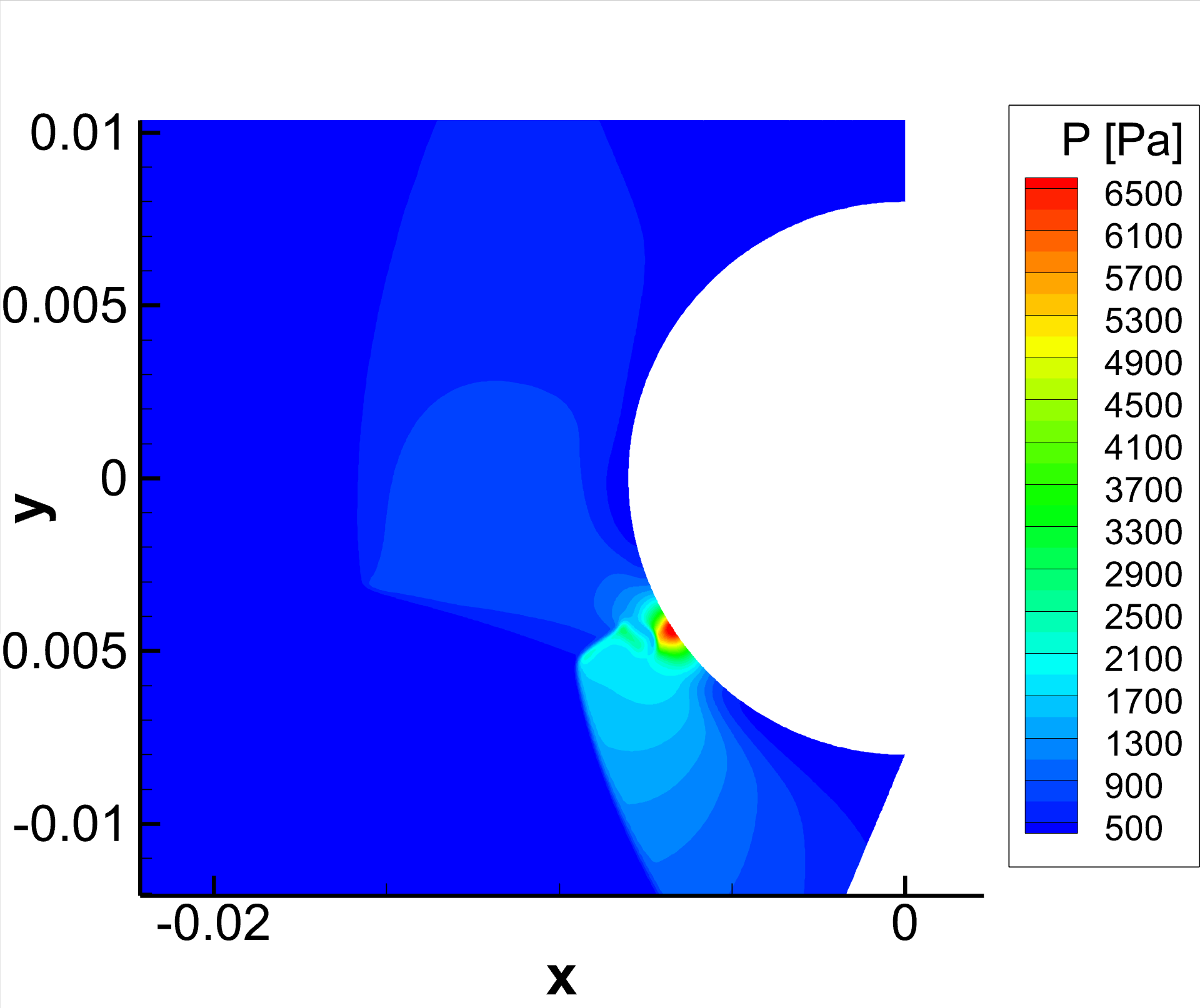}%
	}%
	
	
	
	\subfloat[Translational-rotational temperature\label{shock_interaction_contour_T_tr}]{%
		\includegraphics[width=0.4\textwidth]{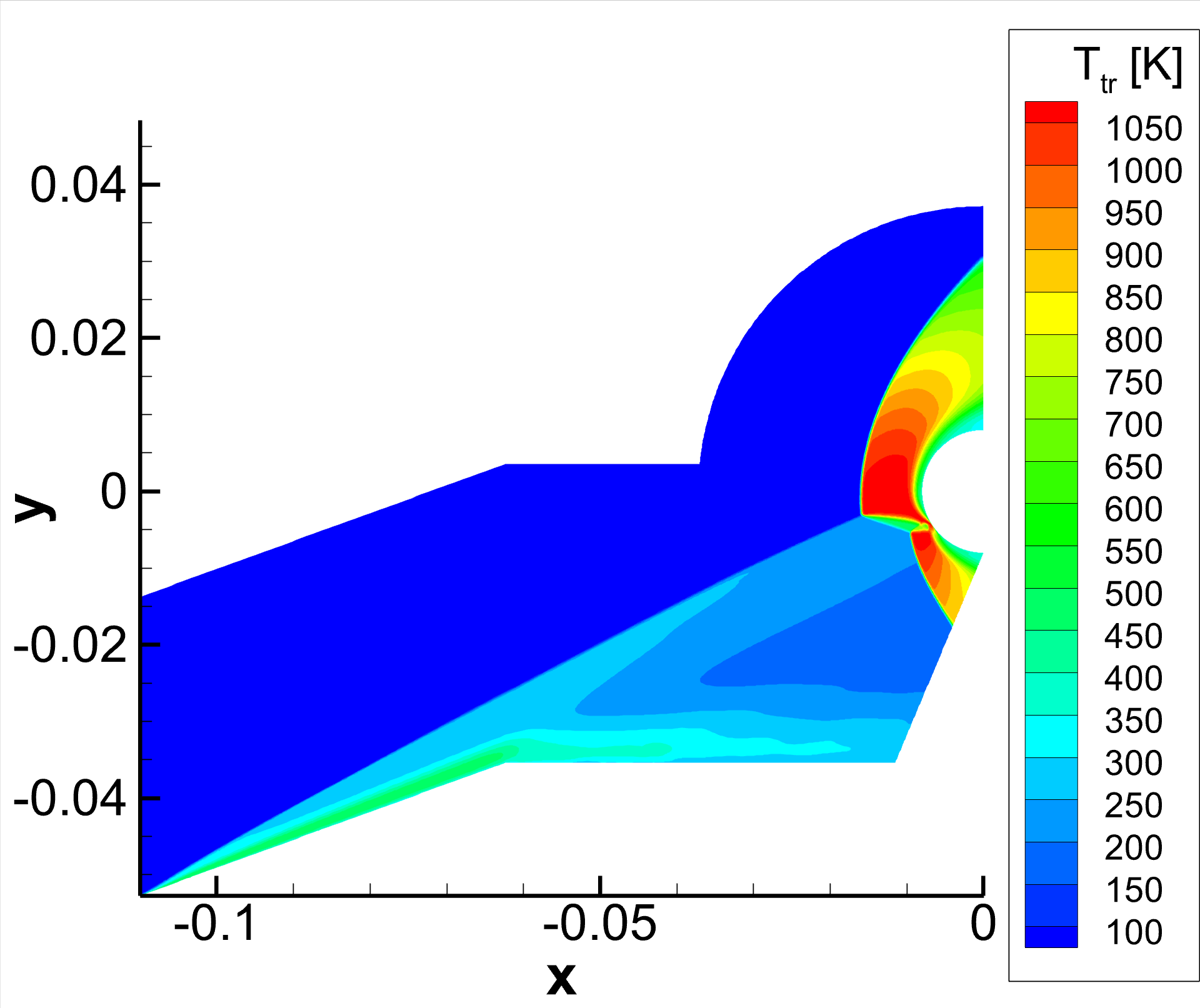}%
	}%
	\hspace{0.05\textwidth} 
	\subfloat[Vibrational temperature\label{shock_interaction_contour_T_v}]{%
		\includegraphics[width=0.4\textwidth]{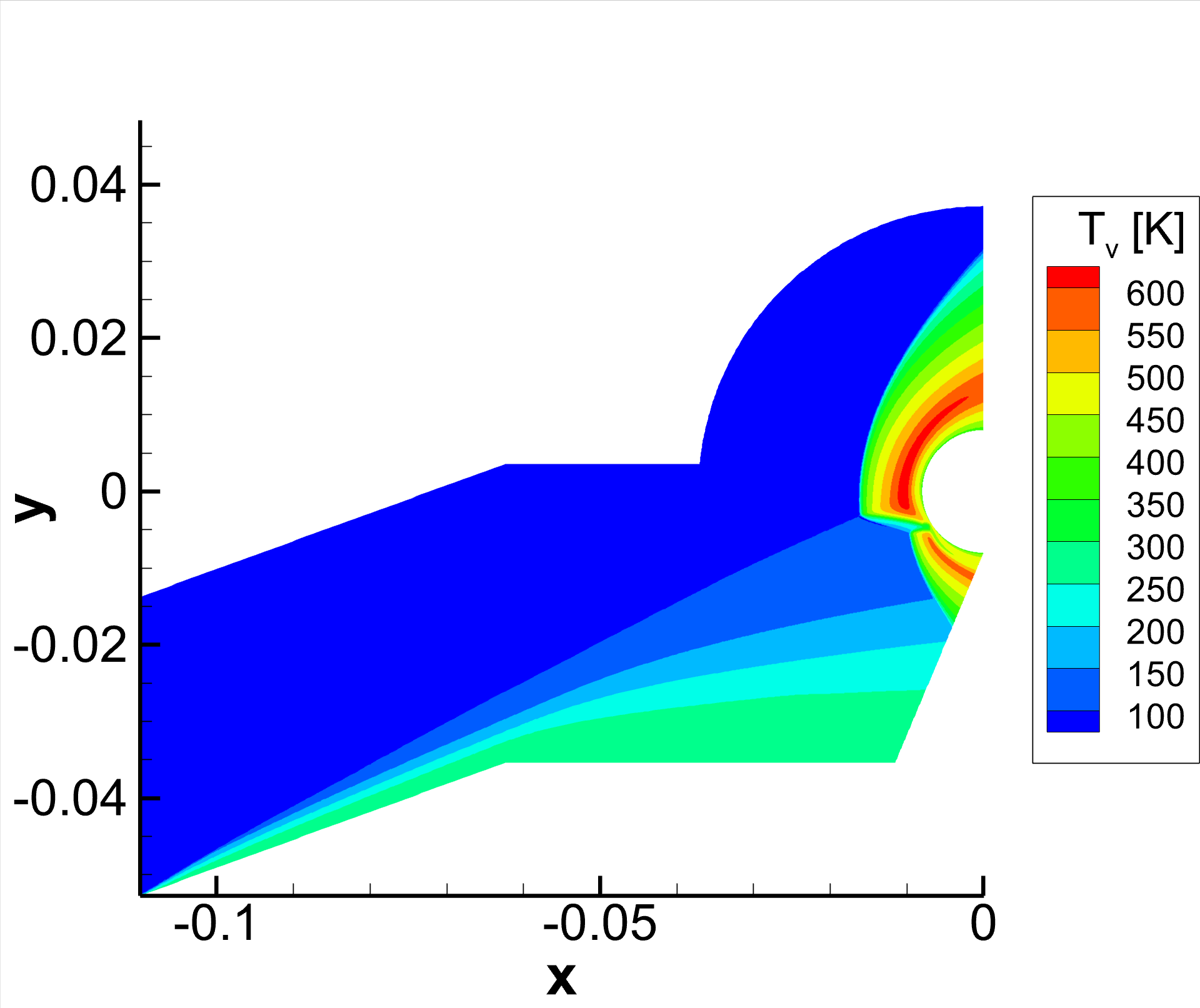}%
	}%
	
	\caption{
	Contours for (a) Mach number, (b) pressure, (c) translational-rotational temperature, and 
	(d) vibrational temperature.}
	\label{shock_interaction_contour}
\end{figure*}

\begin{figure*}[htp]	
	\centering
	\includegraphics[width=0.4\textwidth]{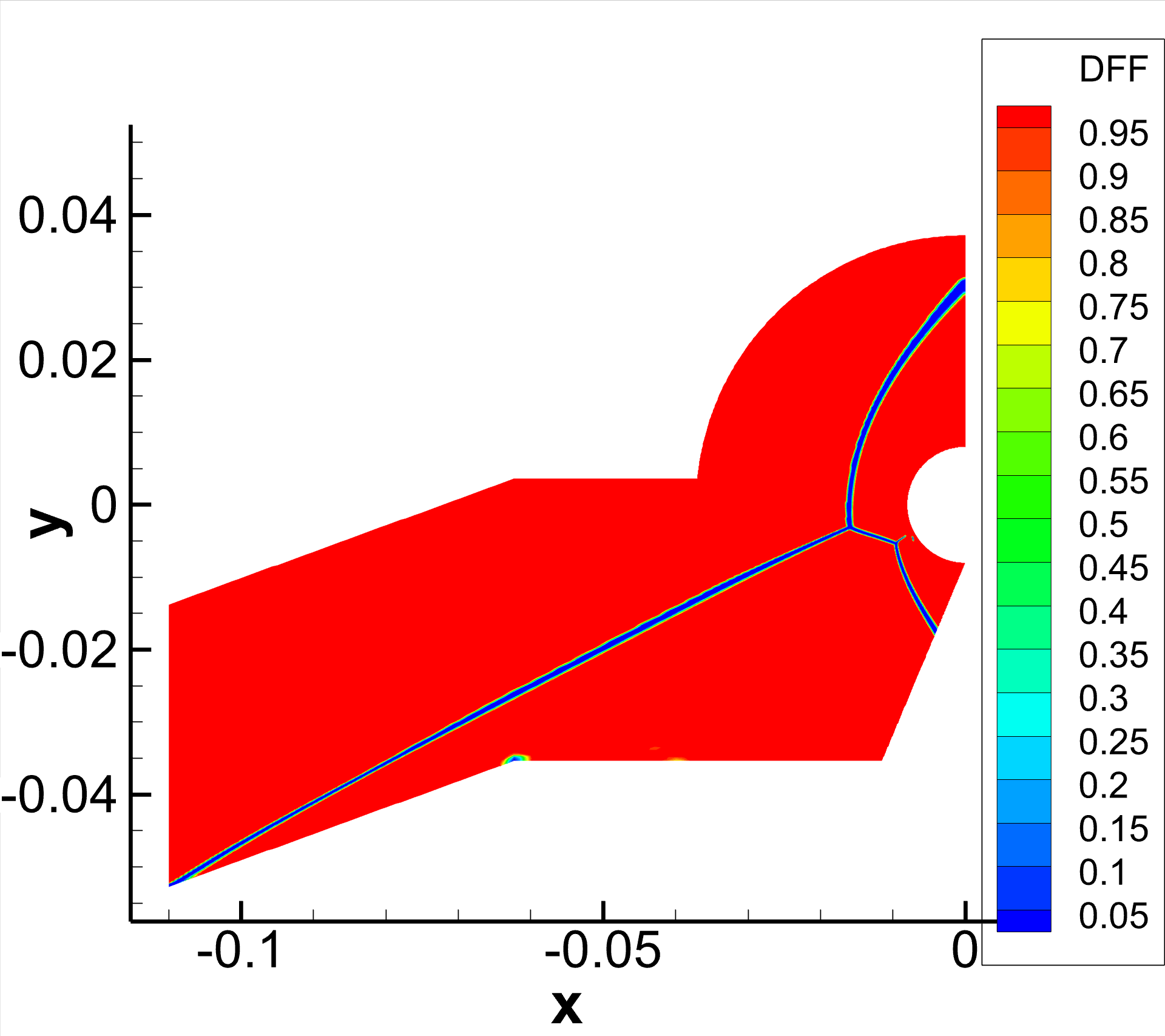}
	\caption{
		Contours for Discontinuity feedback factor DFF.}
	\label{shock_interaction_DFF}
\end{figure*}

Flowfield details are illustrated in Fig.~\ref{shock_interaction_contour}.  
Thermodynamic non-equilibrium effects are clearly observable in  
Fig.~\ref{shock_interaction_contour_T_tr} and Fig.~\ref{shock_interaction_contour_T_v}.
Fig.~\ref{shock_interaction_DFF} shows the contour distribution of the 
discontinuity feedback factor (DFF). From this figure, it can be seen that the DFF clearly 
captures the shock locations and effectively reduces the reconstruction order to first order 
in the shock regions. This visualization provides an intuitive demonstration of how the DFF 
operates in different flow regions.

\subsubsection{Flow Properties on horizontal lines}
\begin{figure*}[htp]	
	\centering
	\subfloat[\label{shock_interaction_line_-2mm_density}]{%
		\includegraphics[width=0.4\textwidth]{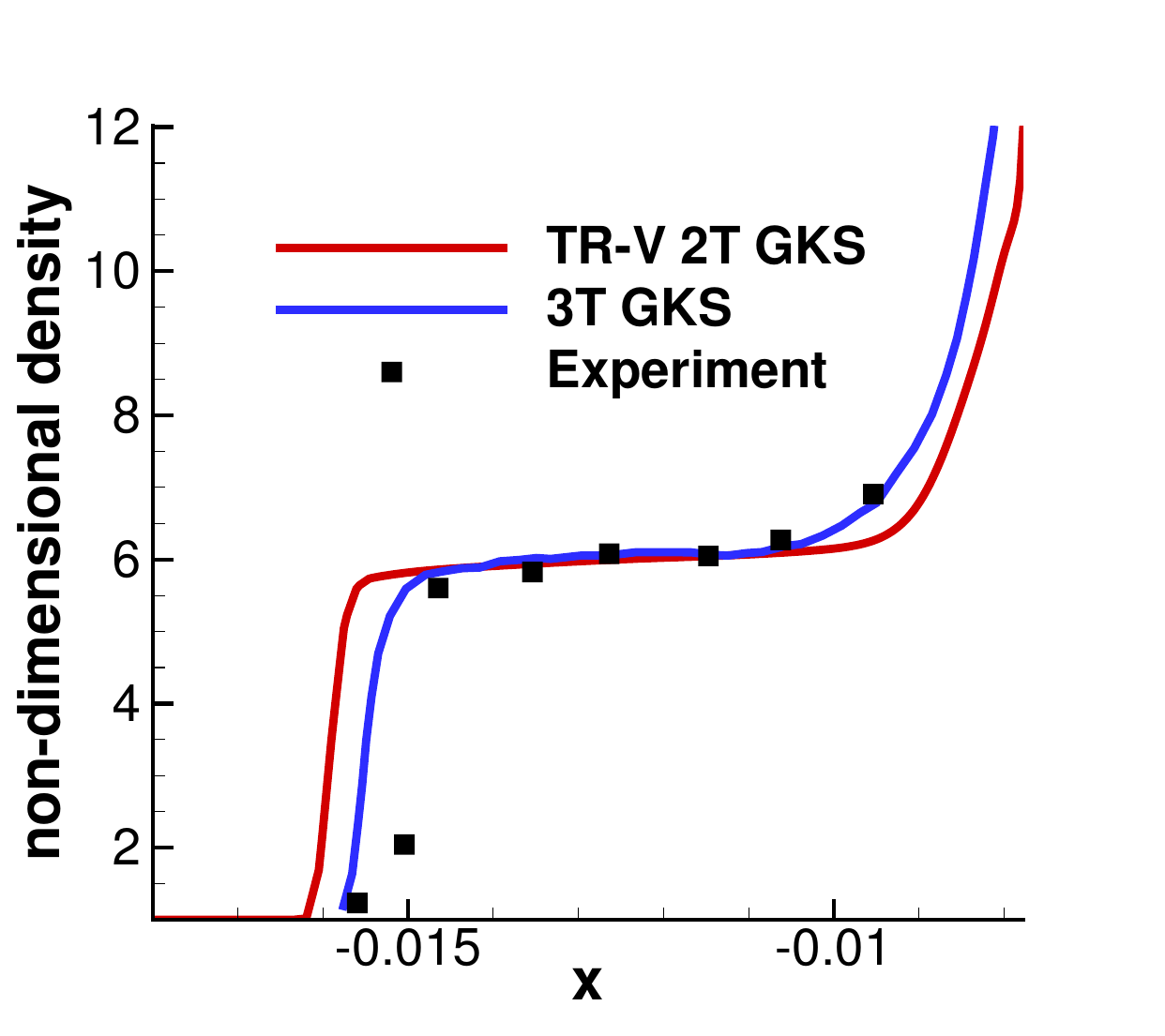}%
	}%
	\hspace{0.05\textwidth} 
	\subfloat[\label{shock_interaction_line_-2mm_T}]{%
		\includegraphics[width=0.4\textwidth]{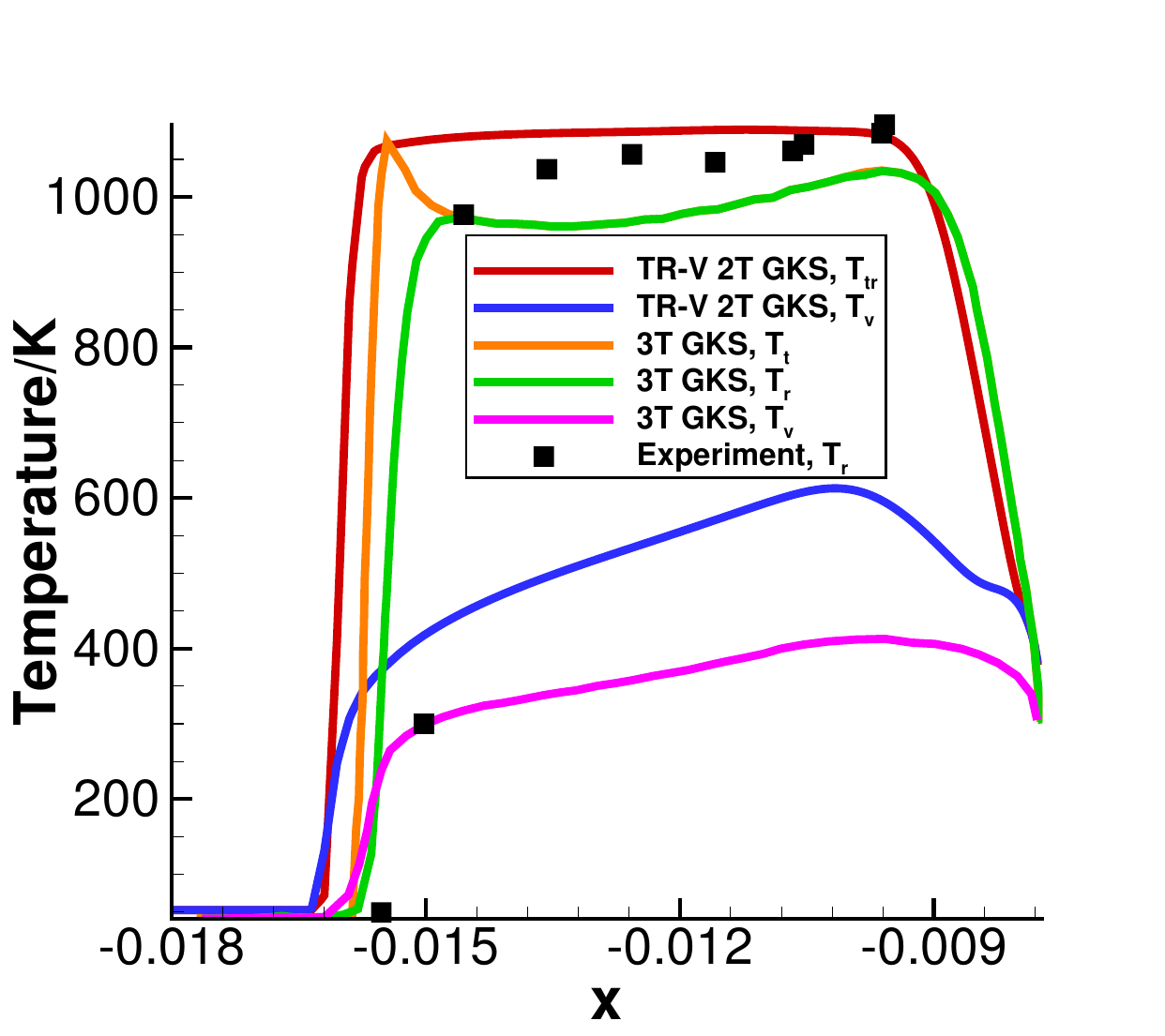}%
	}%
	
	\caption{
	Density (a) and temperature (b) profiles along y=-2mm.}
	\label{shock_interaction_line_-2mm}
\end{figure*}
Fig.~\ref{shock_interaction_line_-2mm} presents the distributions of flowfield properties, 
including temperature and density, along a horizontal line at $y = -2\,\mathrm{mm}$. 
Results from the TR-V 2T GKS, 3T GKS, and experimental data are compared. 
Density is non-dimensionalized by the freestream density. 
This line corresponds to a horizontal cut located $2\,\mathrm{mm}$ below the centerline of the cylinder.

The results indicate that the TR-V 2T GKS model exhibits good agreement with both the experimental data 
and the 3T GKS results in terms of density variation along $y = -2\,\mathrm{mm}$. 
For the rotational temperature along the same line, the TR-V 2T GKS shows slightly less accurate 
shock position prediction compared to the 3T GKS. 
However, it offers a reasonable estimation of the post-shock temperature.

\subsubsection{Surface properties}

\begin{figure*}[htp]	
	\centering
	\subfloat[\label{shock_interaction_surface_pressure}]{%
		\includegraphics[width=0.4\textwidth]{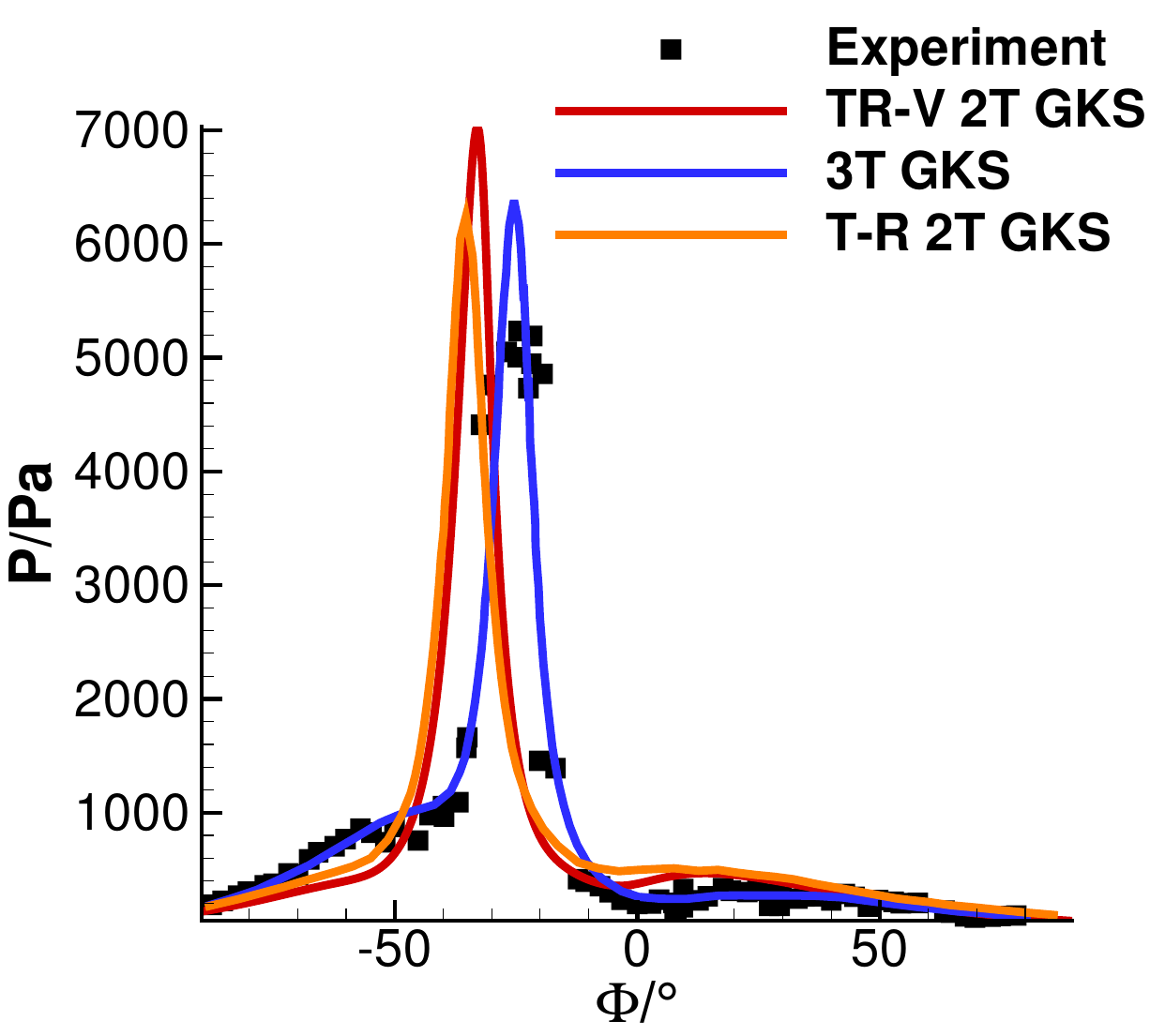}%
	}%
	\hspace{0.05\textwidth} 
	\subfloat[\label{shock_interaction_surface_heat_flux}]{%
		\includegraphics[width=0.4\textwidth]{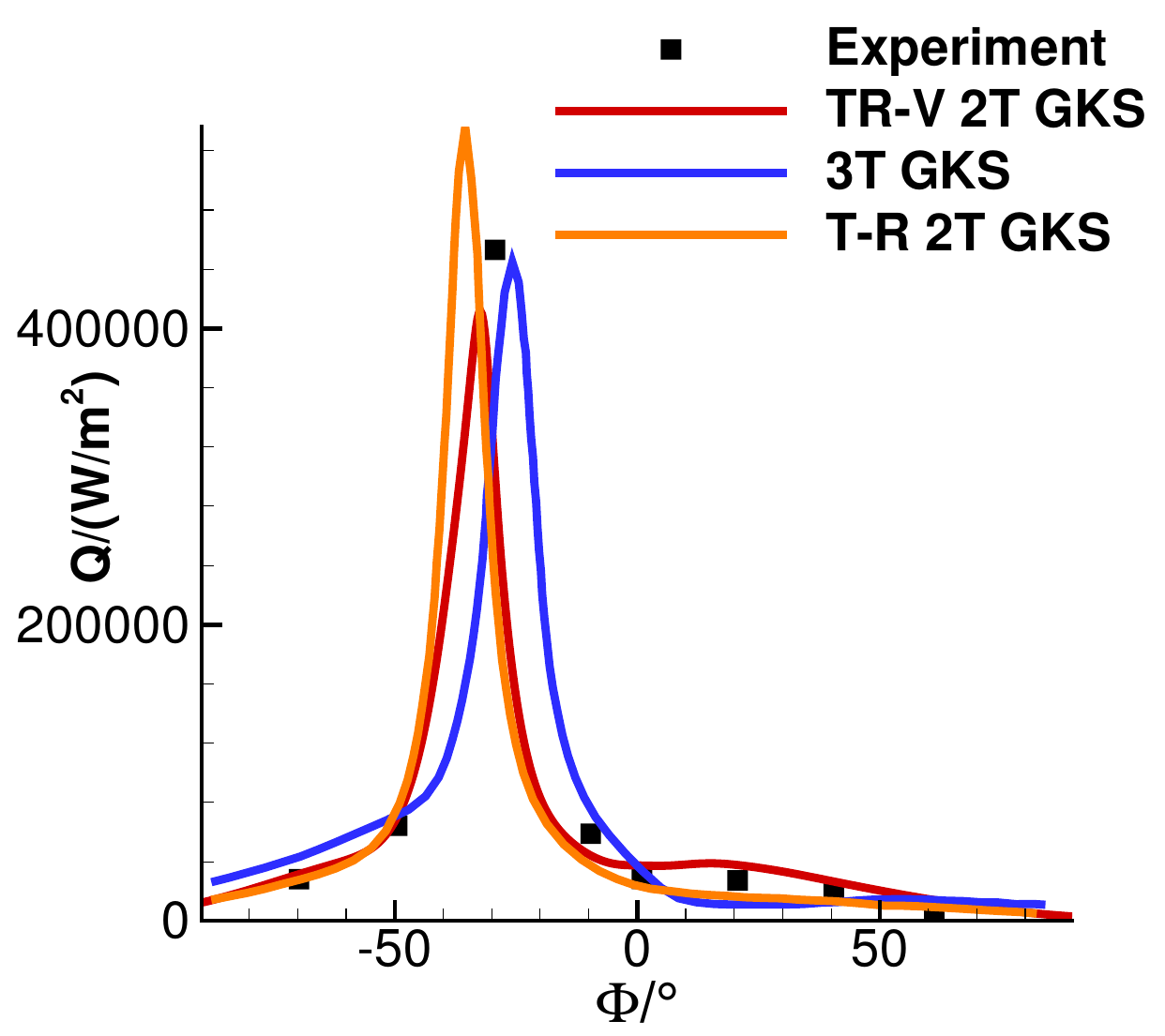}%
	}%
	
	\caption{
	Comparison of TR-V 2T GKS, 3T GKS, and T-R 2T GKS for surface pressure (a) 
	and surface heat flux (b) on the cylinder.}
	\label{shock_interaction_surface_properties}
\end{figure*}

Fig.~\ref{shock_interaction_surface_properties} presents the wall pressure and heat flux distributions 
along the cylinder surface obtained from three models. 
The x-axis indicates the rotation angle measured clockwise from $y = 0$ on the cylinder surface. 
The three methods include the present TR-V 2T GKS, the 3T GKS, and the T-R 2T GKS developed by 
Liu \textit{et al.}~\cite{Liu2021-3T}, where the T-R 2T GKS employs two separate temperatures to 
represent translational and rotational energy modes, while vibrational energy is neglected.
From the wall pressure distribution, the TR-V 2T GKS predicts a higher pressure peak; in terms of the peak 
location, its deviation is smaller than that of the T-R 2T GKS. For the heat flux distribution, the TR-V 2T 
GKS is able to reasonably capture both the peak location and the overall trend, but its predicted peak value 
is slightly lower than the experimental data, making its overall performance slightly inferior to the 3T GKS.

A more plausible explanation for this difference lies not in the physical modeling itself, but in the numerical 
sensitivity of the Edney Type IV interaction. This case involves strong shock/shock interactions and is 
extremely sensitive to numerical dissipation and boundary conditions. Even very small differences can shift 
shock positions and alter shock-layer thicknesses, ultimately leading to noticeable discrepancies in surface 
pressure and heat flux. Since the reference work does not provide sufficient details regarding its numerical 
implementation (e.g., reconstruction scheme, limiter, and boundary conditions), a strict one-to-one comparison 
cannot be made.

\subsection{Computational Cost Comparison}

To quantitatively evaluate the computational efficiency of the proposed TR-V 2T GKS relative to 
the conventional one-temperature (1T) and three-temperature (3T) models, we performed 
simulations of the wedge flow case described in Section~\ref{subsec:wedge}, 
using $10{,}800$ grid cells for $3{,}000$ time steps on the 
same workstation. All computations were executed in parallel using 12 CPU cores under identical 
conditions. The total wall-clock times required by the three models are summarized in 
Table~\ref{tab:cpu_time_comparison}.

\begin{table}[htp]
\centering
\caption{Wall-clock time comparison for the wedge flow case ($10{,}800$ cells, $3{,}000$ steps, 12-core parallel).}
\label{tab:cpu_time_comparison}
\begin{tabular}{lccc}
\hline
Model     & 1T GKS & TR--V 2T GKS & 3T GKS \\
\hline
Wall-clock time (s) & 19  & 35  & 55  \\
\hline
\end{tabular}
\end{table}

As shown in Table~\ref{tab:cpu_time_comparison}, the proposed TR--V 2T GKS requires 
only about $64\%$ of the computational time of the 3T GKS, while providing more accurate 
thermal nonequilibrium modeling than the 1T approach.  
This clearly demonstrates that the TR--V 2T formulation achieves a favorable balance between 
computational cost and physical fidelity.

\section{Conclusion}
Numerical investigations using the translational-rotational vibrational two-temperature GKS were 
performed for 1D shock structures, 2D hypersonic wedge and cylinder flows, and Edney Type 
IV shock/shock interactions.
For the 1D shock structure case, the predicted shock thickness (\textasciitilde 3 mean free paths) falls 
within the typical range (3\textasciitilde10 mean free paths) reported in previous DSMC simulations, theoretical 
estimates, and experimental measurements. In the wedge and 
\textcolor{blue}{2D cylinder flow at 4000 K (chemically frozen)} cases, 
comparisons with DSMC show that the approach accurately predicts shock locations, with surface 
heat flux deviations kept within 10\% of DSMC results. For the 
\textcolor{blue}{2D cylinder flow at 9500 K (limitation case)}, while surface 
pressure predictions remain accurate, noticeable discrepancies appear in shock standoff distance and surface 
heat flux. For the Edney Type IV shock/shock interaction case, the surface heat flux error is within 20\%, 
demonstrating that the present scheme can reasonably capture complex thermal non-equilibrium effects.

Despite these encouraging results, the present study has several limitations. Based on the two cylinder test cases, 
the method provides sufficient accuracy for flow conditions below approximately 4000 K, but deviations grow at 
higher temperatures and the approach becomes unsuitable at around 9500 K. Chemical reactions and dissociation 
were not considered, which further limits applicability to very high-enthalpy or reactive flows. Furthermore, 
only 2D cases were examined, leaving three-dimensional configurations and turbulent hypersonic flows 
for future evaluation.

Future work will therefore focus on extending the present framework to three-dimensional simulations, incorporating 
chemical reaction models to capture coupled thermal and chemical non-equilibrium effects, and testing additional 
configurations such as double-cone and bi-ellipsoid flows for shock/shock and shock wave-boundary-layer interaction 
problems. Further optimization of the numerical implementation will also be pursued to improve computational efficiency.

\section*{Acknowledgments}
This work was supported by the National Natural Science Foundation of China (Nos.~12302378 and 92371201), 
and the Funding of National Key Laboratory of Computational Physics, and the Natural Science Basic Research 
Plan in Shaanxi Province of China (No.~2025SYS-SYSZD-070).

\section*{Data Availability statement}
The data that support the findings of this study are available within the article and its supplementary 
material.

\appendix
\renewcommand{\theequation}{A\arabic{equation}}
\section*{Appendix A: Moments of the Maxwellian distribution function}\label{appendixA}
The second-order time dependent gas distribution function at a cell interface is
\begin{equation}\label{f_all_A}
	\begin{aligned}
f&\left( {{x}_{i + 1/2},{y}_{j},t,u,v,{\xi }_{r},{\xi }_{v}}\right) 
= \left( {1 - {e}^{-t/\tau }}\right) {g}^{c} \\
&+ \left( {\left( {t + \tau }\right) {e}^{-t/\tau } 
- \tau }\right) {a}^{c}(u+v){g}^{c} + \left( {t - \tau  + \tau {e}^{-t/\tau }}\right) 
{A}^{c}{g}^{c} \\
&+{e}^{-t/\tau }{g}^{l}\left\lbrack  {1 - \left( {\tau  + t}\right) 
{a}^{l}(u+v) - \tau {A}^{l}}\right\rbrack  \mathbb{H}\left( {u}_{1}\right) \\ 
&+{e}^{-t/\tau }{g}^{r}\left\lbrack  {1 - \left( {\tau  + t}\right) {a}^{r}(u+v)
- \tau {A}^{r}} \right\rbrack  \left( {1 - \mathbb{H}\left( {u}_{1}\right) }\right),
\end{aligned}
\end{equation}
and the flux at the cell interface$(x_{i+1/2},y_j)$
can be obtained by integrating Eq.~\eqref{f_all_A} during a time step $\Delta t$
\begin{equation}\label{flux_A}
	\begin{pmatrix}F_\rho\\F_{\rho U}\\F_{\rho V}\\F_{\rho E}\\F_{\rho E_{v}}\end{pmatrix}
	=\int_0^{\Delta t}\int u{\psi_\alpha}f\mathrm{d}u\mathrm{d}
	v\mathrm{d}{\xi}_{t}\mathrm{d}{\xi}_{r}\mathrm{d}{\xi}_{v}\mathrm{d}t.
\end{equation}
According to Eq.~\eqref{flux_A}, we need to evaluate the complex combination
of different moments of the Maxwellian distribution functions $g^c, g^l$ and $g^r$. 
To proceed with the evaluation, the general formulas of moment evaluations are given first.

For a two-dimensional translational-rotational vibrational two-temperature intermediate equilibrium
Maxwellian distribution
\begin{equation}
g = \rho {\left( \frac{{\lambda }_{tr}}{\pi }\right) }^{\frac{3+K_r}{2}} 
{e}^{-{\lambda }_{tr}\left\lbrack  {{\left( u - U\right) }^{2} + {\left( v - V\right) }^{2}
+{\xi }_{t}^{2}+{\xi }_{r}^{2} }\right\rbrack}  
{\left( \frac{{\lambda }_{v}}{\pi }\right) }^{\frac{K_v}{2}}{e}^{-{\lambda }_{v}{\xi }_{v}^{2}},
\end{equation}
and the moments of $g$ is defined as
\begin{equation}\label{moment_A}
\rho\langle| \dots |\rangle=\int(\dots)g\mathrm{d}u \mathrm{d}v \mathrm{d}\xi_t 
\mathrm{d}\xi_r \mathrm{d}\xi_v.
\end{equation}
All cases of moment evaluation can be represented in the following form, and a 
detailed derivation is provided below.
\begin{equation}
	\begin{aligned}
	\rho&\langle| u^i v^j \xi_t^k \xi_r^l \xi_v^m |\rangle
	=\int(u^i v^j \xi_t^k \xi_r^l \xi_v^m)g\mathrm{d}u \mathrm{d}v \mathrm{d}\xi_t 
	\mathrm{d}\xi_r \mathrm{d}\xi_v \\
	&=\int(u^i v^j \xi_t^k \xi_r^l \xi_v^m)
	\rho {\left( \frac{{\lambda }_{tr}}{\pi }\right) }^{\frac{3+K_r}{2}} 
	{e}^{-{\lambda }_{tr}\left\lbrack  {{\left( u - U\right) }^{2} + {\left( v - V\right) }^{2}
	+{\xi }_{t}^{2}+{\xi }_{r}^{2} }\right\rbrack} \times \\
	&\quad {\left( \frac{{\lambda }_{v}}{\pi }\right) }^{\frac{K_v}{2}}{e}^{-{\lambda }_{v}{\xi }_{v}^{2}}
	\mathrm{d}u \mathrm{d}v \mathrm{d}\xi_t \mathrm{d}\xi_r \mathrm{d}\xi_v.
    \end{aligned}
\end{equation}
Since the integral variables are independent of each other, 
the multiple integral in the above equation can be transformed into a series of single integrals. 
Moreover, the macroscopic quantities $\lambda$ and $\rho$ are independent of the microscopic variables 
$u$ and $\xi$, allowing them to be factored out of the integral.
\begin{equation}
	\begin{aligned}
	\rho&\langle| u^i v^j \xi_t^k \xi_r^l \xi_v^m |\rangle
	=\rho {\left( \frac{{\lambda }_{tr}}{\pi }\right) }^{\frac{3+K_r}{2}} 
	{\left( \frac{{\lambda }_{v}}{\pi }\right) }^{\frac{K_v}{2}} \times \\
	&\int_{-\infty}^{+\infty} u^i e^{-\lambda_{tr}(u-U)^2}\mathrm{d}u
	\int_{-\infty}^{+\infty} v^j e^{-\lambda_{tr}(v-V)^2}\mathrm{d}v \times \\
	&\int_{-\infty}^{+\infty} \xi_t^k e^{-\lambda_{tr}\xi_t^2}\mathrm{d}\xi_t
	\int_{-\infty}^{+\infty} \xi_r^l e^{-\lambda_{tr}\xi_r^2}\mathrm{d}\xi_r
	\int_{-\infty}^{+\infty} \xi_v^m e^{-\lambda_{v}\xi_v^2}\mathrm{d}\xi_v.
    \end{aligned}
\end{equation}
Each integral term in the above equation is a Gaussian-type integral, 
which generally takes the following form:
\begin{equation}
I=\int_{-\infty}^\infty x^ne^{-a(x+b)^2}dx,\quad a>0.
\end{equation}
Its solution can be obtained using Gaussian integration or the Gamma function. 

From the above formulation, the evaluation of moments involving multiple degrees of freedom can 
be decomposed as follows
\begin{equation}
\langle|u^i v^j \xi_t^{2k} \xi_r^{2l} \xi_v^{2m}|\rangle=\langle|u^i|\rangle\langle|v^j|\rangle
\langle|\xi_t^{2k}|\rangle \langle|\xi_r^{2l}|\rangle \langle|\xi_v^{2m}|\rangle,
\end{equation}
where $i, j, k, l, m$ are integers (owing to the symmetrical property of $\xi_r$ and $\xi_v$, 
the moments of $\xi_r$ and $\xi_v$ are always even-order). With the integral from $-\infty$ 
to $+\infty$, we have
\begin{equation}
    \begin{gathered}
	\langle|u^{0}|\rangle=\mathrm{1,}\\
	\langle|u^{1}|\rangle=U,\\
	\cdots \\
	\langle|u^{n+2}|\rangle=U\langle|u^{n+1}|\rangle+\frac{n+1}{2\lambda_{tr}}|\langle u^{n}|\rangle.
    \end{gathered}
\end{equation}
The moments of $\langle|\xi_t^{2k}|\rangle$ , $\langle|\xi_r^{2l}|\rangle$ 
and $\langle|\xi_v^{2m}|\rangle$ from $-\infty$ to $+\infty$ are
\begin{equation}
	\begin{aligned}
	&\langle|\xi_t^{0}|\rangle=1,\\
	&\langle|\xi_t^{2}|\rangle=(\frac{K_t}{2\lambda_{tr}}),\\
	\langle|\xi_t^{2k}|\rangle=&\frac{K_t+2(k-1)}{2\lambda_{tr}}\langle|\xi_t^{2(k-1)}|\rangle,
    \end{aligned}
\end{equation}
and
\begin{equation}
	\begin{aligned}
	&\langle|\xi_r^{0}|\rangle=1,\\
	&\langle|\xi_r^{2}|\rangle=(\frac{K_r}{2\lambda_{tr}}),\\
	\langle|\xi_r^{2l}|\rangle=&\frac{K_r+2(l-1)}{2\lambda_{tr}}\langle|\xi_r^{2(l-1)}|\rangle,
    \end{aligned}
\end{equation}
and
\begin{equation}
	\begin{aligned}
	&\langle|\xi_v^{0}|\rangle=1,\\
	&\langle|\xi_v^{2}|\rangle=(\frac{K_v}{2\lambda_{v}}),\\
	\langle|\xi_v^{2m}|\rangle=&\frac{K_v+2(m-1)}{2\lambda_{v}}\langle|\xi_v^{2(m-1)}|\rangle.
    \end{aligned}
\end{equation}
Due to the presence of the Heaviside function $\mathbb{H}$ in the integral terms, 
integrals over the velocity ranges from 0 to $+\infty$ and from $-\infty$ to 0 arise. 
The error function is used for representation. 
The integral from 0 to $+\infty$ is denoted by the symbol $\langle|\dots|\rangle_{>0}$, 
and the result is as follows
\begin{equation}
\begin{aligned}
	\langle|u^{0}|\rangle_{>0}&=\frac{1}{2}\mathrm{erfc}(-\sqrt{\lambda_{tr}}\mathrm{U}),\\
	\langle|u^{1}|\rangle_{>0}&=U\langle|u^{0}|\rangle_{>0}+\frac{1}{2}\frac{e^{-\lambda_{tr} U^{2}}}
	{\sqrt{\pi\lambda_{tr}}}, \\
	&\cdots\\
	\langle|u^{n+2}|\rangle_{>0}&=U\langle|u^{n+1}|\rangle_{>0}+\frac{n+1}{2\lambda_{tr}}\langle|u^{n}|
	\rangle_{>0},
\end{aligned}
\end{equation}
and from $-\infty$ to 0 as $\langle|\dots|\rangle_{<0}$, 
\begin{equation}
	\begin{aligned}
		\langle|u^{0}|\rangle_{<0}&=\frac{1}{2}\mathrm{erfc}(\sqrt{\lambda_{tr}}\mathrm{U}),\\
		\langle|u^{1}|\rangle_{<0}&=U\langle|u^{0}|\rangle_{<0}-\frac{1}{2}\frac{e^{-\lambda_{tr} U^{2}}}
		{\sqrt{\pi\lambda_{tr}}},\\
		&\cdots\\
		\langle|u^{n+2}|\rangle_{<0}&=U\langle|u^{n+1}|\rangle_{<0}+\frac{n+1}{2\lambda_{tr}}\langle|u^{n}|
		\rangle_{<0},
	\end{aligned}
\end{equation}
where erfc is the standard complementary error function, which can be expressed as follows
\begin{equation}
	\begin{aligned}
		\mathrm{erf}\left( x\right)  &= \frac{2}{\sqrt{\pi }}\int_{0}^{x}{e}^{-{t}^{2}}{dt}, \\
		\mathrm{erfc}\left( x\right)  &= 1-\mathrm{erf}\left( x\right).
	\end{aligned}
\end{equation}

\renewcommand{\theequation}{B\arabic{equation}}
\setcounter{equation}{0}
\section*{Appendix B: Connection between translational-rotational vibrational two-temperature 
BGK model and macroscopic governing equations}\label{appendixB}
To derive the NS equations, let 
\begin{equation}
\tau = \epsilon \widehat{\tau},
\end{equation}
where $\epsilon$ is a small dimensionless quantity. 
The distribution function can be expanded in power of $\epsilon$
\begin{equation}
f = {f}_{0} + \epsilon {f}_{1} + {\epsilon }^{2}{f}_{2}\cdots.
\end{equation}
We define the material derivative as
\begin{equation}
{D}_{\mathbf{u}} = \frac{\partial }{\partial t} + {u}_{i}\frac{\partial }{\partial {x}_{i}}.
\end{equation}
The non-equilibrium BGK equation is restated below for clarity:
\begin{equation}
	\frac{\partial f}{\partial t} + u\frac{\partial f}{\partial x} + v\frac{\partial f}{\partial y} 
	= \frac{{f}^{eq} - f}{\tau } + \frac{g - {f}^{eq}}{Z_v\tau } 
	= \frac{{f}^{eq} - f}{\tau } + {Q}_{s}.
\end{equation}
Assuming frozen vibrational energy exchange, i.e., \( Z_v \to \infty \), 
the second term on the right-hand side vanishes, and the equation reduces to
\begin{equation}
\epsilon\widehat{\tau}D_{\mathbf{u}}f=g-f.                                                                                                                                                                                                                                                                             
\end{equation}
Expanding the equation in powers of $\epsilon$, we have
\begin{equation}\label{CE}
f= g - \epsilon \widehat{\tau }{D}_{\mathbf{u}}g + {\epsilon }^{2}\widehat{\tau }
{D}_{\mathbf{u}}\left( {\widehat{\tau }{D}_{\mathbf{u}}g}\right)  + \cdots.
\end{equation}
which is known as the Chapman--Enskog expansion. Truncating at first order gives
\begin{equation}
	f = g - \epsilon \widehat{\tau }{D}_{\mathbf{u}}g.
\end{equation}
Taking moments of the BGK equation with respect to \( \psi_\alpha \), 
and applying the compatibility condition Eq.~\eqref{cc}, we obtain
\begin{equation}\label{L=R+S}
	\int\psi_{\alpha}D_{\mathbf{u}}g\mathrm{d}\Xi=\epsilon\int\psi_{\alpha}D_{\mathbf{u}}
	(\widehat{\tau}D_{\mathbf{u}}g)\mathrm{d}\Xi+S+\mathcal{O}(\epsilon^{2}),
\end{equation}
where the source term is
\begin{equation}\label{source_C}
S=(0,0,0,0,s)^{T}=\frac{(\rho E_v)^{eq}-\rho E_v}{Z_v\tau}.
\end{equation}
Define \( \mathcal{L}_\alpha \) and \( \mathcal{R}_\alpha \) as the left- and right-hand integrals:
\begin{equation}\label{L=R+S2}
\mathcal{L}_{\alpha}=\epsilon\mathcal{R}_{\alpha}+S+\mathcal{O}(\epsilon^{2}).
\end{equation}
By dropping \( \mathcal{O}(\epsilon) \) terms, the Euler equations are recovered; 
retaining \( \mathcal{O}(\epsilon) \) terms yields the NS equations. 
To simplify notation, define the expectation operator as
\begin{equation}
<\psi_{\alpha}(\dots)>\equiv\int\psi_{\alpha}(\dots)gd\Xi.
\end{equation}
Here, the expectation operator is defined differently from that in 
\hyperref[appendixA]{Appendix A} (Eq.~\eqref{moment_A}). 
As a result, we adopt a different notation using angle brackets without \( \rho \) on the left-hand side.
This definition relates to that in \hyperref[appendixA]{Appendix A} through the following identity:
\begin{equation}
<\psi_{\alpha}(\dots)> =  \rho \langle \psi_{\alpha}(\dots) \rangle.
\end{equation}
Since the microscopic velocities $\mathbf{u}$ and internal degrees of freedom $\boldsymbol{\xi}$ are 
independent of macroscopic coordinates $x_i$ and time $t$, the derivatives with respect to $x_i$ 
and $t$ can be taken outside the integral. For example,
\begin{equation}
<\psi_{\alpha}>_{,t} = \left(\int \psi_{\alpha} g\, \mathrm{d}\Xi\right)_{,t} = 
\int (\psi_{\alpha} g)_{,t}\, \mathrm{d}\Xi = \int \psi_{\alpha} g_{,t}\, \mathrm{d}\Xi.
\end{equation}
Based on which we define
\begin{equation}
	\begin{aligned}
	\mathcal{L}_{\alpha} &\equiv \int \psi_{\alpha} D_{\mathbf{u}}g\, \mathrm{d}\Xi \\[6pt]
	&= \int \psi_{\alpha}(g_{,t} + u_l g_{,l})\, \mathrm{d}\Xi \\[6pt]
	&= <\psi_{\alpha}>_{,t} + <\psi_{\alpha} u_l>_{,l}.
	\end{aligned}
\end{equation}
Then, according to Eq.~\eqref{L=R+S2}, we obtain the following relation:
\begin{equation}\label{L}
<\psi_{\alpha}>_{,t} + <\psi_{\alpha} u_l>_{,l} = S+\mathcal{O}(\epsilon)
\end{equation}
for all $\alpha$. Since $\mathcal{R}_\alpha$ on the right-hand side of Eq.~\eqref{L=R+S2} is already 
$\mathcal{O}(\epsilon)$, we may neglect terms of order $\mathcal{O}(\epsilon)$ and their derivatives 
during its evaluation.

In practice, we first simplify $\mathcal{L}_\alpha$ by assuming $\mathcal{L}_\alpha = S$ 
(for $\alpha=1,\dots,5$), which yields the Euler equations. Then, using the fact that 
$\mathcal{L}_\alpha = S+\mathcal{O}(\epsilon)$, we further simplify $\mathcal{R}_\alpha$ to derive the 
NS equations. The term $\mathcal{R}_\alpha$ is defined as:
\begin{equation}\label{R}
\begin{aligned}
\mathcal{R}_\alpha &= \int \psi_\alpha D_{\mathbf{u}}(\widehat{\tau} D_{\mathbf{u}} g)\, \mathrm{d}\Xi \\
&= \int \psi_\alpha D_{\mathbf{u}} \left[\widehat{\tau} (g_{,t} + u_l g_{,l}) \right]\, \mathrm{d}\Xi \\
&= \int \psi_\alpha \left\{ \widehat{\tau} \left[ g_{,tt} + (u_l g_{,l})_{,t} + u_k g_{,tk} + u_k (u_l g_{,l})_{,k} \right] \right. \\
&\quad \left. + (\widehat{\tau}_{,t} + u_k \widehat{\tau}_{,k})(g_{,t} + u_l g_{,l}) \right\}\, \mathrm{d}\Xi \\
&= \widehat{\tau} \left[ <\psi_\alpha>_{,tt} + 2<\psi_\alpha u_k>_{,tk} + <\psi_\alpha u_k u_l>_{,lk} \right] \\
&\quad + \widehat{\tau}_{,t} \left[ <\psi_\alpha>_{,t} + <\psi_\alpha u_l>_{,l} \right] \\
&+ \widehat{\tau}_{,k} \left[ <\psi_\alpha u_k>_{,t} + <\psi_\alpha u_k u_l>_{,l} \right].
\end{aligned}
\end{equation}
To eliminate the time derivatives in $\mathcal{R}_\alpha$, note that the term multiplied by 
$\widehat{\tau}_{,t}$ is $\mathcal{O}(\epsilon)$ due to Eq.~\eqref{L}, and may thus be neglected. 
For the leading term, consider:
\begin{equation}
\begin{aligned}
\frac{\partial}{\partial t}\left[<\psi_\alpha>_{,t} + <\psi_\alpha u_k>_{,k}\right] 
&= <\psi_\alpha>_{,tt} + <\psi_\alpha u_k>_{,kt} \\
&= \mathcal{L}_{\alpha,t} = \mathcal{O}(\epsilon).
\end{aligned}
\end{equation}
Then the first term in Eq.~\eqref{R} becomes:
\begin{equation}
\widehat{\tau} \frac{\partial}{\partial x_k} \left[ <\psi_\alpha u_k>_{,t} + <\psi_\alpha u_k u_l>_{,l} \right] + \mathcal{O}(\epsilon),
\end{equation}
which, when combined with the third term, gives the simplified form:
\begin{equation}\label{R_simp}
\mathcal{R}_\alpha = \frac{\partial}{\partial x_k} \left\{ \widehat{\tau} \left[ <\psi_\alpha u_k>_{,t} + <\psi_\alpha u_k u_l>_{,l} \right] \right\} + \mathcal{O}(\epsilon).
\end{equation}
This eliminates the second-order time derivatives in $\mathcal{R}_\alpha$. 

The Euler equations are obtained by setting $\mathcal{L}_\alpha = S$. For instance, for $\alpha = 1$,
\begin{equation}
\mathcal{L}_1 = <\psi_1>_{,t} + <\psi_1 u_k>_{,k} = \rho_{,t} + (\rho U_k)_{,k},
\end{equation}
where we used $\psi_1 = 1$. Neglecting $\mathcal{O}(\epsilon)$ yields the continuity equation.

For $\alpha = 2,3$, we define $\mathcal{L}_i$ and $\mathcal{R}_i$ with $i = \alpha - 1$, and let $w_i = u_i - U_i$. Then,
\begin{equation}
\begin{aligned}
\mathcal{L}_i &= <u_i>_{,t} + <u_i u_k>_{,k} \\
&= (\rho U_i)_{,t} + <(w_i + U_i)(w_k + U_k)>_{,k} \\
&= (\rho U_i)_{,t} + [\rho U_i U_k + <w_i U_k> + <w_k U_i> + <w_i w_k>]_{,k} \\
&= (\rho U_i)_{,t} + [\rho U_i U_k + <w_i w_k>]_{,k}.
\end{aligned}
\end{equation}

Since $w_l$ represents the deviation of microscopic velocity from the macroscopic velocity due to molecular 
thermal motion, which averages out in the macroscopic sense, all odd-order moments of $g$ in $w_l$ vanish. 
That is, 
\begin{equation}
    <w_i{}^m> = 0, \quad m = 1, 3, 5, \cdots,
\end{equation}
The pressure tensor is defined as
\begin{equation}
p_{ik} = <w_i w_k> \equiv p_{tr} \delta_{ik},
\end{equation}
where $p_{tr} = \rho R T_{tr}$. The validity of the above equation can be shown as
\begin{equation}
\begin{aligned}
<w^2> &= <(u - U)^2> \\
&= <u^2 - 2uU + U^2> \\
&= \rho(U^2 + \frac{1}{2\lambda_{tr}}) - 2\rho U^2 + \rho U^2 \\
&= \frac{\rho}{2\lambda_{tr}}.
\end{aligned}
\end{equation}
Using the definitions $\lambda_{tr} = \frac{m}{2kT_{tr}}$, $R = \frac{k}{m}$, and $p_{tr} = \rho R T_{tr}$, we obtain
\begin{equation}
<w^2> = p_{tr}.
\end{equation}
Substituting the above result into the expression of $\mathcal{L}_i$, we get
\begin{equation}\label{L2,3}
\mathcal{L}_i = (\rho U_i)_{,t} + (\rho U_i U_k + p_{tr} \delta_{ik})_{,k}.
\end{equation}
From the derivation above, we summarize the following relations:
\begin{equation}\label{relation}
\begin{aligned}
<w_i{}^m> &= 0, \quad m = 1, 3, 5, \cdots, \\
<u_i u_k> &= \rho U_i U_k + p_{tr} \delta_{ik},
\end{aligned}
\end{equation}
and $\mathcal{L}_i = 0$ corresponds to the momentum conservation in Euler equations.

For the energy equation, we define
\begin{equation}
\mathcal{L}_4 = \frac{1}{2}<{u_n}^2 + \xi^2>_{,t} 
+ \frac{1}{2}<u_l ({u_n}^2 + \xi^2)>_{,l},
\end{equation}
where ${u_n}^2 = u^2 + v^2$ and $\xi^2 = \xi_t^2 + \xi_r^2 + \xi_v^2$.
Expanding the convective term in the $x$-direction, we have
\begin{equation}
\begin{aligned}
&\frac{1}{2}<u({u_n}^2 + \xi^2)>_{,x} 
= \frac{1}{2}<u^3 + u v^2 + u \xi_t^2 + u \xi_r^2 + u \xi_v^2>_{,x} \\
&= \frac{1}{2} \left< u^3 + u v^2 + u \xi_t^2 + u \xi_r^2 + u \xi_v^2 \right>_{,x} \\
&= \frac{1}{2} \left\{ \rho \left[ U^3 + \frac{3U}{2\lambda_{tr}} 
+ U \left( V^2 + \frac{1}{2\lambda_{tr}} + \frac{K_t}{2\lambda_{tr}} 
+ \frac{K_r}{2\lambda_{tr}} + \frac{K_v}{2\lambda_v} \right) \right] \right\}_{,x} \\
&= \frac{1}{2} \left[ \rho U \left( U^2 + V^2 
+ \frac{4 + K_t + K_r}{2\lambda_{tr}} + \frac{K_v}{2\lambda_v} \right) \right]_{,x} \\
&= \left( \frac{1}{2} \rho U (U^2 + V^2) 
+ \frac{K_r + 5}{2} p_{tr} U + \frac{K_v}{2} p_v U \right)_{,x}.
\end{aligned}
\end{equation}
Applying the same procedure in the $y$-direction and combining, we obtain
\begin{equation}\label{L4_der}
\begin{aligned}
\mathcal{L}_4 
=& \left( \frac{1}{2} \rho U_n^2 + \frac{K_r + 3}{2} p_{tr} + \frac{K_v}{2} p_v \right)_{,t} \\
&+ \left( \frac{1}{2} \rho U_k U_n^2 + \frac{K_r + 5}{2} p_{tr} U_k + \frac{K_v}{2} p_v U_k \right)_{,k},
\end{aligned}
\end{equation}
where $U_n^2 = U^2 + V^2$, and $p_v = \rho R T_v$.
Setting $\mathcal{L}_4 = 0$ yields the conservative form of the energy equation without dissipation.

For the vibrational energy equation, we write
\begin{equation}
	\begin{aligned}
	\mathcal{L}_{5} 
	&= \frac{1}{2} < \xi_v{}^{2} >_{,t} 
	+ \frac{1}{2} < u_{l} \xi_v{}^{2} >_{,l} \\
	&= \left( \frac{K_v}{2} p_v \right)_{,t} 
	+ \left( \frac{K_v}{2} p_v U_{l} \right)_{,l}.
	\end{aligned}
\end{equation}
This completes the derivation of the Euler equations. We now proceed to the NS equations.

To derive the NS equations, we further simplify $\mathcal{R}_\alpha$ by eliminating the time derivatives. 
In particular, the time derivatives in $\mathcal{R}1$ can be neglected by noting that 
$\mathcal{L}{i} = \mathcal{O}(\epsilon)$. From Eq.~\eqref{R_simp}, for $\alpha = 1$, we have
\begin{equation}
	\mathcal{R}_1 = \left\{ \hat{\tau} \left[ < u_k >_{,t} + < u_k u_l >_{,l} \right] \right\}_{,k}.
\end{equation}
The quantity inside the brackets has the same form as $\mathcal{L}_{2,3}$, implying that 
$\mathcal{R}_1 = \mathcal{O}(\epsilon)$ and hence 
$\mathcal{L}_1 = \epsilon \mathcal{R}_1 = \mathcal{O}(\epsilon^2)$. 
Therefore, to the order retained, we have $\mathcal{R}_1 = 0$ and $\mathcal{L}_1 = 0$, 
which yields the continuity equation:
\begin{equation}
	\rho_{,t} + \left( \rho U_k \right)_{,k} = 0.
\end{equation}
This continuity equation can be used to simplify the momentum, total energy, and vibrational energy 
equations by eliminating corresponding time derivative terms in the RHS $\mathcal{R}_\alpha$ of 
Eq.~\eqref{L=R+S2}. 
By multiplying the continuity equation by $U_i$ and subtracting the result from $\mathcal{L}_i$, we 
get (from Eq.~\eqref{L2,3}):
\begin{equation}\label{L2,3_simp}
	\begin{aligned}
	\mathcal{L}_i 
	&= (\rho U_i)_{,t} + (\rho U_i U_k + p_{tr} \delta_{ik})_{,k} \\
	&= \rho_{,t} U_i + \rho U_{i,t} + \rho_{,k} U_i U_k + \rho U_{i,k} U_k + \rho U_i U_{k,k} + p_{tr,i} \\
	&= U_i (\rho_{,t} + \rho_{,k} U_k + \rho U_{k,k}) + \rho U_{i,t} + \rho U_{i,k} U_k + p_{tr,i} \\
	&= U_i \big[ \rho_{,t} + (\rho U_k)_{,k} \big] + \rho U_{i,t} + \rho U_{i,k} U_k + p_{tr,i} \\
	&= \rho U_{i,t} + \rho U_{i,k} U_k + p_{tr,i}.
	\end{aligned}
\end{equation}
For $\mathcal{L}_4$, we group terms as follows:
\begin{equation}
	\begin{aligned}
	\mathcal{L}_4 
	&= \left( \frac{1}{2} \rho U_n^2 + \frac{K_r + 3}{2} p_{tr} + \frac{K_v}{2} p_v \right)_{,t} \\
	&\quad + \left( \frac{1}{2} \rho U_k U_n^2 + \frac{K_r + 5}{2} p_{tr} U_k + \frac{K_v}{2} p_v U_k \right)_{,k} \\
	&= \frac{1}{2} \rho_{,t} U_n^2 + \rho U_n U_{n,t} + \frac{K_r + 3}{2} p_{tr,t} + \frac{K_v}{2} p_{v,t} \\
	&\quad + \frac{1}{2} (\rho U_k)_{,k} U_n^2 + \rho U_k U_n U_{n,k} + \frac{K_r + 5}{2} p_{tr,k} U_k \\
	&\quad + \frac{K_r + 5}{2} p_{tr} U_{k,k} + \frac{K_v}{2} p_{v,k} U_k + \frac{K_v}{2} p_v U_{k,k} \\
	&= \frac{1}{2} U_n^2 \mathcal{L}_1 + U_n \mathcal{L}_n + \rho U_k U_n U_{n,k} + U_k p_{tr,k} \\
	&\quad + \frac{K_r + 3}{2} (p_{tr,t} + U_k p_{tr,k}) + \frac{K_r + 5}{2} p_{tr} U_{k,k} \\
	&\quad + \frac{K_v}{2} (p_{v,t} + p_{v,k} U_k + p_v U_{k,k}).
	\end{aligned}
\end{equation}
Here, the first term is $\frac{1}{2} U_n^2 \mathcal{L}_1 = \mathcal{O}(\epsilon^2)$, and the second 
is $U_n \mathcal{L}_n = \mathcal{O}(\epsilon^2)$.
Then
\begin{equation}\label{L4}
	\begin{aligned}
	\mathcal{L}_{4}&=\frac{K_r+3}{2}[p_{tr,t}+U_{k}p_{tr,k}]+\frac{K_r+5}{2}p_{tr}U_{k,k} \\
	&\quad +\frac{K_v}{2}(p_{v,t}+p_{v,k}U_k+p_{v}U_{k,k})+U_n \mathcal{L}_{n}.
	\end{aligned}
\end{equation}
We may omit the last term in the reduction of $\mathcal{R}_{\alpha}$;  
however, the term $U_n \mathcal{L}_n$ must be retained when explicitly expressing  
$\mathcal{L}_{4} = \epsilon \mathcal{R}_{4}$.  

For $\mathcal{L}_5$, we have
\begin{equation}
	\begin{aligned}
		\mathcal{L}_5
		&= {\left(\frac{K_v}{2} p_v\right)}_{,t} 
		+ {\left(\frac{K_v}{2} p_v U_{l}\right)}_{,l} \\
		&= \frac{K_v}{2} (p_{v,t} + p_{v,l} U_l + p_v U_{l,l}).
	\end{aligned}
\end{equation}

To simplify the right-hand side term $\mathcal{R}_{\alpha}$,  
considering that the vibrational energy equation contains a source term $S$ on its RHS,  
we treat $\mathcal{L}_5 - S$ as a whole,  
i.e., $\mathcal{L}_5 - S = \epsilon \mathcal{R}_5 = \mathcal{O}(\epsilon)$.  
Thus, $\mathcal{L}_5 - S$ can be expressed as follows,  
with the source term given by Eq.~\eqref{source} and Eq.~\eqref{source_cal}:
\begin{equation}\label{L5-S}
	\begin{aligned}
		&\mathcal{L}_5 - S
		= \frac{K_v}{2}(p_{v,t} + p_{v,l} U_l + p_v U_{l,l}) - \frac{(\rho E_v)^{eq} - \rho E_v}{Z_v \tau} \\
		&= \frac{K_v}{2}(p_{v,t} + p_{v,l} U_l + p_v U_{l,l}) \\
		&\quad - \frac{1}{Z_v \tau} \left( \frac{K_v}{2} \rho R \frac{(3 + K_r) T_{tr} + K_v T_v}{3 + K_r + K_v} - \rho E_v \right) \\
		&= \frac{K_v}{2}(p_{v,t} + p_{v,l} U_l + p_v U_{l,l}) \\
		&\quad - \frac{1}{Z_v \tau} \left( \frac{K_v}{2} \frac{(3 + K_r) p_{tr} + K_v p_v}{3 + K_r + K_v} - \frac{K_v}{2} p_v \right) \\
		&= \frac{K_v}{2} \left( p_{v,t} + p_{v,l} U_l + p_v U_{l,l} - \frac{1}{Z_v \tau} \frac{(3 + K_r)(p_{tr} - p_v)}{3 + K_r + K_v} \right).
	\end{aligned}
\end{equation}
For the right-hand sides of the momentum equations, consider 
$\mathcal{R}_{j} = (\widehat{\tau} F_{jk})_{,k}$, where
\begin{equation}
	F_{jk} \equiv <u_j u_k>_{,t} + <u_j u_k u_l>_{,l}.
\end{equation}
Using Eq.~\eqref{relation} and noting that all moments odd in $w_k$ vanish,  
and letting $u_i = w_i + U_i$, we have
\begin{equation}
	\begin{aligned}
		<u_j u_k u_l>_{,l} &= <(U_j + w_j)(U_k + w_k)(U_l + w_l)>_{,l} \\
		&= <(U_j U_k + U_j w_k + U_k w_j + w_j w_k)(U_l + w_l)>_{,l} \\
		&= <U_j U_k U_l + U_j p_{tr} \delta_{kl} + U_k p_{tr} \delta_{jl} + U_l p_{tr} \delta_{jk}>_{,l} \\
		&= U_j (\rho U_k U_l)_{,l} + U_{j,l} \rho U_k U_l \\
		&\quad + U_{j,k} p_{tr} + U_j p_{tr,l} \delta_{kl} \\
		&\quad + U_{k,j} p_{tr} + U_k p_{tr,j} + U_{l,l} p_{tr} \delta_{jk} + U_l p_{tr,l} \delta_{jk}.
	\end{aligned}
\end{equation}
Substituting into $F_{jk}$, we get
\begin{equation}
	\begin{aligned}
		F_{jk} &= (\rho U_j U_k + p_{tr} \delta_{jk})_{,t} + <u_j u_k u_l>_{,l} \\
		&= U_j \left[ (\rho U_k)_{,t} + (\rho U_k U_l + p_{tr} \delta_{kl})_{,l} \right] \\
		&\quad + \rho U_k U_{j,t} + p_{tr,t} \delta_{jk} + U_{j,l} \rho U_k U_l + U_{j,k} p_{tr} \\
		&\quad  + U_{k,j} p_{tr} + U_k p_{tr,j} + U_{l,l} p_{tr} \delta_{jk} + U_l p_{tr,l} \delta_{jk}.
	\end{aligned}
\end{equation}
The term in square brackets multiplying $U_j$ is $\mathcal{L}_k = \mathcal{O}(\epsilon)$ and can 
be neglected. Collecting terms, we have
\begin{equation}
\begin{aligned}
	F_{jk} &= U_k \left[ \rho U_{j,t} + \rho U_l U_{j,l} + p_{tr,j} \right] + p_{tr} \left[ U_{k,j} + U_{j,k} + U_{l,l} \delta_{jk} \right] \\
	&\quad + \delta_{jk} \left[ p_{tr,t} + U_l p_{tr,l} \right].
\end{aligned}
\end{equation}
According to Eq.~\eqref{L2,3_simp}, the coefficient of $U_k$ is $\mathcal{L}_j = \mathcal{O}(\epsilon)$ and 
can be neglected.
To eliminate $p_{tr,t}$ from the last term, we substitute Eq.~\eqref{L4} for $\mathcal{L}_4$ to obtain
\begin{equation}\label{p_{tr,t}}
\begin{aligned}
{p}_{tr,t} + {U}_{k}{p}_{tr,k} &= -\frac{K_r + 5}{K_r + 3}p_{tr}{U}_{k,k} \\
&\quad - \frac{K_v}{K_r+3}\left(p_{v,t}+p_{v,k}U_k+p_vU_{k,k}\right) + \mathcal{O}\left( \epsilon \right).
\end{aligned}
\end{equation}
To eliminate $p_{v,t}$ from the above, we invoke Eq.~\eqref{L5-S} for $\mathcal{L}_5$, yielding
\begin{equation}\label{p_{v,t}}
p_{v,t} + p_{v,l}U_l + p_v U_{l,l}
= \frac{1}{Z_v \tau}\frac{(3+K_r)(p_{tr}-p_v)}{3+K_r+K_v} + \mathcal{O}(\epsilon).
\end{equation}
Substituting Eq.~\eqref{p_{v,t}} into Eq.~\eqref{p_{tr,t}}, the expression for ${p}_{tr,t}$ becomes
\begin{equation}\label{p_{tr,t}_2}
{p}_{tr,t} + {U}_{k}{p}_{tr,k} = -\frac{K_r + 5}{K_r + 3}p_{tr}{U}_{k,k} 
- \frac{K_v(p_{tr}-p_v)}{Z_v \tau(3+K_r+K_v)} + \mathcal{O}\left( \epsilon \right).
\end{equation}
Substituting the above expression into the calculation of $F_{jk}$, and decomposing the velocity gradient 
tensor $U_{k,j}$ into its symmetric trace-free (shear) and dilational parts, we obtain
\begin{equation}
\begin{aligned}
F_{jk} &= p_{tr}\left[ U_{k,j} + U_{j,k} - \frac{2}{3}U_{l,l} \right] \\
&\quad + \delta_{jk} \left[ \frac{2}{3}\left( \frac{K_r}{K_r + 3} \right)p_{tr} U_{l,l}
- \frac{K_v(p_{tr} - p_v)}{Z_v \tau (3 + K_r + K_v)} \right].
\end{aligned}
\end{equation}
The second term on the right-hand side corresponds to the bulk viscosity, which vanishes for monoatomic 
gases ($K_r = 0$).

For $\alpha = 4$, we write
\begin{equation}
{\mathcal{R}}_{4} = {\left( \widehat{\tau }{N}_{k}\right) }_{,k},
\end{equation}
where
\begin{equation}
{N}_{k} \equiv   < {u}_{k}\frac{\left( {u}_{n}^{2} + {\xi }^{2}\right) }{2}{ > }_{,t} 
+  < {u}_{k}{u}_{l}\frac{\left( {u}_{n}^{2} + {\xi }^{2}\right) }{2}{ > }_{,l},
\end{equation}
where $\xi^2 = \xi_t^2 + \xi_r^2 + \xi_v^2$.  
Let $u_i = w_i + U_i$, then the above equation can be written as  
$N_k = N_k^{(1)} + N_k^{(2)}$, where
\begin{equation}
{N}_{k}^{\left( 1\right) } 
= {[  {U}_{k}\frac{ < {u}_{n}{}^{2} + {\xi }^{2} > }{2}]  }_{,t} 
+ {[ {U}_{k} < {u}_{l}\frac{\left( {u}_{n}{}^{2} + {\xi }^{2}\right) }{2} > ]  }_{,l},
\end{equation}
and
\begin{equation}
{N}_{k}{}^{\left( 2\right) } =  < {w}_{k}\frac{{u}_{n}{}^{2} + {\xi }^{2}}{2}{ > }_{,t} 
+  < {w}_{k}{u}_{l}\frac{\left( {u}_{n}{}^{2} + {\xi }^{2}\right) }{2}{ > }_{,l}.
\end{equation}
For $N_k{}^{(1)}$, which can be simplified using $\mathcal{L}_4$ 
in Eq.~\eqref{L4_der}, we have
\begin{equation}
\begin{aligned}
	{N_{k}}^{(1)}
	&= U_{k}[\frac{<{u_{n}}^{2}+\xi^{2}>_{,t}}{2}+\frac{<u_{l}({u_{n}}^{2}+\xi^{2})>_{,l}}{2}] \\
	&\quad+U_{k,t}[\frac{<{u_{n}}^{2}+\xi^{2}>}{2}]+U_{k,l}[\frac{<u_{l}({u_{n}}^{2}+\xi^{2})>}{2}] \\
	&= U_{k}[\frac{<{u_{n}}^{2}+\xi^{2}>_{,t}}{2}+\frac{<u_{l}({u_{n}}^{2}+\xi^{2})>_{,l}}{2}] \\
	&\quad+U_{k,t}{\left( \frac{1}{2}\rho {U}_{n}{}^{2} + \frac{K_r + 3}{2}p_{tr}
	+ \frac{K_v}{2}p_v\right) } \\
	&\quad + U_{k,l}{\left( \frac{1}{2}\rho {U}_{l}{U}_{n}{}^{2} + \frac{K_r + 5}{2}p_{tr}{U}_{l}
	+ \frac{K_v}{2}p_v{U}_{l}\right) }.
\end{aligned}
\end{equation}
The coefficient of $U_k$ in the equation above is $\mathcal{L}_4$ in Eq.~\eqref{L4_der}
, and can therefore be neglected, and the remaining terms can be rewritten as
\begin{equation}
	(U_{k,t}+U_lU_{k,l})(\frac{1}{2}\rho{U_n}^2+\frac{K_r+3}{2}p_{tr}+\frac{K_v}{2}p_v)+p_{tr}U_lU_{k,l}.
\end{equation}
Then, using the fact that $\mathcal{L}_k = \mathcal{O}(\epsilon)$ and Eq.~\eqref{L2,3_simp}, we obtain
\begin{equation}
N_k{}^{(1)}=-p_{tr,k}(\frac{1}{2}U_n{}^2+\frac{K_r+3}{2}\frac{p_{tr}}{\rho}
+\frac{K_v}{2}\frac{p_v}{\rho})+p_{tr}U_lU_{k,l}.
\end{equation}
For $N_k^{(2)}$, using Eq.~\eqref{relation}, and noting that moments odd in $w_k$ vanish, we obtain
\begin{equation}
	\begin{aligned}
&{N}_{k}{}^{(2)} 
=< {w}_{k}\frac{{u}_{n}{}^{2} + {\xi }^{2}}{2}{ > }_{,t} 
+  < {w}_{k}{u}_{l}\frac{ {u}_{n}{}^{2} + {\xi }^{2} }{2}{ > }_{,l} \\
&=< {w}_{k}\frac{(U+w_x)^2+(V+w_y)^2+\xi^2}{2}{ > }_{,t} \\
&\quad +< {w}_{k}(U_l+w_l)\frac{ (U+w_x)^2+(V+w_y)^2 + {\xi }^{2} }{2}{ > }_{,l} \\
&=< {w}_{k}\frac{(U^2+2Uw_x+w_x^2)+(V^2+2Vw_y+w_y^2)+\xi^2}{2}{ > }_{,t} \\
&+< {w}_{k}(U_l+w_l)\frac{(U^2+2Uw_x+w_x^2)+(V^2+2Vw_y+w_y^2)+\xi^2}{2}{ > }_{,l} \\
&=< {w}_{k}\frac{2Uw_x+2Vw_y}{2}{ > }_{,t}
+< {w}_{k}U_l\frac{2Uw_x+2Vw_y}{2}{ > }_{,l} \\
&\quad +< {w}_{k}w_l\frac{U^2+w_x^2+V^2+w_y^2+\xi^2}{2}{ > }_{,l} \\
&=< {w}_{k}^2U_k{ > }_{,t}
+< {w}_{k}^2U_kU_l{ > }_{,l} \\
&\quad +< {w}_{k}^2\frac{U^2+V^2}{2}{ > }_{,k}+< {w}_{k}^2\frac{w_x^2+w_y^2}{2}{ > }_{,k} 
+< {w}_{k}^2\frac{\xi^2}{2}{ > }_{,k}.
    \end{aligned}
\end{equation}
Note that the second-to-last term on the RHS involves the fourth power of $w$. 
Next, we derive the fourth-order moment of $w$:
\begin{equation}
	\begin{aligned}
	<w^4>&=<(u-U)^4> \\
	&=<u^4+U^4+4u^2U^2+2u^2U^2-4u^3U-4uU^3> \\
	&=\rho\big[U^4+\frac{3U^2}{\lambda_{tr}}+\frac{3}{4\lambda_{tr}^2}+U^4
	+6U^2(U^2+\frac{1}{2\lambda_{tr}}) \\
	&\quad -4U(U^3+\frac{3}{2\lambda_{tr}}-4U^4)\big] \\
	&=\frac{3\rho}{4\lambda_{tr}^2} = \frac{3p_{tr}^2}{\rho}.
    \end{aligned}
\end{equation}
Substituting the above into the expression for $N_k^{(2)}$, we obtain:
\begin{equation}
	\begin{aligned}
{N}_{k}{}^{(2)} 
&=(p_{tr}U_k)_{,t}+(p_{tr}U_lU_k)_{,l}+\frac{1}{2}[p_{tr}(U^2+V^2)]_{,k} \\
&\quad +(\frac{3p_{tr}^2}{2\rho})_{,k}+(\frac{p_{tr}^2}{2\rho})_{,k}
+[\frac{(K_t+K_r)p_{tr}^2}{2\rho}]_{,k}+(\frac{K_vp_{tr}p_v}{2\rho})_{,k} \\
&=p_{tr}U_{k,t}+p_{tr,t}U_k+p_{tr,l}U_lU_k+p_{tr}U_{l,l}U_k+p_{tr}U_lU_{k,l} \\
&\quad + \frac{1}{2}p_{tr,k}U_n{}^2+p_{tr}U_lU_{l,k}+[\frac{(5+K_r)p_{tr}^2}{2\rho}]_{,k}
+(\frac{K_vp_{tr}p_v}{2\rho})_{,k} \\
&=p_{tr}[U_{k,t}+U_{l}U_{k,l}+U_{k}U_{l,l}+U_{l}U_{l,k}]\\
&\quad+U_{k}(p_{tr,t}+U_{l}p_{tr,l})+\frac{1}{2}p_{tr,k}U_n{}^2 \\
&\quad +\frac{5+K_r}{2}(\frac{p_{tr}^{2}}{\rho})_{,k}+(\frac{K_vp_{tr}p_v}{2\rho})_{,k}.
    \end{aligned}
\end{equation}
Using $\mathcal{L}_k = \mathcal{O}(\epsilon)$ and Eq.~\eqref{p_{tr,t}_2}, we remove the 
time derivatives (as they contribute only $\mathcal{O}(\epsilon)$), and obtain:
\begin{equation}
	\begin{aligned}
{N}_{k}{}^{(2)}
&=p_{tr}[-U_lU_{k,l}-\frac{p_{tr,k}}{\rho}+U_{l}U_{k,l}+U_{k}U_{l,l}+U_{l}U_{l,k}]\\
&\quad+U_k \big[-\frac{K_r + 5}{K_r + 3}p_{tr}{U}_{l,l} 
-\frac{K_v(p_{tr}-p_v)}{Z_v \tau(3+K_r+K_v)}\big] \\
&\quad+\frac{1}{2}p_{tr,k}U_n{}^2
+\frac{5+K_r}{2}(\frac{p_{tr}^{2}}{\rho})_{,k}+(\frac{K_vp_{tr}p_v}{2\rho})_{,k} \\
&=p_{tr}[-\frac{p_{tr,k}}{\rho}+U_{k}U_{l,l}+U_{l}U_{l,k}]\\
&\quad+U_k \big[-\frac{K_r + 5}{K_r + 3}p_{tr}{U}_{l,l} 
-\frac{K_v(p_{tr}-p_v)}{Z_v \tau(3+K_r+K_v)}\big] \\
&\quad+\frac{1}{2}p_{tr,k}U_n{}^2
+\frac{5+K_r}{2}(\frac{p_{tr}^{2}}{\rho})_{,k}+(\frac{K_vp_{tr}p_v}{2\rho})_{,k} \\
&=p_{tr}[-\frac{p_{tr,k}}{\rho}+U_{k}U_{l,l}+U_{l}U_{l,k}]\\
&\quad+U_k \big[-\frac{K_r + 5}{K_r + 3}p_{tr}{U}_{l,l} 
-\frac{K_v(p_{tr}-p_v)}{Z_v \tau(3+K_r+K_v)}\big] \\
&\quad+\frac{1}{2}p_{tr,k}U_n{}^2
+\frac{5+K_r}{2}(\frac{p_{tr}^{2}}{\rho})_{,k}+(\frac{K_vp_{tr}p_v}{2\rho})_{,k}. \\
    \end{aligned}
\end{equation}
Finally, combining $N_k^{(1)}$ and $N_k^{(2)}$, we have:
\begin{equation}
	\begin{aligned}
{N}_{k} 
&=-p_{tr,k}(\frac{1}{2}U_n{}^2+\frac{K_r+3}{2}\frac{p_{tr}}{\rho}
+\frac{K_v}{2}\frac{p_v}{\rho})+p_{tr}U_lU_{k,l} \\
&\quad +p_{tr}[-\frac{p_{tr,k}}{\rho}+U_{k}U_{l,l}+U_{l}U_{l,k}] \\
&\quad+U_k \big[-\frac{K_r + 5}{K_r + 3}p_{tr}{U}_{l,l} 
-\frac{K_v(p_{tr}-p_v)}{Z_v \tau(3+K_r+K_v)}\big] \\
&\quad+\frac{1}{2}p_{tr,k}U_n{}^2+\frac{5+K_r}{2}(\frac{p_{tr}^{2}}{\rho})_{,k}
+(\frac{K_vp_{tr}p_v}{2\rho})_{,k} \\
&=p_{tr}[U_l(U_{k,l}+U_{l,k})-\frac{2}{K_r+3}U_{k}U_{l,l}]
-U_k\frac{K_v(p_{tr}-p_v)}{Z_v \tau(3+K_r+K_v)} \\
&\quad+\frac{(5+K_r)}{2}p_{tr}(\frac{p_{tr}}{\rho})_{,k}
+\frac{K_vp_{tr}}{2}(\frac{p_v}{\rho})_{,k}. 
    \end{aligned}
\end{equation}
For $\alpha = 5$, we write
\begin{equation}
		\mathcal{R}_5=(\widehat{\tau}M_k)_{,k},
\end{equation}
where
\begin{equation}
\begin{aligned}
	M_k&\equiv   < {u}_{k}\frac{{\xi_v }^{2}}{2}{ > }_{,t} 
	+  < {u}_{k}{u}_{l}\frac{{\xi_v }^{2}}{2}{ > }_{,l} \\
	&=(\frac{K_v}{2}p_vU_k)_{,t}+(\frac{K_v}{2}p_vU_kU_l)_{,l}
	+(\frac{K_v}{2}\frac{p_v}{\rho}p_{tr}\delta_{kl})_{,l} \\
	&=\frac{K_v}{2}(p_{v,t}U_k+p_vU_{k,t}+p_{v,l}U_kU_l+p_vU_{k,l}U_l+p_vU_kU_{l,l} \\
	&\quad +p_{tr}\delta_{kl}(\frac{p_v}{\rho})_{,l}+\frac{p_v}{\rho}p_{tr,k}) \\
	&=\frac{K_v}{2}\big[\frac{p_v}{\rho}(\rho U_{k,t}+\rho U_{k,l}U_l+p_{tr,k}) \\
	&\quad +U_k(p_{v,t}+p_{v,l}U_l+p_vU_{l,l})+p_{tr}(\frac{p_v}{\rho})_{,k} \big].
\end{aligned}
\end{equation}
The coefficient of $p_v/\rho$ is $\mathcal{L}_j$ (see Eq.~\eqref{L2,3_simp}) and can be neglected. 
The coefficient of $U_k$ can be simplified using Eq.~\eqref{p_{v,t}}, yielding:
\begin{equation}
		M_k=\frac{K_v}{2}\big[\frac{U_k}{Z_v \tau}\frac{(3+K_r)(p_{tr}-p_v)}{3+K_r+K_v}
		+p_{tr}(\frac{p_v}{\rho})_{,k}\big].
\end{equation}
All time derivatives have now been eliminated from $\mathcal{R}_\alpha$ for all $\alpha$.
The remaining steps in the derivation of the NS equations can now be summarized concisely as follows:

\noindent 1). Drop the $\mathcal{O}(\epsilon)$ terms in Eq.~\eqref{L=R+S2}.

\noindent 2). Combine $\epsilon$ and $\widehat{\tau}$ to recover $\tau = \epsilon \widehat{\tau}$.

Finally, the NS equations derived from the 2D translational-rotational vibrational two-temperature 
BGK model are
\begin{equation}
    \frac{\partial W}{\partial t}+\frac{\partial F}{\partial x}+\frac{\partial G}{\partial y}
    =\frac{\partial F_{v}}{\partial x}+\frac{\partial G_{v}}{\partial y}+S,
\end{equation}
with
\begin{equation}
	\begin{gathered}
    W=\begin{pmatrix}\rho\\ \rho U\\ \rho V\\ \rho E\\\rho E_v \end{pmatrix}
    F=\begin{pmatrix}\rho U\\ \rho U^2+p\\ \rho U V\\ \left (  \rho E+p \right )U\\
		\rho E_v U\end{pmatrix}
    G=\begin{pmatrix}\rho V\\ \rho U V\\ \rho V^2+p\\ (\rho E+p)V\\\rho E_v V\end{pmatrix} \\
    F_v=\begin{pmatrix}0\\ \tau_{xx}\\ \tau_{xy}\\ U\tau_{xx}+V\tau_{xy}+q_x\\U\tau_{tr-v}+q_{vx}
	\end{pmatrix}
    G_v=\begin{pmatrix}0\\ \tau_{yx}\\ \tau_{yy}\\ U\tau_{yx}+V\tau_{yy}+q_y\\V\tau_{tr-v}+q_{vy},
	\end{pmatrix}
    \end{gathered}
\end{equation}
Here, the total energy is defined as
\begin{equation}
\rho E = \frac{1}{2} \rho \left( U^2 + (3 + K_r) R T_{tr} + K_v R T_v \right),
\end{equation}
and the vibrational energy as
\begin{equation}
\rho E_v = \frac{K_v}{2} \rho R T_v.
\end{equation}
The viscous normal stress terms are
\begin{equation}
    \begin{aligned}
    \tau_{xx} &=\tau p[2\frac{\partial U}{\partial x}
    -\frac{2}{3+K_r}(\frac{\partial U}{\partial x}+\frac{\partial V}{\partial y})] \\
    &\quad -\frac{\rho K_v}{2(K_r+K_v+3)Z_v}(\frac{1}{\lambda_{tr}}-\frac{1}{\lambda_{v}}), \\
    \tau_{yy} &=\tau p[2\frac{\partial V}{\partial y}
    -\frac{2}{3+K_r}(\frac{\partial U}{\partial x}+\frac{\partial V}{\partial y})] \\
    &\quad -\frac{\rho K_v}{2(K_r+K_v+3)Z_v}(\frac{1}{\lambda_{tr}}-\frac{1}{\lambda_{v}}), 
    \end{aligned}
\end{equation}
the viscous shear stress term is
\begin{equation}
	\tau_{xy}=\tau_{yx}=\tau p(\frac{\partial U}{\partial y}+\frac{\partial V}{\partial x}), 
\end{equation}
and the heat conduction terms are
\begin{equation}
    \begin{gathered}
    q_{x}=\tau p [\frac{K_v}{4}\frac{\partial}{\partial x}(\frac{1}{\lambda_v})
    +\frac{5+K_r}{4}\frac{\partial}{\partial x}(\frac{1}{\lambda_{tr}})], \\
    q_{y}=\tau p[\frac{K_v}{4}\frac{\partial}{\partial y}(\frac{1}{\lambda_{v}})
    +\frac{5+K_r}{4}\frac{\partial}{\partial y}(\frac{1}{\lambda_{tr}})].
    \end{gathered}
\end{equation}
The following terms relate to the governing equation of vibrational energy $\rho E_v$:
\begin{equation}
	\begin{gathered}
    \tau_{tr-v}=\frac{(K_r+3)\rho K_v}{4(K_r+K_v+3)Z_v}(\frac{1}{\lambda_{tr}}-\frac{1}{\lambda_v}), \\
    q_{vx}=\tau p \frac{K_v}{4}\frac{\partial}{\partial x}(\frac{1}{\lambda_v}), \\
    q_{vy}=\tau p \frac{K_v}{4}\frac{\partial}{\partial y}(\frac{1}{\lambda_v}).
    \end{gathered}
\end{equation}
The source term is
\begin{equation}
	\mathbf{S}=(0,0,0,0,s)^{T},
\end{equation}
with
\begin{equation}
	\begin{aligned}
		s&=\frac{(\rho E_v)^{eq}-\rho E_v}{Z_v\tau}, \\
	\rho E_{v}^{eq}&=\frac{K_{v}}{2}\rho RT^{eq},\\
	T^{eq}&=\frac{(3+K_{r}) T_{tr}+K_{v} T_{v}}{3+K_{r}+K_{v}}.
    \end{aligned}
\end{equation}

\renewcommand{\theequation}{C\arabic{equation}}
\setcounter{equation}{0}
\section*{Appendix C: Derivation of the microscopic slopes}\label{appendixC}
Once the reconstruction for macroscopic flow derivatives is finished, the microscopic derivatives 
$a^{l,r,c}$ and $A^{l,r,c}$ in Eq.~\eqref{f_all} can be obtained as follows.  
From the Taylor expansion of the Maxwellian distribution, all microscopic derivatives take the form
\begin{equation}
	\begin{aligned}
	a_x&=a_{x1}+a_{x2}u+a_{x3}v+a_{x4}(u^2+v^2+\xi_t^2+\xi_r^2)+a_{x5} \xi_v^2 \\
	&=a_{x \beta}\omega_{\beta}, \\
	a_y&=a_{y1}+a_{y2}u+a_{y3}v+a_{y4}(u^2+v^2+\xi_t^2+\xi_r^2)+a_{y5} \xi_v^2 \\
	&=a_{y \beta}\omega_{\beta}.
    \end{aligned}
\end{equation}
Based on the relation between the macroscopic variables and the microscopic gas distribution function 
given in Eq.~\eqref{macro&micro}, taking the derivative with respect to \( x \) gives:
\begin{equation}
	\frac{\partial W}{\partial x}=\int_{-\infty}^\infty\psi_\alpha a_x g \mathrm{d}u
	\mathrm{d}v \mathrm{d}\xi_t \mathrm{d}\xi_r \mathrm{d}\xi_v,
\end{equation}
where $\psi_\alpha = (1, u, v, \frac{1}{2}(u^2 + v^2 + \xi_t^2 + \xi_r^2 + \xi_v^2), 
\frac{1}{2} \xi_v^2)^T$. This leads to the equation
\begin{equation}
\begin{gathered}
\left( \begin{array}{l} {b}_{1} \\  {b}_{2} \\  {b}_{3} \\  {b}_{4} \\  {b}_{5} \end{array}\right)
= \frac{1}{\rho }\frac{\partial W}{\partial x} = \frac{1}{\rho }
\left( \begin{matrix} 
	\frac{\partial \rho }{\partial x} \\  
	\frac{\partial \left( {\rho U}\right) }{\partial x} \\  
	\frac{\partial \left( {\rho V}\right) }{\partial x} \\  
	\frac{\partial \left( {\rho E}\right) }{\partial x} \\  
	\frac{\partial \left( {\rho E_v}\right) }{\partial x} 
\end{matrix}\right)  \\
= \left\langle  \left| {\psi_{\alpha} a_x}\right| \right\rangle  
= \left\langle  \left| {{\psi }_{\alpha }{\zeta }_{\beta }{\alpha }_{x \beta }}\right| \right\rangle   
= \left\langle  \left| {{\psi }_{\alpha }{\zeta}_{\beta }}\right| \right\rangle  
\left( \begin{array}{l} {a}_{x1} \\  {a}_{x2} \\  {a}_{x3} \\  {a}_{x4} \\  {a}_{x5} \end{array}\right).
\end{gathered}
\end{equation}
Letting $\mathbf{M} = \left\langle \left|\psi_\alpha \zeta_\beta\right| \right\rangle$, 
this becomes a linear system
\begin{equation}\label{Ma=b}
\mathbf{M} \mathbf{a} = \mathbf{b},
\end{equation}
where the matrix $\mathbf{M}$ is given by
\begin{widetext}
\begin{equation}
\mathbf{M}=\begin{pmatrix}\langle|1|\rangle&\langle|u^1|\rangle&\langle|v^1|\rangle&
	\langle|u^2+v^2+\xi_t^2+\xi_r^2|\rangle&\langle|\xi_v^2|\rangle\\
	\langle|u^1|\rangle&\langle|u^2|\rangle&\langle|u^1v^1|\rangle&
	\langle|u^1(u^2+v^2+\xi_t^2+\xi_r^2)|\rangle&\langle|u^1\xi_v^2|\rangle\\
	\langle|v^1|\rangle&\langle|u^1v^1|\rangle&\langle|v^2|\rangle&
	\langle|v^1(u^2+v^2+\xi_t^2+\xi_r^2)|\rangle&\langle|v^1\xi_v^2|\rangle\\
	\langle|\psi_4|\rangle&\langle|\psi_4u^1|\rangle&\langle|\psi_4v^1|\rangle&
	\langle|\psi_4(u^2+v^2+\xi_t^2+\xi_r^2)|\rangle&\langle|\psi_4\xi_v^2|\rangle\\
	\langle|\frac{1}{2}\xi_v^2|\rangle&\langle|\frac{1}{2}\xi_v^2u^1|\rangle&
	\langle|\frac{1}{2}\xi_v^2v^1|\rangle&\langle|\frac{1}{2}\xi_v^2(u^2+v^2+\xi_t^2+\xi_r^2)|\rangle&
	\langle|\frac{1}{2}\xi_v^4|\rangle\end{pmatrix}.
\end{equation}
\end{widetext}
Now define the following quantities:
\begin{equation}
    \begin{aligned}
        &B=2\frac{1}{\rho}\frac{\partial(\rho E-\rho E_{v})}{\partial x}
        -\frac{1}{\rho}(U^{2}+V^{2}+\frac{K_r+3}{2\lambda_{tr}})\frac{\partial\rho}{\partial x},\\
        &A_1=\frac{1}{\rho}\frac{\partial(\rho U)}{\partial x}
        -\frac{U}{\rho}\frac{\partial\rho}{\partial x}, \\
        &A_2=\frac{1}{\rho}\frac{\partial(\rho V)}{\partial x}
        -\frac{V}{\rho}\frac{\partial\rho}{\partial x}.
    \end{aligned}
\end{equation}
Then the solution of Eq.~\eqref{Ma=b} becomes
\begin{equation}
    \begin{aligned}
        &a_{x5} =2\frac{1}{\rho}\frac{\lambda_{v}^{2}}{K_v}(2\frac{\partial(\rho E_{v})}{\partial x}
        -\frac{1}{2}\frac{K_v}{\lambda_{v}}\frac{\partial\rho}{\partial x}),  \\
        &a_{x4} =\frac{2\lambda_{tr}^{2}}{K_r+3}(B-2UA_1-2VA_2),  \\
        &a_{x3} =2\lambda_{tr}A_2-2Va_{x4},  \\
        &a_{x2} =2\lambda_{tr}A_1-2Ua_{x4},  \\
        &a_{x1} =\frac{1}{\rho}\frac{\partial\rho}{\partial x}-a_{x2}U-a_{x3}V \\
        &\quad -a_{x4}(U^{2}+V^{2}+\frac{K_r+3}{2\lambda_{tr}})-a_{x5}\frac{K_v}{2\lambda_{v}}.
        \end{aligned}
\end{equation}
Hence, once the macroscopic variables and their derivatives are reconstructed, the microscopic 
first-order spatial derivatives can be calculated. The $y$-direction derivatives can be obtained 
in a similar manner. We then proceed to compute $A^{l,r,c}$.
For clarity, the non-equilibrium BGK model is restated as:
\begin{equation}
	\frac{\partial f}{\partial t} + u\frac{\partial f}{\partial x} + v\frac{\partial f}{\partial y} 
	= \frac{{f}^{eq} - f}{\tau } + \frac{g - {f}^{eq}}{Z_v\tau } 
	= \frac{{f}^{eq} - f}{\tau } + {Q}_{s}.
\end{equation}
Here, the collision operator on the right-hand side consists of elastic and inelastic terms. Since 
vibrational energy is assumed to remain frozen during a single collision time, the inelastic term is 
neglected. Moreover, elastic collisions preserve internal energy modes, and thus vibrational energy 
is conserved. In this context, mass, momentum, and total energy conservation lead to the following 
compatibility condition:
\begin{equation}\label{cc1}
	\int\frac{f^{eq}-f}{\tau}\psi_\alpha \mathrm{d}u\mathrm{d}v\mathrm{d}\xi_t \mathrm{d}\xi_r 
	\mathrm{d}\xi_v=0.
\end{equation}
Under the zeroth-order Chapman--Enskog expansion, the distribution function is approximated by the 
intermediate equilibrium state:
\begin{equation}\label{CE_B}
f = f^{eq}.
\end{equation}
Taking moments of the BGK equation with respect to $\psi$, and applying the compatibility condition 
in Eq.~\eqref{cc1}, we obtain
\begin{equation}
	\langle a_x u+a_y v+A\rangle=0,
\end{equation}
from which the coefficient \( A \) can be explicitly solved as:
\begin{equation}
	\langle A \rangle = -\langle a_x u+a_y v\rangle.
\end{equation}

\bibliographystyle{elsarticle-num}
\bibliography{gaobib}

\begin{thebibliography}{10}
\expandafter\ifx\csname url\endcsname\relax
  \def\url#1{\texttt{#1}}\fi
\expandafter\ifx\csname urlprefix\endcsname\relax\def\urlprefix{URL }\fi
\expandafter\ifx\csname href\endcsname\relax
  \def\href#1#2{#2} \def\path#1{#1}\fi

\bibitem{Anderson1989}
J.~D. Anderson, Hypersonic and High Temperature Gas Dynamics, AIAA, 1989.

\bibitem{Bose2013}
D.~Bose, J.~L. Brown, D.~K. Prabhu, P.~Gnoffo, C.~O. Johnston, B.~Hollis, Uncertainty assessment of hypersonic aerothermodynamics prediction capability, Journal of Spacecraft and Rockets 50~(1) (2013) 12--18.

\bibitem{Olynick1999}
D.~Olynick, Y.-K. Chen, M.~E. Tauber, Aerothermodynamics of the stardust sample return capsule, Journal of Spacecraft and Rockets 36~(3) (1999) 442--462.
\newblock \href {https://doi.org/10.2514/2.3466} {\path{doi:10.2514/2.3466}}.

\bibitem{Li2017}
H.~Li, A.~Shi, P.~Ma, et~al., Recent advances in hypersonic nonequilibrium flows, in: C.~S. of~Mechanics, B.~I. of~Technology (Eds.), Proceedings of the 2017 Chinese Mechanics Conference: Celebrating the 60th Anniversary of the Chinese Society of Mechanics (Volume B), China Aerodynamics Research and Development Center, Beijing, China, 2017, pp. 56--96, in Chinese.

\bibitem{Kane2022}
A.~A. Kane, R.~K. Peetala, Influence of vibration--dissociation coupling and number of reactions in hypersonic nonequilibrium flows, Journal of Fluids Engineering 144~(8) (2022) 081207.
\newblock \href {https://doi.org/10.1115/1.4053650} {\path{doi:10.1115/1.4053650}}.

\bibitem{Dong1996}
W.~Dong, Numerical simulation and analysis of thermochemical nonequilibrium effects on hypersonic flows, Master's thesis, Beihang University, in Chinese (1996).

\bibitem{Wang2017}
J.~Wang, Numerical Study on Coupled Effects of the Chemical Nonequilibrium and Thermal Radiation in High-Speed and High-Temperature Flows, Beihang University, Beijing, 2017, in Chinese.

\bibitem{Hao2018}
J.~Hao, Modeling of Thermochemical Nonequilibrium Coupling Effects in Hypersonic Flows, Beihang University, Beijing, 2018, in Chinese.

\bibitem{Candler2019}
G.~V. Candler, Rate effects in hypersonic flows, Annual Review of Fluid Mechanics 51~(1) (2019) 379--402.

\bibitem{Appleton1964}
J.~Appleton, K.~Bray, The conservation equations for a non-equilibrium plasma, Journal of Fluid Mechanics 20~(4) (1964) 659--672.

\bibitem{Park1989-1}
C.~Park, Assessment of two-temperature kinetic model for ionizing air, Journal of thermophysics and heat transfer 3~(3) (1989) 233--244.

\bibitem{Park1989-2}
C.~Park, W.~Griffith, Nonequilibrium hypersonic aerothermodynamics, Physics Today 44~(2) (1991) 98--98.
\newblock \href {https://doi.org/10.1063/1.2809999} {\path{doi:10.1063/1.2809999}}.

\bibitem{Li2023}
S.~Li, Z.~Sun, B.~Zha, Y.~Zhu, Y.~Ding, Y.~Xia, A family of spatio-temporal optimized finite difference schemes with adaptive dispersion and critical-adaptive dissipation for compressible flows, Journal of Computational Physics 474 (2023) 111821.

\bibitem{Li2022}
S.~Li, Y.~Hu, Z.~Sun, Y.~Shi, K.~Mao, A high-resolution finite volume scheme based on optimal spectral properties of the fully discrete scheme with minimized dispersion and adaptive dissipation, Computers \& Fluids 233 (2022) 105226.

\bibitem{He2023}
X.~He, K.~Wang, T.~Liu, Y.~Feng, B.~Zhang, W.~Yuan, X.~Wang, Hodg: high-order discontinuous galerkin methods for solving compressible euler and navier-stokes equations-an open-source component-based development framework, Computer Physics Communications 286 (2023) 108660.

\bibitem{Xu1998}
K.~Xu, Gas-kinetic schemes for unsteady compressible flow simulations, Computational Fluid Dynamics, Annual Lecture Series, 29 th, Rhode-Saint-Genese, Belgium (1998).

\bibitem{Xiao2019}
T.~Xiao, K.~Xu, Q.~Cai, A unified gas-kinetic scheme for multiscale and multicomponent flow transport, Applied Mathematics and Mechanics 40~(3) (2019) 355--372.

\bibitem{Liu2019}
C.~Liu, G.~Zhou, W.~Shyy, K.~Xu, Limitation principle for computational fluid dynamics, Shock Waves 29~(8) (2019) 1083--1102.

\bibitem{Pan2020}
L.~Pan, F.~Zhao, K.~Xu, High-order {ALE} gas-kinetic scheme with {WENO} reconstruction, Journal of Computational Physics 417 (2020) 109558.

\bibitem{Li2020}
Q.~Li, An improved gas-kinetic scheme for multimaterial flows, Commun. Comput. Phys. 27~(1) (2020) 145--166.

\bibitem{Pan2017}
L.~Pan, J.~Li, K.~Xu, A few benchmark test cases for higher-order {Euler} solvers, Numerical Mathematics: Theory, Methods and Applications 10~(4) (2017) 711--736.

\bibitem{Xu2014}
K.~Xu, Direct Modeling for Computational Fluid Dynamics: Construction and Application of Unified Gas-Kinetic Schemes, Vol.~4, World Scientific, 2014.

\bibitem{Xu2004}
K.~Xu, Z.~Li, Microchannel flow in the slip regime: gas-kinetic {BGK-Burnett} solutions, Journal of Fluid Mechanics 513 (2004) 87--110.

\bibitem{Xu2017}
K.~Xu, C.~Liu, A paradigm for modeling and computation of gas dynamics, Physics of Fluids 29~(2) (2017).

\bibitem{Hou2018}
S.~Hou, Z.~Li, X.~Jiang, S.~Zeng, Numerical study on two-dimensional micro-channel flows using the gas-kinetic unified algorithm, Commun. Comput. Phys. 23~(5) (2018) 1393--1414.

\bibitem{Li2011}
J.~Li, Q.~Li, K.~Xu, Comparison of the generalized {Riemann} solver and the gas-kinetic scheme for inviscid compressible flow simulations, Journal of Computational Physics 230~(12) (2011) 5080--5099.

\bibitem{Xu2002}
K.~Xu, A well-balanced gas-kinetic scheme for the shallow-water equations with source terms, Journal of Computational Physics 178~(2) (2002) 533--562.

\bibitem{Xu1995}
K.~Xu, L.~Martinelli, A.~Jameson, Gas-kinetic finite volume methods, flux-vector splitting, and artificial diffusion, Journal of computational physics 120~(1) (1995) 48--65.

\bibitem{Li2009}
Z.-H. Li, H.-X. Zhang, Gas-kinetic numerical studies of three-dimensional complex flows on spacecraft re-entry, Journal of Computational Physics 228~(4) (2009) 1116--1138.

\bibitem{Tang2005}
L.~Tang, Y.~Zheng, D.~Liu, Gas-kinetic scheme for hypersonic plasma aerodynamics, in: AIAA/CIRA 13th International Space Planes and Hypersonics Systems and Technologies Conference, 2005, p. 3220.

\bibitem{Liao2007}
W.~Liao, L.~Luo, K.~Xu, Gas-kinetic scheme for continuum and near-continuum hypersonic flows, Journal of Spacecraft and Rockets 44~(6) (2007) 1232--1240.

\bibitem{Ong2006}
J.~C. Ong, A.~Omar, W.~Asrar, Z.~Zaludin, Gas-kinetic {BGK} scheme for hypersonic flow simulation, in: 44th AIAA Aerospace Sciences Meeting and Exhibit, 2006, p. 990.

\bibitem{Li2005}
Q.~Li, S.~Fu, K.~Xu, Application of gas-kinetic scheme with kinetic boundary conditions in hypersonic flow, AIAA Journal 43~(10) (2005) 2170--2176.

\bibitem{Li2010}
Q.~Li, K.~Xu, S.~Fu, A high-order gas-kinetic {Navier-Stokes} flow solver, Journal of Computational Physics 229~(19) (2010) 6715--6731.

\bibitem{Cao2019}
G.~Cao, H.~Su, J.~Xu, K.~Xu, Implicit high-order gas kinetic scheme for turbulence simulation, Aerospace Science and Technology 92 (2019) 958--971.

\bibitem{Cao2019-3T}
G.~Cao, L.~Pan, K.~Xu, Three dimensional high-order gas-kinetic scheme for supersonic isotropic turbulence {I}: criterion for direct numerical simulation, Computers \& Fluids 192 (2019) 104273.

\bibitem{Cao2021}
G.~Cao, L.~Pan, K.~Xu, Three dimensional high-order gas-kinetic scheme for supersonic isotropic turbulence {II}: Coarse-graining analysis of compressible {Ksgs} budget, Journal of Computational Physics 439 (2021) 110402.

\bibitem{Kumar2013}
G.~Kumar, S.~S. Girimaji, J.~Kerimo, {WENO}-enhanced gas-kinetic scheme for direct simulations of compressible transition and turbulence, Journal of Computational Physics 234 (2013) 499--523.

\bibitem{Righi2014}
M.~Righi, A modified gas-kinetic scheme for turbulent flow, Communications in Computational Physics 16~(1) (2014) 239--263.

\bibitem{Li2010-2}
Q.~Li, S.~Tan, S.~Fu, K.~Xu, Numerical simulation of compressible turbulence with gas-kinetic {BGK} scheme, in: 13th Asian Congress of Fluid Mechanics, Dhaka, Bangladesh, 2010, pp. 12--17.

\bibitem{Li2014}
Q.~Li, S.~Fu, High-order accurate gas-kinetic scheme and turbulence simulation, SCIENTIA SINICA Physica, Mechanica \& Astronomica 44~(3) (2014) 278.

\bibitem{Tan2018}
S.~Tan, Q.~Li, S.~Fu, Gas-kinetic scheme for multiscale turbulence simulation, in: Progress in Hybrid RANS-LES Modelling: Papers Contributed to the 6th Symposium on Hybrid RANS-LES Methods, 26-28 September 2016, Strasbourg, France 6, Springer, 2018, pp. 135--142.

\bibitem{Tan2011}
S.~Tan, Q.~Li, S.~Fu, S.~Zeng, Engineering simulation of turbulence with gas-kinetic {BGK} scheme, in: AIP Conference Proceedings, Vol. 1376, American Institute of Physics, 2011, pp. 78--80.

\bibitem{Xu2005-Trans}
K.~Xu, E.~Josyula, Multiple translational temperature model and its shock structure solution, Physical Review E 71~(5) (2005).
\newblock \href {https://doi.org/10.1103/PhysRevE.71.056308} {\path{doi:10.1103/PhysRevE.71.056308}}.

\bibitem{Xu2007-Trans}
K.~Xu, H.~Liu, J.~Jiang, Multiple-temperature kinetic model for continuum and near continuum flows, Physics of Fluids 19~(1) (2007).

\bibitem{Xu2004-2T}
K.~Xu, L.~Tang, Nonequilibrium {Bhatnagar-Gross-Krook} model for nitrogen shock structure, Physics of Fluids 16~(10) (2004) 3824--3827.
\newblock \href {https://doi.org/10.1063/1.1783372} {\path{doi:10.1063/1.1783372}}.

\bibitem{Xu2006-2T}
K.~Xu, E.~Josyula, Continuum formulation for non-equilibrium shock structure calculation, Communications in computational physics 1~(3) (2006) 425--448.

\bibitem{Xu2008-2T}
K.~Xu, X.~He, C.~Cai, Multiple temperature kinetic model and gas-kinetic method for hypersonic non-equilibrium flow computations, Journal of Computational Physics 227~(14) (2008) 6779--6794.
\newblock \href {https://doi.org/10.1016/j.jcp.2008.03.035} {\path{doi:10.1016/j.jcp.2008.03.035}}.

\bibitem{Cao2018-2T}
G.~Cao, H.~Liu, K.~Xu, Physical modeling and numerical studies of three-dimensional non-equilibrium multi-temperature flows, Physics of Fluids 30~(12) (2018).

\bibitem{Cai2008-3T}
C.~Cai, D.~D. Liu, K.~Xu, One-dimensional multiple-temperature gas-kinetic {Bhatnagar-Gross-Krook} scheme for shock wave computation, AIAA Journal 46~(5) (2008) 1054--1062.
\newblock \href {https://doi.org/10.2514/1.27432} {\path{doi:10.2514/1.27432}}.

\bibitem{Liu2021-3T}
H.~Liu, G.~Cao, W.~Chen, Multiple-temperature gas-kinetic scheme for type {IV} shock/shock interaction, Communication in Computational Physics (2021) 853.

\bibitem{Cao2022-3T}
G.~Cao, Y.~Shi, K.~Xu, S.~Chen, Modeling and simulation in supersonic three-temperature carbon dioxide turbulent channel flow, Physics of Fluids 34~(12) (2022).
\newblock \href {https://doi.org/10.1063/5.0129353} {\path{doi:10.1063/5.0129353}}.

\bibitem{Benettin1997}
G.~Benettin, A.~Carati, G.~Gallavotti, A rigorous implementation of the {Jeans - Landau - Teller} approximation for adiabatic invariants, Nonlinearity 10~(2) (1997) 479.
\newblock \href {https://doi.org/10.1088/0951-7715/10/2/011} {\path{doi:10.1088/0951-7715/10/2/011}}.

\bibitem{Bhatnagar1954}
P.~L. Bhatnagar, E.~P. Gross, M.~Krook, A model for collision processes in gases. {I}. small amplitude processes in charged and neutral one-component systems, Physical Review 94~(3) (1954) 511--525.
\newblock \href {https://doi.org/10.1103/PhysRev.94.511} {\path{doi:10.1103/PhysRev.94.511}}.

\bibitem{Bird1994}
G.~A. Bird, Molecular Gas Dynamics and the Direct Simulation of Gas Flows, Oxford University Press, 1994.
\newblock \href {https://doi.org/10.1093/oso/9780198561958.001.0001} {\path{doi:10.1093/oso/9780198561958.001.0001}}.

\bibitem{Wang2017-UGKS}
Z.~Wang, H.~Yan, Q.~Li, K.~Xu, Unified gas-kinetic scheme for diatomic molecular flow with translational, rotational, and vibrational modes, Journal of Computational Physics 350 (2017) 237--259.

\bibitem{Chapman1970}
S.~Chapman, T.~G. Cowling, The Mathematical Theory of Non-Uniform Gases, 3rd Edition, Cambridge University Press, Cambridge, 1970.

\bibitem{Sutherland1893}
W.~Sutherland, {LII}. the viscosity of gases and molecular force, The London, Edinburgh, and Dublin Philosophical Magazine and Journal of Science 36~(223) (1893) 507--531.

\bibitem{Xu2001}
K.~Xu, A gas-kinetic {BGK} scheme for the {Navier-Stokes} equations and its connection with artificial dissipation and {Godunov} method, Journal of Computational Physics 171~(1) (2001) 289--335.
\newblock \href {https://doi.org/10.1006/jcph.2001.6790} {\path{doi:10.1006/jcph.2001.6790}}.

\bibitem{Ji2021}
X.~Ji, W.~Shyy, K.~Xu, A gradient compression-based compact high-order gas-kinetic scheme on {3D} hybrid unstructured meshes, International Journal of Computational Fluid Dynamics 35~(7) (2021) 485--509.

\bibitem{Lockerby2004}
D.~A. Lockerby, J.~M. Reese, D.~R. Emerson, R.~W. Barber, Velocity boundary condition at solid walls in rarefied gas calculations, Physical Review E—Statistical, Nonlinear, and Soft Matter Physics 70~(1) (2004) 017303.

\bibitem{Maccormack1989}
R.~Maccormack, Nonequilibrium effects for hypersonic transitional flows using continuum approach, in: 27th Aerospace sciences meeting, 1989, p. 461.

\bibitem{Maccormack1987}
R.~Maccormack, D.~Chapman, Computational fluid dynamics near the continuum limit, in: 8th Computational Fluid Dynamics Conference, 1987, p. 1115.

\bibitem{Saad1985}
M.~A. Saad, Compressible fluid flow, Englewood Cliffs (1985).

\bibitem{Liepmann2001}
H.~W. Liepmann, A.~Roshko, Elements of Gasdynamics, Courier Corporation, 2001.

\bibitem{Vincenti1966}
W.~G. Vincenti, C.~H. Kruger~Jr, T.~Teichmann, Introduction to physical gas dynamics, American Institute of Physics, 1966.

\bibitem{Guo2007}
C.~Guo, Calculation of shock layer thickness, Master's thesis, Inner Mongolia University, in Chinese (2007).
\newblock \href {https://doi.org/10.7666/d.y1156438} {\path{doi:10.7666/d.y1156438}}.

\bibitem{Lofthouse2008}
A.~J. Lofthouse, \href{https://deepblue.lib.umich.edu/handle/2027.42/58370}{Nonequilibrium hypersonic aerothermodynamics using the direct simulation {Monte Carlo} and {Navier--Stokes} models}, Ph.D. thesis (2008).
\newline\urlprefix\url{https://deepblue.lib.umich.edu/handle/2027.42/58370}

\bibitem{Boyd1995}
I.~D. Boyd, G.~Chen, G.~V. Candler, Predicting failure of the continuum fluid equations in transitional hypersonic flows, Physics of fluids 7~(1) (1995) 210--219.

\bibitem{Lofthouse2007}
A.~J. Lofthouse, I.~D. Boyd, M.~J. Wright, Effects of continuum breakdown on hypersonic aerothermodynamics, Physics of Fluids 19~(2) (2007).

\bibitem{Martinez2002}
J.~Martinez~Schramm, K.~Hannemann, W.~Beck, S.~Karl, Cylinder shock layer density profiles measured in high enthalpy flows in {HEG}, in: 22nd AIAA Aerodynamic Measurement Technology and Ground Testing Conference, 2002, p. 2913.

\bibitem{Li2013}
L.~Hua, Numerical Simulation of Hypersonic Slip Flow Aerothermodynamics with Parallel Computing, National Defense Industry Press, Beijing, 2013, in Chinese.

\bibitem{Edney1968}
B.~Edney, Anomalous heat transfer and pressure distributions on blunt bodies at hypersonic speeds in the presence of an impinging shock, Report, Flygtekniska Forsoksanstalten, Stockholm (Sweden) (1968).

\bibitem{Hedde1995}
T.~Hedde, D.~Guffond, {ONERA} three-dimensional icing model, AIAA Journal 33~(6) (1995) 1038--1045.

\bibitem{Thibert2000}
J.~Thibert, D.~Arnal, A review of {ONERA} aerodynamic research in support of a future supersonic transport aircraft, Progress in Aerospace Sciences 36~(8) (2000) 581--627.

\end{thebibliography}

\end{document}